\documentclass[10pt]{article}
\usepackage{graphicx}
\usepackage{amsfonts}
\usepackage{amsmath}
\usepackage{amssymb}
\usepackage{url}
\usepackage{xcolor}
\usepackage{dsfont}
\usepackage{fancyhdr}
\usepackage{indentfirst}
\usepackage{enumerate}
\usepackage{tabularx}
\usepackage{booktabs,siunitx,multirow}
\usepackage[normalem]{ulem}
\usepackage{mathrsfs}
\usepackage{multirow}
\usepackage{diagbox}
\usepackage[margin=1in]{geometry}
 \usepackage[colorlinks=true,citecolor=blue]{hyperref}
\usepackage{amsthm}
\usepackage{color}
\definecolor{DarkGreen}{rgb}{0.2,0.6,0.2}

\definecolor{purple}{rgb}{0.6,0.3,0.8}

\usepackage{natbib}
\usepackage{comment}

\def\d{\mathrm{d}}

\newcommand{\E}{\mathbb{E}}

\newcommand{\R}{\mathbb{R}}

\newcommand{\p}{\mathbb{P}}

\newcommand{\A}{\mathcal{A}}

\newcommand{\M}{\mathcal{M}}

\newcommand{\id}{\mathds{1}}

\renewcommand{\ge}{\geqslant}

\renewcommand{\geq}{\geqslant}
\renewcommand{\leq}{\leqslant}
\renewcommand{\epsilon}{\varepsilon}

\renewcommand{\cdots}{\dots}
\theoremstyle{plain}
\newtheorem{theorem}{Theorem}
\newtheorem{corollary}{Corollary}
\newtheorem{lemma}{Lemma}
\newtheorem{proposition}{Proposition}
\theoremstyle{definition}
\newtheorem{definition}{Definition}
\newtheorem{example}{Example}

\theoremstyle{remark}
\newtheorem{remark}{Remark}

\DeclareMathOperator*{\argmin}{arg\,min}

\newcommand{\VaR}{\mathrm{VaR}}

\newcommand{\ES}{\mathrm{ES}}
\newcommand{\GlueVaR}{\mathrm{GlueVaR}}

\newcommand{\GD}{\mathrm{GD}}
\newcommand{\MMD}{\mathrm{MMD}}
\newcommand{\IQD}{\mathrm{IQD}}
\newcommand{\RVaR}{\mathrm{RVaR}}

\newcommand{\thedate}{\today}

\usepackage{setspace}

\usepackage{footmisc}
\begin{document}  
\title{Robust Distortion Risk Metrics with Wasserstein and Unimodal Constraints}
\author{
Peng Liu\thanks{Department of Mathematics, Statistics and Actuarial Science, University of Essex, UK. Email: peng.liu@essex.ac.uk}\and Steven Vanduffel \thanks{Department of Economics and Political Science, Vrije Universiteit Brussel, Belgium. Email: Steven.Vanduffel@vub.be} \and Yi Xia\thanks{Department of Mathematics, Statistics and Actuarial Science, University of Essex, UK. Email: yx21416@essex.ac.uk}\and Jing Yao\thanks{Center for Financial Engineering, Soochow University, Suzhou, People's Republic of China. Email:j.yao@suda.edu.cn.}
}

  \date{\thedate}

 \maketitle
\begin{abstract}
We establish sharp upper and lower bounds for distortion risk metrics under distributional uncertainty. The uncertainty sets are 
characterized by four key features of the underlying 
distribution: mean, variance, unimodality, and Wasserstein distance to a reference distribution.

We first examine a broad class of distortion functions, assuming only finite variation and imposing neither continuity nor monotonicity. This class includes important examples such as the Gini deviation, the mean–median deviation, and inter-quantile differences. When the uncertainty set is defined by a fixed mean, variance, and Wasserstein distance, we derive the worst- and best-case values of the distortion risk metric and identify the corresponding extremal distributions. 

We then impose an additional unimodality constraint. 
In this case, for absolutely continuous distortion functions, we again characterize the worst- and best-case values and explicitly determine the optimal distributions attaining these bounds. Moreover, an algorithm is developed to compute sharp upper bounds and the worst-case quantiles.

We apply our results to a robust portfolio optimization problem of interest. 

\begin{bfseries}Key-words\end{bfseries}: Robust distortion risk metrics; Mean-variance; Wasserstein metrics; Unimodality; Portfolio optimization.
\end{abstract}
\newpage

\section{Introduction}\label{sec:1}

In traditional decision-making frameworks, a decision-maker assumes a known distribution function $F$ for the random variable $X$ at hand, and uses a law-invariant functional $\mathcal{V}$ such as variance,  expected (dis)utility or value-at-risk to assess the risk $\mathcal{V}(X):=\mathcal{V}(F)$. This approach, however, relies heavily on the correctness of a single probabilistic model, and it is well understood that it may lead to poor decisions if the true probabilities are uncertain or misspecified. As a result, the question of how to account for distributional ambiguity in decision making has become a central concern in a number of fields, including economics, finance, engineering and operations research. A major modeling paradigm to address ambiguity is distributionally robust optimization (DRO). In its standard form, DRO amounts to dealing with a problem of the form
\begin{align}\label{P1}
\min_{{\mathbf\theta} \in 
\Theta}
\max_{F_{{\mathbf{X}}} \in \mathcal{F}} 
\mathcal{V}(f({{\mathbf\theta}}, \mathbf{X})).
\end{align}

Here, $f$ is a loss function, $\theta$ denotes a decision vector (e.g.,  weights), $F_{\mathbf{X}}$ is an admissible distribution function for a given risk vector $\mathbf{X}$ and $\mathcal{F}$ is the uncertainty set containing all admissible distribution functions. A DRO thus captures the basic idea of making decisions that perform optimally under worst-case scenarios. In this paper, we contribute to the literature by solving the inner maximization problem for uncertainty sets that are reflective of real-world ambiguity, for a very broad class of decision functionals $\mathcal{V}$ that have been used in practical decision making. 
We apply the results to deal with novel robust portfolio selection problems that are of high practical interest.  

The min-max formulation for optimal decision making under ambiguity can be traced back to \cite{S58}, who studied the newsvendor problem under the distributional ambiguity of the demand function. Its theoretical appeal stems from the axiomatization of \cite{GB1987} for the case of expected utility under ambiguity, where choices are based on the worst-case expected utility over a set of plausible probability distributions; see also  \cite{HS01} who propose the min-max formulation for robust control problems.  The arrival of modern robust optimization techniques in the last decade, see e.g., \cite{BT09}   
has further contributed to the successful application of DROs in various areas including engineering, finance, operations research and economics. 

As for the choice of the decision functional $\mathcal{V}$, adopting the expected (dis)utility framework appears to be the most natural choice given its prominent place in the academic literature. There are, however, several shortcomings to its use (
\cite{S00}). First, it is not obvious at all for a decision maker to specify his utility function. In the context of optimal portfolio strategies, \cite{BS81} point out that “from a practical
point of view, it may well prove easier for the investor to choose directly his optimal
quantile function than it would be for him to communicate his utility function to a portfolio manager”. The same observation has led \cite{GJS08}  to introduce a tool, called the distribution builder, which makes it possible for investors to analyse their desired payoff function and to elicit a utility to explain their choice. Second, there is ample empirical evidence that real world decision making cannot be reconciled with the use of utility functions (e.g., \cite{A53}). In response to this criticism, numerous alternatives, such as Yaari's dual theory
(\cite{Y87}), rank-dependent utility theory (\cite{Q82}) and cumulative prospect theory (\cite{TK92}), have been proposed. 

All these theories have been justified by proposing axioms that are considered “sensible". While providing a prescriptive foundation for a decision theory—as a practical guide to making choices—is important, the real issue lies in understanding how people actually choose, and observed real-world behavior should not be dismissed simply because it violates conventional choice axioms (\cite{S00}).
In this context, Yaari’s dual theory has a lot of appeal, because it aligns more closely with observed decision-making behavior. Indeed, this theory gives rise to quantile-based functionals, called distortion risk measures, such as value-at-risk (VaR), range value at risk (RVaR), tail value at risk (TVaR), all of which are used in real-world decision making, mainly because they reflect the human tendency to ask questions such as “What if this happens?” or “What would I lose under this scenario?”\footnote{\cite{R095} attribute this to mental simulation and counterfactual thinking. That is, people naturally engage in mental simulations of (extreme) events, imagining what could happen in those scenarios; see also Tversky and Kahneman \cite{KT73} who explain this behaviour because extreme events are more memorable.} Our set-up is however broader in that we do not require the distortion function to be non-decreasing. This makes it possible to extend the scope and to also include the inter-quantile difference (IQD), the Gini deviation (GD), the mean-median deviation or the difference of two distortion risk measures in our framework. Specifically, the 
functionals that we consider are referred to as distortion risk metrics\footnote{In the paper of \cite{WWW20a}, they are called distortion riskmetrics, which is slightly different from ours.} and will be denoted by $\rho_g$ where $g$ reflects the underlying distortion function. They were first proposed in \cite{WWW20}. 

We are not the first to pursue DRO using distortion risk measures and its generalization distortion risk metrics. This was initiated by \cite{BT09} for the case of VaR, extended in \cite{CHZ11} to include the case of TVaR, and significantly generalized by \cite{L18} to the entire class of convex distortion risk measures, and by \cite{CLM23} to general distortion risk measures. In all these works, the ambiguity set $\mathcal{F}$ is  characterized by the mean and covariance matrix of the random vector $\mathbf{X}$. 

To deal with a DRO, one needs to first solve an inner problem of the form 
\begin{align}\label{P2}
\max_{F_{{{Y}}} \in \widetilde{\mathcal{F}}} 
\rho_g({Y}).
\end{align}

Problem \ref{P2} specifically deals with the extent to which measurements can be affected by model misspecification. This problem is of particular relevance in statistics—where it has been studied under the notion of robust statistics—and in the financial industry, where the assessment of model uncertainty became a regulatory priority in the aftermath of the 2008–2009 financial crisis. For instance, in February 2017, the European Central Bank published its Guide to the Targeted Review of Internal Models, in which it declared that every institution “should have a model risk management framework in place that allows it to identify, understand, and manage its model risk” (\cite{ECB17}). 
An early contribution in this regard is Cantelli's bound on tail risk (survival probabilities); by inversion, this  result yields a sharp bound on Value-at-Risk (VaR). An explicit formula can be traced back to \cite{E03}. Interestingly, the bound on VaR is achieved by a two-point distribution, and it follows that the same bound also applies to Tail Value-at-Risk (TVaR), which can be viewed as the concave distortion risk measure closest to VaR. Since then, it has become apparent that this correspondence between VaR and its concavification TVaR carries over to general distortion risk measures. Indeed, \cite{L18}, \cite{CLM23} and \cite{PWW20} show that for suitable uncertainty sets Problem \ref{P2} is equivalent to the case in which $g$ is replaced by the smallest concave function $g^*$ that dominates $g$. This is a very relevant result, as when in the DRO (problem \ref{P1}) $f$ is concave in $\theta$, it becomes a convex-concave optimisation problem for which powerful computational methods are available. As shown in \cite{PWW20}, the stated equivalence does not hold true if the Wasserstein distance is involved, which is the case in our study. For more studies of worst-case problems in operations research and its applications, see e.g., \cite{CHZ11} and \cite{zym13}.

The ambiguity set is a key component of any distributionally robust optimization model. Rather than letting mathematical convenience drive the choice of the set, it should be primarily guided by available data, possibly complemented by expert opinion. It should be large enough to reasonably include the true distribution but not so broad that it admits implausible distributions, as this could lead to overly conservative decision-making. For example, the distribution function that maximizes VaR and TVaR (i.e., the Cantelli bound) under the sole knowledge of mean and variance has only two mass points, which is not plausible in practice, also highlighting that more (possibly qualitative) information should be used.  
In this paper, we study bounds on distortion risk metrics 
for ambiguity sets that are arguably highly relevant in practice, including the case in which the ambiguity set in addition to containing distributions with given first two moments is also restricted to  unimodal distributions that are close to some reference distribution. The assumption of unimodality is very reasonable in that it is typically consistent with data, which also explains why unimodal
distributions are routinely used in engineering, operations research and in insurance and financial risk modeling. For instance, Pareto, Gamma, Normal, Log-Normal,  Beta, Weibull, and Student t-distributions are all unimodal. The literature on risk bounds for unimodal distribution functions is limited. \cite{P05} proposes semidefinite programming to determine best-possible bounds on tail (survival) probabilities under mean, variance and a unimodality constraint from which by numerical inversion $\VaR$ bounds can be obtained. \cite{LSWY18} and \cite{BKV2020} derive explicit bounds on Range Value-at-Risk (RVaR), but results for general distortion risk metrics appear to be missing.   

In this paper, we develop some algorithms to approximate the bounds and the corresponding worst-case quantiles on robust distortion risk metrics with the uncertainty sets characterized by the unimodality.  To address the natural requirement that admissible distributions are ``close enough" to a reference distribution, we use the 2-Wasserstein distance. This choice, aside from offering mathematical convenience, allows us to handle distributions with differing supports.

\subsection{Our contributions}
For the uncertainty sets with fixed mean, variance and 2-Wasserstein distance, we obtain the worst-case value for general distortion risk metrics, where the distortion function is only required to have finite variation. In particular, our results also cover non-monotone and discontinuous distortion functions. If the distortion function is upper semicontinuous, we also derive the worst-case distribution. Our results extend those in \cite{BPV24}, where the increasing and absolutely continuous distortion functions were considered. The projection method used in \cite{BPV24} does not work for discontinuous distortion functions, and a different technique is required. The method we employ is a variant of the concave envelope approach, differing from the one adopted in \cite{CLM23} and \cite{PWW20}, and requiring  non-trivial technical proofs. In these papers, the envelope is constructed on the distortion function. However, to handle the Wasserstein constraint we construct the concave envelope on a linear combination of the distortion function and a functional of the reference distribution with a mixing parameter. We then choose the best parameter, where the continuity of the concave envelope with respect to the parameter is crucial and its proof is non-trivial.
Applying our results, we derive the explicit sharp bounds  for  $\GlueVaR$\footnote{
Introduced by \cite{BGS13}, this measure balances two widely used risk measures, VaR and ES, as the former tends to underestimate risk while the latter is often overly conservative. \cite{CLM23} further highlight the strong practical demand for such intermediate measures in industry.
},  inter-quantile difference and the discrepancy between expected shortfall ($\ES$)  and $\VaR$, all of which are not available in the literature.   

When the uncertainty sets are characterized by four features, namely mean, variance, Wasserstein distance and unimodality, we derive bounds on distortion risk metrics for absolutely continuous distortion functions. 
We start with the uncertainty sets characterized by fixed mean, variance and unimodality with a fixed inflection point. Employing the projection of a function on the set of  all increasing and concave-convex functions with a fixed inflection point on $(0,1)$, we obtain the worst-case value and the worst-case distribution for the distortion risk metrics under this uncertainty set. In the literature, only some special risk measures (Range-Value-at-Risk or $\VaR$) are considered under a similar uncertainty set (although the inflection point is not fixed), but in  our paper, we build up a general theory to involve unimodality in the uncertainty set. Although the projection method is powerful,  it appears often difficult to obtain the explicit expression of the projection. To address this, we design an efficient algorithm to approximate the worst-case distribution and the worst-case value with any desired degree of precision. Based on this result and employing the projection method, we also find the worst-case value and distribution for the distortion risk metrics when the uncertainty set is characterized by the features of mean, variance, unimodality and Wasserstein distance. Finally, we also discuss the case of unimodality with unknown inflection point, i.e.,  we provide bounds when the inflection point is in an interval.

We apply our results to robust portfolio optimization under an uncertainty set defined by a mean–variance constraint combined with a probability constraint based on the Wasserstein metric for the random return vector. Our results make it possible to reduce this problem to a standard (single-level) optimization problem. A numerical example illustrates the behavior of the resulting optimal robust portfolios for several risk measures of interest.

\section{Preliminary}

Let $(\Omega,\mathcal A,\p)$ be a given atomless probability space. Denote by $L^2$ the set of all real-valued random variables with finite second moments, and by $\mathcal M^2$ the set of all corresponding distribution functions. We interpret a positive value of a random variable as a financial loss. All random variables and distribution functions mentioned hereafter are assumed to belong to $L^2$ and $\mathcal M^2$, respectively.

The left quantile and right quantile of a distribution function $G$ are defined as
$$
G^{-1}(p)=\inf\{x\in \R: G(x) \ge p \},~p\in (0,1],
$$
and
$$
G_+^{-1}(p)=\inf\{x\in \R: G(x) > p \},~p\in [0,1),
$$
respectively, with the convention that $\inf\emptyset =\infty$. A left quantile is often also called a Value-at-Risk (VaR). We also use the notation $\VaR$ to denote a left quantile, and $\VaR^+$ to denote a right quantile. The expected shortfall ($\ES$) is another regulatory risk measure widely used in banking and finance, defined as
$$
\ES_\alpha(G)=\frac{1}{1-\alpha}\int_{\alpha}^1 G^{-1}(t)\d t,~0\leq \alpha<1.
$$

We denote by $\mathcal H$ the set of functions $g:[0,1]\to\mathbb R$ with bounded variation satisfying $g(0)=g(0+)=0$ and $g(1)=g(1-)$. For $g\in\mathcal H$, define $\hat{g}(x)=\max\{g(x-), g(x), g(x+)\}$ for $x\in (0,1)$ and $\hat{g}(x)=g(x)$ for $x=0,1$. Hence, $\hat{g}$ is the upper semicontinuous version of $g$. For $g\in\mathcal H$, a distortion risk metric $\rho_g:\M^2\to\mathbb R$ is defined as
\begin{align}
\rho_g(G)=\int_{0}^{\infty} g(1-G(x))\d x+\int_{-\infty}^{0} (g(1-G(x))-g(1))\d x.
\end{align}

In this paper, we aim to determine the \emph{worst-case} and \emph{best-case} values of a distortion risk metric $\rho_g$ over certain distributional uncertainty sets $\mathcal{F} \subseteq {\mathcal M^2}$. That is, we consider problems of the form

\begin{subequations}\label{eq: prob opt1}
\begin{minipage}{0.45\textwidth}
\begin{equation}\label{eq:problem sup}
\sup_{G \in \mathcal{F}} \rho_g(G)
\end{equation}
\end{minipage}
\begin{minipage}{0.45\textwidth}
\begin{equation}\label{eq:problem inf}
\inf_{G \in \mathcal{F}} \rho_g(G)\,.
\end{equation}
\end{minipage}%
\end{subequations}\\%

\noindent
The sets $\mathcal{F}$ will contain all distribution functions with a given mean and variance that lie within a Wasserstein ball around a given reference distribution $F$ and/or that are unimodal. The set of quantile functions corresponding to the cdfs contained in $\mathcal{F}$ will be denoted by $\mathcal{F}^{-1}$.

In addition to the worst- and best-case values, we also study \emph{worst-case} and \emph{best-case distribution functions}, if they exist---that is, the distribution functions attaining \eqref{eq:problem sup} and \eqref{eq:problem inf}, respectively.

Note that
\begin{align}\label{best-case}
\inf_{G \in \mathcal{F}} \rho_g(G)=-\sup_{G \in \mathcal{F}} \bigl(-\rho_g(G)\bigr)=-\sup_{G \in \mathcal{F}}\rho_{-g}(G),
\end{align}
where $\rho_{-g}$ is also a distortion risk metric. Moreover, $G^*$ is the worst-case distribution for $\sup_{G \in \mathcal{F}}\rho_{-g}(G)$ if and only if it is the best-case distribution for $\inf_{G \in \mathcal{F}} \rho_g(G)$. Hence, the problem \eqref{eq:problem inf} can be fully reduced to problem \eqref{eq:problem sup}. This is one of the motivations to study distortion risk metrics rather than distortion risk measures, as stated in \cite{PWW20}. In this paper, we mainly focus on \eqref{eq:problem sup}.


We end the section with a discussion of several distortion risk metrics of interest. We first present three variability measures that are also distortion risk metrics, namely the Gini deviation ($\GD$), the mean--median deviation ($\MMD$), and the inter-quantile difference ($\IQD$).

The $\GD$ of a distribution function $G$ is defined as
$$
\GD(G)=\frac{1}{2}\mathbb E(|X-Y|):=\rho_{g_{\GD}}(G),
$$
where $X\sim G$ and $Y\sim G$ are independent and $g_{\GD}(t)=t-t^2,~t\in [0,1]$. The Gini deviation thus measures the average absolute difference between two randomly chosen realizations from $G$. After normalization, it becomes the Gini coefficient, which is widely used to measure income inequality. In finance, it was proposed by \cite{SY84} as a substitute for variance as a measure of risk within Markowitz's portfolio selection model. Specifically, these authors develop a portfolio selection approach based on the mean and the Gini deviation as measures of return and risk, respectively. Apart from being more robust, the use of the Gini deviation also enables the derivation of necessary conditions for stochastic dominance, allowing agents to eliminate from the efficient set any feasible solutions that are stochastically dominated by others.

Furthermore, the $\MMD$ of $G$ is defined as
$$
\MMD(G)=\min_{x\in\R}\mathbb E(|X-x|)=\E(|X-G^{-1}(1/2)|):=\rho_{g_{\MMD}}(G),
$$
where $X\sim G$ and $g_{\MMD}(t)=t\wedge(1-t),~t\in [0,1]$.

As for $\IQD$, we define
$$
\IQD_{\alpha}^+(G)=G_+^{-1}(1-\alpha)-G^{-1}(\alpha):=\rho_{g_{\IQD^+}}(G),~\alpha\in (0,1/2],
$$
and
$$
\IQD_{\alpha}^-(G)=G^{-1}(1-\alpha)-G_+^{-1}(\alpha):=\rho_{g_{\IQD^-}}(G),~\alpha\in (0,1/2),
$$
where $g_{\IQD^+}(t)=\id_{[\alpha,1-\alpha]}(t),~t\in [0,1]$ and $g_{\IQD^-}(t)=\id_{(\alpha,1-\alpha)}(t),~t\in [0,1]$.
Note that the definitions of $\IQD_{\alpha}^+$ and $\IQD_{\alpha}^-$ can be found in \cite{BFWW22} and \cite{LLW23}, respectively. The measures $\MMD$ and $\IQD$ play a prominent role in robust statistics, and are therefore also useful in portfolio selection and risk management, where resilience to outliers (data contamination) is desired. Their application to the problem of risk sharing can be found in \cite{LLW23}.

The $\mathrm{GlueVaR}$ of a distribution $G$ was introduced in \cite{BGS13} and is defined as the distortion risk metric $\rho_g(G)$ for which the distortion $g:=g^{h_1,h_2}_{\beta,\alpha}$ is given by
\begin{equation}
g^{h_1,h_2}_{\beta,\alpha} (t) =
\begin{cases}
\frac{h_1}{1-\beta}  t, & 0 \leq t < 1-\beta, \\
h_1 + \frac{h_2 - h_1}{\beta - \alpha} [t - (1-\beta)], & 1-\beta \leq t \leq 1-\alpha, \\
1, & 1-\alpha< t \leq 1,
\end{cases}
\end{equation}
where $\alpha, \beta \in [0, 1]$ satisfy $\alpha \leq \beta$, $h_1 \in [0, 1]$, and $h_2 \in [h_1, 1]$.

Note that $\VaR_{\alpha}$, $\ES_{\alpha}$, and Range-Value-at-Risk ($\RVaR$) are particular cases of this family of risk measures, with the corresponding distortion functions $g^{0,0}_{\alpha,\alpha}(u)$, $g^{1,1}_{\alpha,\alpha}(u)$, and $g^{0,1}_{\beta,\alpha}(u)$ with $\alpha<\beta$, respectively. Here, $\RVaR$, introduced by \cite{CONT10} as a family of robust risk measures, is defined as
$$
\RVaR_{\alpha,\beta}(G)=\frac{1}{\beta-\alpha}\int_{\alpha}^\beta G^{-1}(t)\d t,~0<\alpha<\beta<1.
$$
Furthermore, $\GlueVaR$ can be rewritten as a linear combination of $\ES$ and $\VaR$. If $\frac{h_1}{1-\beta}\geq \frac{h_2 - h_1}{\beta - \alpha}$, then $\GlueVaR_{\beta,\alpha}^{h_1,h_2}=w_1\ES_{\alpha}+w_2\ES_{\beta}+w_3\VaR_\alpha$ with some $w_1,w_2,w_3\geq 0$ satisfying $w_1+w_2+w_3=1$; see \cite{BGS13} for more details.

Finally, for $0<\alpha_1<\alpha_2<1$, the discrepancy between $\ES$ and $\VaR$ is defined as
$$
\rho_{g_{\alpha_1,\alpha_2}}=\ES_{\alpha_1}-\VaR_{\alpha_2},
$$
where $g_{\alpha_1,\alpha_2}(t)=\frac{t}{1-\alpha_1}\wedge 1-\id_{(1-\alpha_2,1]}(t)$. In practice, one often uses the parameter values $\alpha_1=0.975$ and $\alpha_2=0.99$.

In what follows, the notation $V$ denotes a standard uniformly distributed random variable.

\section{Bounds for distortion risk metrics under Wasserstein distance constraints}\label{Sec:general}

One popular notion used in mass transportation and distributionally robust optimization is the Wasserstein metric. For more details, we refer to \cite{EK18} and \cite{BM19}. For two random variables $X$ and $Y$ with respective distributions $F$ and $G$, the one-dimensional Wasserstein metric of order $2$ is given by
$$
d_W(X,Y)=d_W(F,G)=d_{W}(F^{-1},G^{-1})=\left(\int_{0}^{1}|F^{-1}(x)-G^{-1}(x)|^2\d x\right)^{1/2}.
$$
In this section, we study problem \eqref{eq:problem sup} when the uncertainty set $\mathcal{F}$ is given by
$$
\mathcal{F}:=\mathcal M_\epsilon(\mu, \sigma)=\left\{G\in \mathcal M^2: \int_{\R} x\d G=\mu, \int_{\R}x^2\d G=\mu^2+\sigma^2, d_{W}(G,F)\leq \sqrt{\epsilon}\right\},
$$
where $\mu\in\R$, $\sigma>0$, $\epsilon>0$, and $F\in \mathcal M^2$. Here, the distribution function $F$ is the center of a Wasserstein ball, and we denote its mean and standard deviation by $\mu_F$ and $\sigma_F>0$, respectively. Note that
$$
\mathcal M_\infty(\mu, \sigma)=\left\{G\in \mathcal M^2: \int_{\R} x\d G=\mu, \int_{\R}x^2\d G=\mu^2+\sigma^2\right\}.
$$

For $g\in\mathcal H$, let $g^*$  denote the concave  envelope of $g$, i.e.,
$$g^*=\inf\{h\in\mathcal H: h~\text{is concave on}~[0,1]~\text{and}~ h\geq g\}.$$
For any concave or convex function $h\in\mathcal H$, let $h'(t):=\partial_+ h(t)$.
Let $c_0=Corr(F^{-1}(V), (g^*)'(1-V))$, with the convention that $c_0=0$ if $(g^*)'$ is constant, where $V$ is a random variable following the uniform distribution over $[0,1]$, i.e.,  $V\sim U[0,1]$. Note that $c_0\geq 0$.
Moreover, let $g_{\lambda}(t)=g(t)+\lambda \int_{1-t}^{1} F^{-1}(s)\d s$ for $t\in [0,1]$ and $\lambda\geq 0$, and
\begin{align}\label{worst-case-quantile}
h_{\lambda}(t)=\mu+\sigma\frac{(g_{\lambda}^*)'(1-t)-a_{\lambda}}{b_{\lambda}},
\end{align}
where $a_{\lambda}= E((g_{\lambda}^*)'(V))=g(1)+\lambda\mu_F$ and $b_{\lambda}=\sqrt{Var((g_{\lambda}^*)'(V))}$. We denote the distribution function corresponding to $h_\lambda$ by $H_\lambda$, with $H_{\lambda}^{-1}=h_{\lambda}$.

To ensure that $h_{\lambda}$ is well defined throughout the paper, we make the following assumption:

\vspace{12pt}

\noindent
{\bf Assumption A}: $\int_{0}^{1} ((g^*)'(t))^2\d t<\infty$ and $(g_{\lambda}^*)'$ is not constant for all $\lambda\geq 0$.

\vspace{12pt}
\noindent
For $G\in \mathcal M_\epsilon(\mu, \sigma)$ and $g\in\mathcal H$, it follows from {\bf Assumption A} that
$$
\rho_g(G)\leq \rho_{g^*}(G)=\int_{0}^{1} (g^{*})'(1-t)G^{-1}(t)\d t\leq \sqrt{\sigma^2+\mu^2} \left(\int_{0}^{1} ((g^*)'(t))^2\d t\right)^{1/2}<\infty.
$$
Hence, {\bf Assumption A} also guarantees that $\rho_g(G)<\infty$ for all $G\in \mathcal M_\epsilon(\mu, \sigma)$. The following lemma is crucial to the main results of the paper (see Theorem \ref{Th:main} below). Its proof is highly non-trivial, and the underlying arguments also play a key role in the proof of Corollary \ref{Cor:2} below. We postpone the proof to the Appendix.
\begin{lemma}\label{lem:1}
The function $Corr(F^{-1}(V), (g_{\lambda}^*)'(1-V))$ is continuous in $\lambda$ on $[0,\infty)$, and
$$
\lim_{\lambda\to\infty} Corr(F^{-1}(V), (g_{\lambda}^*)'(1-V))=1.
$$
\end{lemma}

Next, we present our first main result, which gives the worst-case distribution and the worst-case  distortion risk metrics over the uncertainty set $\mathcal M_{\epsilon}(\mu, \sigma)$.

\begin{theorem}\label{Th:main}
Suppose $g\in \mathcal H$ and $g=\hat{g}$.
\begin{enumerate}
\item[(i)] If $(\mu_F-\mu)^2+(\sigma_F-\sigma)^2<\epsilon<(\mu_F-\mu)^2+(\sigma_F-\sigma)^2+2\sigma_F\sigma(1-c_0)$, then
$$
\sup_{G\in \mathcal M_{\epsilon}(\mu, \sigma)} \rho_g(G)=\rho_g(H_{\lambda_\epsilon}),
$$
where $H_{\lambda_\epsilon}$ is the unique worst-case distribution function, determined by $d_{W}(F,H_{\lambda_\epsilon})=\sqrt{\epsilon}$ for some $\lambda_\epsilon>0$.
\item[(ii)]
If $\epsilon\geq (\mu_F-\mu)^2+(\sigma_F-\sigma)^2+2\sigma_F\sigma(1-c_0)$, then case (i) applies with $\lambda_\epsilon =0$.
\end{enumerate}
\end{theorem}

The conclusions in Theorem \ref{Th:main} hold for very general distortion functions, covering many results in the literature, such as \cite{BPV24} (increasing and absolutely continuous $g$), \cite{SZ23} and \cite{SZ24} (increasing $g$ and $\epsilon=\infty$), \cite{LSWY18} (Range-Value-at-Risk and $\epsilon=\infty$), \cite{L18} (concave and increasing $g$ and $\epsilon=\infty$), and \cite{CLM23} and \cite{PWW20} (general $g$ and $\epsilon=\infty$). Compared with the results in the literature, a main novelty of Theorem \ref{Th:main} is that it covers the case where $g$ can be non-monotone and discontinuous, in particular including the distortion functions of $\GD$, $\MMD$, $\VaR^+$, and $\IQD^+$ as special cases. We emphasize that Theorem \ref{Th:main} extends the results in \cite{BPV24} from increasing and absolutely continuous distortion functions to upper semicontinuous distortion functions with finite variation. The projection method used in \cite{BPV24} requires $g$ to be absolutely continuous and cannot be applied in the general case considered here.

The method used to prove Theorem \ref{Th:main} is a variant of the concave envelope technique. It differs from the approaches in \cite{CLM23} and \cite{PWW20}, where the envelope is constructed for the distortion function itself. Here, the concave envelope is taken for a linear combination of the distortion function and a functional of the reference distribution $F$, with combination parameter $\lambda$. We then choose the best parameter. The existence of the best parameter is based on a continuity property of this envelope, as shown in Lemma \ref{lem:1}. This continuity property is key to obtaining Theorem \ref{Th:main}, and its proof is highly non-trivial.

To find the worst-case distribution, an explicit expression for $(g_\lambda^*)'(1-t)$ is crucial. Note that if $g$ is concave, then $(g_\lambda^*)'(1-t)=g'(1-t)+\lambda F^{-1}(t)$ for $t\in (0,1)$, which covers the cases of $\GD$ and $\MMD$. If $g$ is nonconcave, it becomes cumbersome to compute $(g_\lambda^*)'(1-t)$. Nevertheless, Corollary \ref{Cor:1} below provides explicit expressions for the cases of $\VaR_\alpha^+$, $\IQD_{\alpha}^+$, and the discrepancy $\rho_{g_{\alpha_1,\alpha_2}}$ between $\ES$ and $\VaR$, respectively.
To this end, for $\alpha\in (0,1)$ and $\lambda\geq 0$, let us define
\begin{equation}\label{t0}
t_{\alpha,\lambda}=\inf\left\{t\in [0,\alpha): \frac{1+\lambda\int_{1-\alpha}^{1-t} F^{-1}(s)\d s}{\alpha-t}\geq\lambda F^{-1}(1-t)\right\},
\end{equation}
and
\begin{align}\label{t12}
\hat{t}_{\alpha,\lambda}=\sup\left\{t\in (1-\alpha,1]: \frac{\lambda\int_{1-t}^{\alpha} F^{-1}(s)\d s-1}{t-1+\alpha}\leq\lambda F^{-1}(1-t)\right\}.
\end{align}
For $0<\alpha_1<\alpha_2<1$ and $\lambda\geq 0$, let
\begin{align}\label{u0}
u_{\alpha_1,\alpha_2,\lambda}&=\sup\left\{t\in (1-\alpha_2,1]: \frac{\frac{(t-1+\alpha_2)\wedge (\alpha_2-\alpha_1)}{1-\alpha_1}+\lambda\int_{1-t}^{\alpha_2} F^{-1}(s)\d s-1}{t-1+\alpha_2}\right.\nonumber\\
&~~~~~~~~~~~~~~~~~~~~~~~~~~~~~~~~~~~~~~~~~~~~\left.\leq \frac{1}{1-\alpha_1}\id_{(0,1-\alpha_1)}(t)+\lambda F^{-1}(1-t)\right\}.
\end{align}

We formulate the following corollary.

\begin{corollary}\label{Cor:1}
Suppose $(\mu_F-\mu)^2+(\sigma_F-\sigma)^2<\epsilon<(\mu_F-\mu)^2+(\sigma_F-\sigma)^2+2\sigma_F\sigma(1-c_0)$.
\begin{enumerate}[(i)]
\item If $g\in\mathcal H$ is concave, then we have
$$
\sup_{G\in \mathcal M_{\epsilon}(\mu, \sigma)} \rho_g(G)=\frac{\sigma}{b_{\lambda}}\int_0^1(g'(t)+\lambda_\epsilon F^{-1}(1-t))g'(t)\d t+\left(\mu-\frac{a_{\lambda_\epsilon}\sigma}{b_{\lambda_\epsilon}}\right)g(1),
$$
and the worst-case quantile is
$$
h_{\lambda_\epsilon}=\frac{\sigma(g'(1-t)+\lambda_\epsilon F^{-1}(t))}{b_{\lambda_\epsilon}}+\mu-\frac{a_{\lambda_\epsilon}\sigma}{b_{\lambda_\epsilon}},
$$
where $\lambda_\epsilon$ is the solution to $d_{W}(F,H_\lambda)=\sqrt{\epsilon}$.
\item For $\alpha\in (0,1)$, we have
$$
\sup_{G\in \mathcal M_{\epsilon}(\mu, \sigma)} \VaR_{\alpha}^+(G)=\mu+\sigma\frac{\frac{1+\lambda_\epsilon\int_{\alpha}^{1-t_{1-\alpha,\lambda_\epsilon}} F^{-1}(s)\d s}{1-\alpha-t_{1-\alpha,\lambda_\epsilon}}-a_{\lambda_\epsilon}}{b_{\lambda_\epsilon}},
$$
and the worst-case quantile is $h_{\lambda_\epsilon}$ given by \eqref{worst-case-quantile} with
$$
(g_\lambda^*)'(1-t)=\lambda F^{-1}(t)\id_{(0,\alpha] \cup (1-t_{1-\alpha,\lambda},1)}(t)+\frac{1+\lambda\int_{\alpha}^{1-t_{1-\alpha,\lambda}}F^{-1}(s)\d s}{1-\alpha-t_{1-\alpha,\lambda}}\id_{(\alpha,1-t_{1-\alpha,\lambda}]},~t\in (0,1),
$$
where $\lambda_\epsilon$ is the solution to $d_{W}(F,H_\lambda)=\sqrt{\epsilon}$.
\item For $\alpha\in (0,1/2)$, we have
\begin{align*}
\sup_{G\in \mathcal M_{\epsilon}(\mu, \sigma)} \IQD_{\alpha}^+(G)
=\left(\frac{1+\lambda_\epsilon\int_{1-\alpha}^{1-t_{\alpha,\lambda_\epsilon}} F^{-1}(s)\d s}{\alpha-t_{\alpha,\lambda_\epsilon}}-\frac{\lambda_\epsilon\int_{1-\hat{t}_{\alpha,\lambda_\epsilon}}^{\alpha} F^{-1}(s)\d s-1}{\hat{t}_{\alpha,\lambda_\epsilon}-1+\alpha}\right)\frac{\sigma}{b_{\lambda_\epsilon}}
\end{align*}
and the worst-case quantile is $h_{\lambda_\epsilon}$ given by \eqref{worst-case-quantile} with
\begin{align*}
(g_\lambda^*)'(1-t)&=\frac{1+\lambda\int_{1-\alpha}^{1-t_{\alpha,\lambda}} F^{-1}(s)\d s}{\alpha-t_{\alpha,\lambda}}\id_{(1-\alpha, 1-t_{\alpha,\lambda})}(t)+\frac{\lambda\int_{1-\hat{t}_{\alpha,\lambda}}^{\alpha} F^{-1}(s)\d s-1}{\hat{t}_{\alpha,\lambda}-1+\alpha}\id_{(1-\hat{t}_{\alpha,\lambda},\alpha)}(t)\\
&+\lambda F^{-1}(t)\id_{(0,1-\hat{t}_{\alpha,\lambda})\cup(\alpha,1-\alpha)\cup (1-t_{\alpha,\lambda},1)},~t\in (0,1),
\end{align*}
where $\lambda_\epsilon$ is the solution to $d_{W}(F,H_\lambda)=\sqrt{\epsilon}$.
\item For $0<\alpha_1<\alpha_2<1$, we have
\begin{align*}
&\sup_{G\in \mathcal M_{\epsilon}(\mu, \sigma)} \rho_{g_{\alpha_1,\alpha_2}}(G)\\
&=\frac{\sigma\lambda_\epsilon}{b_{\lambda_\epsilon}(1-\alpha_1)}\left(\int_{\alpha_1}^{\alpha_1\vee(1-u_{\alpha_1,\alpha_2,\lambda_\epsilon})} F^{-1}(s)\d s+\int_{\alpha_2}^{1} F^{-1}(s)\d s\right)\\
&~+\frac{\sigma}{b_{\lambda_\epsilon}}\frac{1-\alpha_2+(1-\alpha_1-u_{\alpha_1,\alpha_2,\lambda_\epsilon})_+}{(1-\alpha_1)^2}+\frac{\sigma c_{\alpha_1,\alpha_2,\lambda_\epsilon}}{b_{\lambda_\epsilon}}\left(\frac{(\alpha_2-\alpha_1)\wedge (u_{\alpha_1,\alpha_2,\lambda_\epsilon}-1+\alpha_2)}{1-\alpha_1}-1\right),
\end{align*}
and the worst-case quantile is $h_{\lambda_\epsilon}$ given by \eqref{worst-case-quantile} with
\begin{align*}
(g_\lambda^*)'(1-t)&=\left(\frac{1}{1-\alpha_1}\id_{(\alpha_1,1)}(t)+\lambda F^{-1}(t)\right)\id_{(0,1-u_{\alpha_1,\alpha_2,\lambda})\cup(\alpha_2,1)}(t)\\
&+c_{\alpha_1,\alpha_2,\lambda}\id_{(1-u_{\alpha_1,\alpha_2,\lambda},\alpha_2)}(t),~t\in (0,1),
\end{align*}
where $\lambda_\epsilon$ is the solution to $d_{W}(F,H_\lambda)=\sqrt{\epsilon}$ and
$$
c_{\alpha_1,\alpha_2,\lambda}=\frac{\frac{(u_{\alpha_1,\alpha_2,\lambda}-1+\alpha_2)\wedge (\alpha_2-\alpha_1)}{1-\alpha_1}+\lambda\int_{1-u_{\alpha_1,\alpha_2,\lambda}}^{\alpha_2} F^{-1}(s)\d s-1}{u_{\alpha_1,\alpha_2,\lambda}-1+\alpha_2}.
$$
\end{enumerate}
\end{corollary}

Note that (i) of Corollary \ref{Cor:1} is valid for distortion risk metrics with non-monotone distortion functions, including all spectral risk measures, as well as $\MMD$ and $\GD$ as special cases. Note also that (ii) of Corollary \ref{Cor:1} is given in Proposition 4.6 of \cite{BPV24}, where a detailed and complicated analysis is required. However, by applying Theorem \ref{Th:main}, we can obtain the worst-case value and the worst-case distribution for $\VaR^+$ immediately. We observe that the corresponding worst-case quantile is a linear function of the quantile of the reference distribution on the tail parts, and it is not constant in general. This contrasts with the case without a Wasserstein constraint, where the worst-case quantile is a step function with two steps (Cantelli bound). The worst-case value and quantile for $\VaR^+$ with only a Wasserstein constraint can be found in \cite{LMWW22}. Moreover, (iii) of Corollary \ref{Cor:1} is entirely new, and shows the worst-case value of the variability (or statistical dispersion) of the data with given mean, variance, and Wasserstein ball. The corresponding worst-case quantile is a linear function of the quantile of the reference distribution on the middle part and tail parts. Finally, the worst-case value of the discrepancy between $\ES$ and $\VaR$ is given in (iv) of Corollary \ref{Cor:1}, and corresponds to the largest possible additional capital requirement when moving from $\VaR$ to $\ES$; see the Basel regulatory framework \cite{BASEL19}.

Theorem \ref{Th:main} requires that $g$ be upper semicontinuous, which excludes the distortion functions for $\VaR$, $\IQD^-$, and GlueVaR. Moreover, following \eqref{best-case}, deriving the best-case value for $\rho_g$ is equivalent to finding the worst-case value for $\rho_{-g}$. If $g$ is upper semicontinuous, then $-g$ is lower semicontinuous, which may not be covered by Theorem \ref{Th:main}, as in the cases of $\VaR^+$, $\IQD^+$, and $\rho_{g_{\alpha_1,\alpha_2}}$. Hence, we next eliminate the restriction of upper semicontinuity and consider all $g\in\mathcal H$.
\begin{theorem}\label{Th:main1}
Suppose $g\in \mathcal H$.
\begin{enumerate}
\item[(i)] If $(\mu_F-\mu)^2+(\sigma_F-\sigma)^2<\epsilon<(\mu_F-\mu)^2+(\sigma_F-\sigma)^2+2\sigma_F\sigma(1-c_0)$ and $\rho_{\hat{g}}(H_\lambda)$ is continuous in $\lambda$ over $(0,\infty)$, then
$$
\sup_{G\in \mathcal M_{\epsilon}(\mu, \sigma)} \rho_g(G)=\sup_{G\in \mathcal M_{\epsilon}(\mu, \sigma)} \rho_{\hat{g}}(G)=\rho_{\hat{g}}(H_{\lambda_\epsilon}),
$$
where $\lambda_\epsilon$ is the solution to $d_{W}(F,H_\lambda)=\sqrt{\epsilon}$.
\item[(ii)]
If $\epsilon\geq (\mu_F-\mu)^2+(\sigma_F-\sigma)^2+2\sigma_F\sigma(1-c_0)$, then
$$
\sup_{G\in \mathcal M_{\epsilon}(\mu, \sigma)} \rho_g(G)=\sup_{G\in \mathcal M_{\epsilon}(\mu, \sigma)} \rho_{\hat{g}}(G)=\rho_{\hat{g}}(H_0).
$$
\end{enumerate}
\end{theorem}

In (i) of Theorem \ref{Th:main1}, we impose an additional assumption: continuity of $\rho_{\hat{g}}(H_\lambda)$ with respect to $\lambda$ over $(0,\infty)$. This assumption is due to the Wasserstein constraint and cannot be removed in our technical proof of Theorem \ref{Th:main1}; see Section \ref{Sec3:proof} in the Appendix. However, this assumption holds for the distortion risk metrics of interest in Corollary \ref{Cor:2} below. It is worth mentioning that the arguments used in the proof of Lemma \ref{lem:1} play a key role in verifying this assumption.

For $0<\alpha<\beta<1$, $0<h_1<h_2<1$, and $\lambda\geq 0$, let
\begin{align}\label{u1}
u_{\alpha,\beta,\lambda}^{h_1,h_2}&=\inf\left\{t\in [0,1-\alpha): \frac{1-g_{\alpha,\beta}^{h_1,h_2}(t)+\lambda\int^{1-t}_{\alpha} F^{-1}(s)\d s}{1-\alpha-t}\right.\nonumber\\
&~~~~~~~~~~~~~~~~~~~~~~~~~~~\left.\geq \frac{h_1}{1-\beta}\id_{(0,1-\beta)}(t)+\frac{h_2 - h_1}{\beta - \alpha}\id_{[1-\beta,1-\alpha)}(t)+\lambda F^{-1}(1-t)\right\}.
\end{align}

\begin{corollary}\label{Cor:2}
Suppose $(\mu_F-\mu)^2+(\sigma_F-\sigma)^2<\epsilon<(\mu_F-\mu)^2+(\sigma_F-\sigma)^2+2\sigma_F\sigma(1-c_0)$.
\begin{enumerate}[(i)]
\item For $\alpha\in (0,1)$, $\VaR_{\alpha}^+(H_\lambda)$ is continuous in $\lambda$ over $(0,\infty)$, and
$$
\sup_{G\in \mathcal M_{\epsilon}(\mu, \sigma)} \VaR_{\alpha}(G)=\sup_{G\in \mathcal M_{\epsilon}(\mu, \sigma)} \VaR_{\alpha}^+(G)=\mu+\sigma\frac{\frac{1+\lambda_\epsilon\int_{\alpha}^{1-t_{1-\alpha,\lambda_\epsilon}} F^{-1}(s)\d s}{1-\alpha-t_{1-\alpha,\lambda_\epsilon}}-a_{\lambda_\epsilon}}{b_{\lambda_\epsilon}},
$$
where $\lambda_\epsilon$ is the solution to $d_{W}(F,H_\lambda)=\sqrt{\epsilon}$.
\item For $\alpha\in (0,1/2)$, $\IQD_{\alpha}^+(H_\lambda)$ is continuous in $\lambda$ over $(0,\infty)$, and
\begin{align*}
\sup_{G\in \mathcal M_{\epsilon}(\mu, \sigma)} \IQD_{\alpha}^-(G)
&=\sup_{G\in \mathcal M_{\epsilon}(\mu, \sigma)} \IQD_{\alpha}^+(G)\\
&=\left(\frac{1+\lambda_\epsilon\int_{1-\alpha}^{1-t_{\alpha,\lambda_\epsilon}} F^{-1}(s)\d s}{\alpha-t_{\alpha,\lambda_\epsilon}}-\frac{\lambda_\epsilon\int_{1-\hat{t}_{\alpha,\lambda_\epsilon}}^{\alpha} F^{-1}(s)\d s-1}{\hat{t}_{\alpha,\lambda_\epsilon}-1+\alpha}\right)\frac{\sigma}{b_{\lambda_\epsilon}},
\end{align*}
where $\lambda_\epsilon$ is the solution to $d_{W}(F,H_\lambda)=\sqrt{\epsilon}$.
\item For $0<\alpha< \beta<1$ and $0<h_1<h_2<1$ satisfying $\frac{h_1}{1-\beta}\geq \frac{h_2 - h_1}{\beta - \alpha}$, $\rho_{\hat{g}_{\beta,\alpha}^{h_1,h_2}}(H_\lambda)$ is continuous in $\lambda$ over $(0,\infty)$, and
\begin{align*}
&\sup_{G\in \mathcal M_{\epsilon}(\mu, \sigma)} \GlueVaR_{\beta,\alpha}^{h_1,h_2}(G)
=\sup_{G\in \mathcal M_{\epsilon}(\mu, \sigma)} \rho_{\hat{g}_{\beta,\alpha}^{h_1,h_2}}(G)\\
&=\mu-\frac{\sigma(1+\lambda_\epsilon\mu_F)}{b_{\lambda_\epsilon}}+\frac{\sigma(1-h_2)}{b_{\lambda_\epsilon}}c_{\alpha,\beta,\lambda_\epsilon}^{h_1,h_2}\\
&+\frac{\sigma h_1}{b_{\lambda_\epsilon}(1-\beta)}\left(c_{\alpha,\beta,\lambda_\epsilon}^{h_1,h_2}(1-u_{\alpha,\beta,\lambda_\epsilon}^{h_1,h_2}-\beta)_++\frac{h_1((1-\beta)\wedge u_{\alpha,\beta,\lambda_\epsilon}^{h_1,h_2})}{1-\beta}+\lambda_\epsilon \int_{\beta\vee (1-u_{\alpha,\beta,\lambda_\epsilon}^{h_1,h_2})}^1 F^{-1}(s)\d s\right)\\
&+\frac{\sigma (h_2-h_1)}{b_{\lambda_\epsilon}(\beta-\alpha)}\left(c_{\alpha,\beta,\lambda_\epsilon}^{h_1,h_2}(\beta\wedge(1-u_{\alpha,\beta,\lambda_\epsilon}^{h_1,h_2})-\alpha)+\frac{(h_2-h_1)(\beta-1+ u_{\alpha,\beta,\lambda_\epsilon}^{h_1,h_2})_+}{\beta-\alpha}\right.\\
&~~~~~~~~~~~~~~~~~~~~~~\left.+\lambda_\epsilon \int_{\beta\wedge (1-u_{\alpha,\beta,\lambda_\epsilon}^{h_1,h_2})}^\beta F^{-1}(s)\d s\right)
\end{align*}
and the worst-case quantile for $\sup_{G\in \mathcal M_{\epsilon}(\mu, \sigma)} \rho_{\hat{g}_{\beta,\alpha}^{h_1,h_2}}(G)$ is $h_{\lambda_\epsilon}$ given by \eqref{worst-case-quantile} with
\begin{align*}
(g_\lambda^*)'(1-t)&=c_{\alpha,\beta,\lambda}^{h_1,h_2}\id_{(\alpha,1-u_{\alpha,\beta,\lambda}^{h_1,h_2})}(t)+\frac{h_1}{1-\beta}\id_{(\beta\vee (1-u_{\alpha,\beta,\lambda}^{h_1,h_2}),1)}(t)\\
&+\frac{h_2-h_1}{\beta-\alpha}\id_{(\beta\wedge (1-u_{\alpha,\beta,\lambda}^{h_1,h_2}),\beta)}(t)
 +\lambda F^{-1}(t)\id_{(0,\alpha)\cup (1-u_{\alpha,\beta,\lambda}^{h_1,h_2},1)},~t\in (0,1),
\end{align*}
where $\lambda_\epsilon$ is the solution to $d_{W}(F,H_\lambda)=\sqrt{\epsilon}$ and
$$
c_{\alpha,\beta,\lambda}^{h_1,h_2}=\frac{1-g_{\alpha,\beta}^{h_1,h_2}(u_{\alpha,\beta,\lambda}^{h_1,h_2})+\lambda\int^{1-u_{\alpha,\beta,\lambda}^{h_1,h_2}}_{\alpha} F^{-1}(s)\d s}{1-\alpha-u_{\alpha,\beta,\lambda}^{h_1,h_2}}.
$$
\end{enumerate}
\end{corollary}

Under $\frac{h_1}{1-\beta}\geq \frac{h_2 - h_1}{\beta - \alpha}$, $\GlueVaR_{\beta,\alpha}^{h_1,h_2}=w_1\ES_{\alpha}+w_2\ES_{\beta}+w_3\VaR_\alpha$ with some $w_1,w_2,w_3\geq 0$ satisfying $w_1+w_2+w_3=1$. By choosing parameters, $\GlueVaR_{\beta,\alpha}^{h_1,h_2}$ can lie between $\VaR$ and $\ES$, corresponding to an attitude that is more conservative than $\VaR$ and less conservative than $\ES$. Hence, (iii) of Corollary \ref{Cor:2} gives the worst-case value of a combination of $\VaR$ and $\ES$.
Applying \eqref{best-case} and Theorem \ref{Th:main1}, we can also derive explicit expressions for the best-case values of $\VaR^+$, $\IQD^+$, and $\rho_{g_{\alpha_1,\alpha_2}}$ in a similar way as in Corollary \ref{Cor:2}. The best-case value of GlueVaR can be derived using \eqref{best-case} and Theorem \ref{Th:main}.

Note that in Theorem \ref{Th:main1}, we do not provide the worst-case distribution. This is because the existence of a worst-case distribution is not guaranteed if $g$ is not upper semicontinuous, as can be seen from the following argument. Under the assumptions of Theorem \ref{Th:main1}, suppose that a worst-case distribution exists for $\sup_{G\in \mathcal M_{\epsilon}(\mu, \sigma)} \rho_g(G)$, denoted by $G_0$. Then Theorem \ref{Th:main1} implies that $G_0$ is also a worst-case distribution for $\sup_{G\in \mathcal M_{\epsilon}(\mu, \sigma)} \rho_{\hat{g}}(G)$. The uniqueness of the worst-case distribution for $\sup_{G\in \mathcal M_{\epsilon}(\mu, \sigma)} \rho_{\hat{g}}(G)$, shown in Theorem \ref{Th:main}, implies $G_0=H_{\lambda_\epsilon}$. Using the worst-case distributions for $(\mu_F-\mu)^2+(\sigma_F-\sigma)^2<\epsilon<(\mu_F-\mu)^2+(\sigma_F-\sigma)^2+2\sigma_F\sigma(1-c_0)$ given in Corollaries \ref{Cor:1} and \ref{Cor:2}, direct computation shows
$$
\VaR_{\alpha}(G_0)=\mu+\sigma\frac{\lambda F^{-1}(\alpha)-a_{\lambda_\epsilon}}{b_{\lambda_\epsilon}}, \qquad
\IQD_{\alpha}^-(G_0)=\sigma\frac{\lambda F^{-1}(1-\alpha)-\lambda F_+^{-1}(\alpha)}{b_{\lambda_\epsilon}},
$$
and
$$
\GlueVaR_{\beta,\alpha}^{h_1,h_2}(G_0)=\rho_{\hat{g}_{\beta,\alpha}^{h_1,h_2}}(H_{\lambda_\epsilon})+\frac{\sigma(1-h_2)}{b_{\lambda_\epsilon}}(\lambda F^{-1}(\alpha)-c_{\alpha,\beta,\lambda_\epsilon}^{h_1,h_2}),
$$
leading to contradictions. Hence, worst-case distributions for $\VaR_{\alpha}$, $\IQD_{\alpha}^-$, and $\GlueVaR_{\beta,\alpha}^{h_1,h_2}$ do not exist. A discussion of a similar issue for $\VaR$ with $\epsilon=\infty$ can be found, e.g., in Corollary 4.1 of \cite{BPV24} and Example 17 of \cite{PWW20}.

Although it appears difficult to obtain an explicit expression for $\lambda_\epsilon$ for a general $g\in \mathcal H$, this is possible in some special cases. In particular, we obtain an explicit expression for $\lambda_\epsilon$ when $g$ is concave.
\begin{proposition}\label{Prop:1}
If $g\in \mathcal H$ is concave and $\int_{0}^{1} (g'(t))^2\d t<\infty$, then for $(\mu_F-\mu)^2+(\sigma_F-\sigma)^2<\epsilon<(\mu_F-\mu)^2+(\sigma_F-\sigma)^2+2\sigma_F\sigma(1-c_0)$, the solution to $d_{W}(F,H_\lambda)=\sqrt{\epsilon}$ is given by
$$
\lambda_\epsilon=\frac{-C_{g,F}+\sqrt{C_{g,F}^2-\sigma_F^2\frac{V_{g} C_{\epsilon,F}^2-\sigma^2 C_{g,F}^2}{C_{\epsilon,F}^2-\sigma^2\sigma_F^2}}}{\sigma_F^2},
$$
where $C_{\epsilon,F}=\frac{\mu_F^2+\sigma_F^2+\mu^2+\sigma^2-2\mu\mu_F-\epsilon}{2}\geq 0$, $V_{g}=Var(g'(V))$, and $C_{g,F}=Cov(F^{-1}(V),g'(1-V))\geq 0$.
\end{proposition}

\section{Bounds for Unimodal Distribution Functions with Wasserstein constraint}\label{Sec:unimodel}

We assume in this section that $g\in\mathcal H$ is such that the distortion risk metrics $\rho_g(G)$ admits the representation
\begin{align}
	\rho_g(G)=\int_{0}^{1} \gamma(u)G^{-1}(u)\d u
\end{align}
with \textit{weight function} $\gamma(u) = \partial_- g(x)|_{x = 1 - u}, ~ 0 < u < 1$, 
and where $\partial_-$ denotes the derivative from the left. Moreover, we assume that $\int_0^1 | \gamma (u)|^2 \mathrm{d}u  < + \infty$. It is clear from the previous that a distortion risk metrics can also be expressed in terms of quantile functions. Specifically, we may  also write $\rho_g(G^{-1})$ instead of $\rho_g(G)$.

\begin{definition}
	\label{def_unimodality}
	\textit{A cdf G $\in \mathcal M^2$ is unimodal if $G$ is convex-concave, i.e., there exists $ x_m \in \mathbb{R}$ (called mode) such that  $G$ is convex on $  (-\infty,x_m) $ and concave on $(x_m, +\infty)$.}
\end{definition}

In what follows, we say that a (left) quantile function $G^{-1}$ is concave-convex if there exists an inflection point $\xi \in [0,1]$ such that $G^{-1}$ is concave on $(0,\xi)$ and convex on $(\xi,1)$. The following lemma shows how unimodality of a cdf can be expressed in terms of its quantile function. Its proof can be found in \cite{BKV2020}.

\begin{lemma}
	\label{lemma-continuous-quantiles}
	A cdf $G$  is unimodal if and only if $G^{-1}$ is continuous on $(0,1)$ and is either concave, convex, or concave-convex.
\end{lemma}

In what follows, quantile functions for which the corresponding distribution function is unimodal will be called unimodal quantile functions. We consider the uncertainty sets  $$\mathcal{F}_{U}=\left\{G\in \mathcal M^2, G\text{ is unimodal}\right\},~\mathcal{F}_{U,\xi}=\left\{G\in \mathcal{F}_{U}, \text{the inflection point is}~ \xi\right\},$$ and  $$\mathcal F_{U,\xi}(\mu, \sigma)=\left\{G\in \mathcal{F}_{U,\xi}: \int_{\R} x\d G=\mu, \int_{\R}x^2\d G=\mu^2+\sigma^2 \right\},$$
$$\mathcal F_{U,\xi}(\mu, \sigma, \epsilon)=\left\{G\in \mathcal{F}_{U,\xi}: \int_{\R} x\d G=\mu, \int_{\R}x^2\d G=\mu^2+\sigma^2, d_W(F,G)\leq \sqrt{\epsilon} \right\},$$where $F\in\mathcal M^2$, $\mu\in\R$, $\sigma>0$ and $\epsilon>0$. 
We also denote the set of all unimodal  distributions with non-fixed inflection point between $[\xi_1,\xi_2]$ with $0\leq \xi_1<\xi_2\leq 1$  by  $\mathcal F_{U,[\xi_1,\xi_2]}$, $\mathcal F_{U, [\xi_1,\xi_2]}(\mu, \sigma)$ and $\mathcal F_{U,[\xi_1,\xi_2]}(\mu, \sigma, \epsilon)$, respectively. Note that $\mathcal F_{U,[\xi_1,\xi_2]}=\cup_{\xi\in [\xi_1,\xi_2]}\mathcal F_{U,\xi}$, $\mathcal F_{U,[\xi_1,\xi_2]}(\mu, \sigma)=\cup_{\xi\in [\xi_1,\xi_2]}\mathcal F_{U,\xi}(\mu, \sigma)$ and $\mathcal F_{U,[\xi_1,\xi_2]}(\mu, \sigma, \epsilon)=\cup_{\xi\in [\xi_1,\xi_2]}\mathcal F_{U,\xi}(\mu, \sigma, \epsilon)$.


In this section, we study bounds for distortion risk metrics in the case of uncertainty sets that are Wasserstein balls containing unimodal distributions with given mean and variance, and with either given or non-fixed inflection points:

\begin{subequations}\label{eq: prob opt2}
	\begin{minipage}{0.45\textwidth}
		\begin{equation}\label{eq:problem sup_1}
			\sup_{G \in \mathcal{F}_{U,\xi}(\mu, \sigma)} \rho_g(G)
		\end{equation}
	\end{minipage}
	\begin{minipage}{0.45\textwidth}
		\begin{equation}\label{eq:problem sup_11}
			\sup_{G \in {F}_{U,\xi}(\mu, \sigma,\epsilon)} \rho_g(G)\,,
		\end{equation}
	\end{minipage}%
\end{subequations}\\%

as well as

\begin{subequations}\label{eq:unknown}
	\begin{minipage}{0.45\textwidth}
		\begin{equation}\label{eq:problem sup_2}
			\sup_{G \in \mathcal{F}_{U,[\xi_1,\xi_2]}(\mu, \sigma)} \rho_g(G)
		\end{equation}
	\end{minipage}
	\begin{minipage}{0.45\textwidth}
		\begin{equation}\label{eq:problem sup_22}
			\sup_{G \in {F}_{U,[\xi_1,\xi_2]}(\mu, \sigma,\epsilon)} \rho_g(G)\,.
		\end{equation}
	\end{minipage}%
\end{subequations}\\%
\subsection{Fixed inflection point} 
Recall that $\mathcal F^{-1}_{U,\xi}$ is the collection of all quantile functions of distribution functions in $\mathcal F_{U,\xi}$.
Note that $\mathcal F^{-1}_{U,\xi}$ is a closed convex cone, which implies that the $L_2$-projection of a function (with domain $(0,1)$) on this set is well defined and unique; see e.g., Theorems 2.1 of \cite{B65}. 
Denote by $\gamma^\uparrow_\xi$ the $L_2$-projection of $\gamma$ on  ${F}^{-1}_{U,\xi}$ and let $\hat{a}_\xi=E(\gamma^\uparrow_\xi(V))$ and $\hat{b}_\xi=\sqrt{Var(\gamma^\uparrow_\xi(V))}$ with $V \sim U(0,1)$. In light of Corollary 2.3 of \cite{B65}, we have $\hat{a}_\xi=\int_0^1\gamma_\xi^\uparrow(u)\d u=g(1)$ and $\hat{b}_\xi=\sqrt{\int_0^1\left(\gamma_\xi^\uparrow(u)-g(1)\right)^2\d u}$.  

\begin{proposition}[Bounds for unimodal distribution functions with a given inflection point]\label{Prop:2}  Suppose $\gamma^\uparrow_\xi$ is not a constant.  Then it holds that $$\sup_{G \in \mathcal{F}_{U,\xi}(\mu, \sigma)} \rho_g(G)=\mu g(1)+\sigma \sqrt{\int_0^1\left(\gamma_\xi^\uparrow(u)-g(1)\right)^2\d u},$$ and   
$h^{\uparrow}_\xi(u):= \mu+ \sigma \left(\frac{\gamma^\uparrow_\xi-\hat{a}_\xi}{\hat{b}_\xi} \right)$ is the unique worst-case quantile. 
\end{proposition}

Note that if unimodality is removed, then the uncertainty set becomes $\mathcal M_{\infty}(\mu, \sigma)$ . Let $\gamma^\uparrow$ be the projection of $\gamma$ on $\mathcal M_{\infty}^{-1}(\mu, \sigma)$. Then Corollay 3.9 in \cite{BPV24} and Theorem 5 and Remark 2 of \cite{PWW20} show that  $$\sup_{G \in \mathcal M_{\infty}(\mu, \sigma)} \rho_g(G)=\mu g(1)+\sigma \sqrt{\int_0^1\left(\gamma^\uparrow(u)-g(1)\right)^2\d u}$$ if  $\gamma^\uparrow$ is not a constant. Moreover, the worst-case  quantile is given by $\mu+ \sigma \left(\frac{\gamma^\uparrow-g(1)}{\sqrt{\int_0^1\left(\gamma^\uparrow(u)-g(1)\right)^2\d u}} \right)$.

By Theorem 2.8 of \cite{B65}, we have $$\int_0^1\left(\gamma^\uparrow(u)-g(1)\right)^2\d u\geq \int_0^1\left(\gamma_\xi^\uparrow(u)-g(1)\right)^2\d u+\int_0^1\left(\gamma^\uparrow(u)-\gamma_\xi^\uparrow(u)\right)^2\d u.$$
Hence, if $\gamma^\uparrow \notin\mathcal{F}^{-1}_{U,\xi}(\mu, \sigma)$, the information on unimodality can effectively reduce the worst-case value of the distortion risk metrics. 

Moreover, the worst-case distributions without the constraint of unimodality for $\ES$ and $\RVaR$ are two-point distributions, which is not desirable practically; see e.g.  \cite{BPV24} or \cite{PWW20}. However, with the constraint of unimodality, the worst-case distribution is typically not discrete.

In \cite{LSWY18} and \cite{BKV2020}, the Range-Value-at-Risk ($\gamma(u)=\frac{1}{\beta-\alpha}\id_{(\alpha,\beta)}$ with $0\leq \alpha<\beta\leq 1$) was considered with mean, variance and  unimodal constraints without fixing the inflection point. Our result in Proposition \ref{Prop:2} is valid for all distortion risk metrics with absolutely continuous distortion functions $g$, and it is more accurate with a fixed inflection point. By maximizing the worst-case values with different inflection points, we can immediately derive the results with unknown inflection points, which is discussed in Section \ref{Sec:unknown}.

Optimal solutions to problem \eqref{eq:problem sup_1} are thus obtained once the projection $\gamma^\uparrow_\xi$ of the function $\gamma$ on the set ${F}^{-1}_{U,\xi}$ is established. Whilst computing such projection in closed-form is in general very difficult, it can always be numerically obtained with any desired degree of precision.

The next proposition shows that for a step function $\gamma$, its projection on ${F}^{-1}_{U,\xi}$ is a piecewise linear function.
\begin{proposition}\label{stepfunction} Let $\gamma$ be a step function with $n$ steps, i.e., $\gamma(u)=\sum_{i=1}^n y_i\id_{(x_{i-1},x_i)}(u)$ with $0=x_0<x_1<\dots<x_n=1$ and $y_i\in \R$. Then $\gamma^\uparrow_\xi$ is a piecewise linear function with at most $2n+2$ pieces. More specifically, if $\xi=x_{i_0}$ for some $i_0=0,1,\cdots,n$, then $\gamma^\uparrow_\xi$ satisfies
 \begin{align}\label{stepprojection}\partial \gamma^\uparrow_\xi(u)/\partial u&=\sum_{i=1}^n\left(c_{i}^-\id_{(x_{i-1},a_i)}(u)+c_{i}^+\id_{(a_i,x_i)}(u)\right)
 \end{align}
 and 
 \begin{align}\label{step0}\gamma^\uparrow_\xi(1)&=g(1)+\sum_{i=1}^n \frac{c_i^-(a_i-x_{i-1})^2+c_i^+(x_i-a_i)^2}{2}
 \end{align}
 with the parameters constrained in $\mathcal D_n:=\{(\mathbf a, \mathbf c): a_{i}\in [x_{i-1},x_i],~i=1,\cdots,n$,  $c_1^-\geq c_{1}^+\geq \cdots \geq c_{i_0}^-\geq c_{i_0}^+\geq 0$, $0\leq c_{i_0+1}^-\leq c_{i_0+1}^+\leq \cdots c_n^-\leq c_n^+\}$ with $\mathbf a=(a_1,\dots,a_n)$ and $\mathbf c=(c_1^-,c_1^+,\dots, c_n^-,c_n^+)$.

The optimal parameters $(\mathbf a^*, \mathbf c^*)$ are given by
 \begin{align}\label{Stepoptimization}
    \argmin_{(\mathbf a, \mathbf c)\in \mathcal D_n} \sum_{i=1}^n&\left\{(a_i-x_{i-1})\left[(e_i^+)^2+e_i^+e_i^-+(e_i^-)^2\right]\right.\nonumber\\
   &~\left. +(x_i-a_i)\left[(e_{i}^++c_i^+(x_i-a_i))^2+(e_{i}^++c_i^+(x_i-a_i))e_i^++(e_i^+)^2\right]\right\},
 \end{align}
 where $e_i^-=g(1)+\sum_{j=1}^n \frac{c_j^-(a_j-x_{j-1})^2+c_j^+(x_j-a_j)^2}{2}-\sum_{j=i}^{n} \left(c_j^-(a_j-x_{j-1})+c_j^+(x_j-a_j)\right)-y_i$, $e_i^+=e_i^-+c_i^-(a_i-x_{i-1}),~i=1,\dots,n$. 
 
 If $\xi\in (x_{i_0-1},x_{i_0})$ for some $i_0=1,\cdots,n$, then  we rewrite $\gamma(u)=\sum_{i=1}^{n+1} y_i'\id_{(x_{i-1}',x_i')}(u)$ with $0=x_0'<x_1'<\dots<x_{n+1}'=1$, $y_i'\in \R$ and $x_{i_0}'=\xi$. Then it is transferred to the previous case. 
\end{proposition}

As it is shown in Proposition \ref{stepfunction}, for a step function $\gamma$ with $n$ steps, finding its projection $\gamma^\uparrow_\xi$ amounts to solving an optimization problem \eqref{Stepoptimization} with $3n$ or $3n+3$  parameters depending on the location of $\xi$. In light of \eqref{stepprojection} and \eqref{step0}, we have
\begin{align*}\gamma^\uparrow_\xi(u)&=e_i^-+y_i+c_i^-(u-x_{i-1}),~u\in [x_{i-1},a_i),\\
\gamma^\uparrow_\xi(u)&=e_i^++y_i+c_i^+(u-a_i),~u\in [a_i,x_i),~i=1,\dots,n.
\end{align*}

For a general $\gamma$, it can be approximated with any precision using step functions $\gamma_n$.  Furthermore, let $\gamma^\uparrow_{\xi,n} $ denote the projection of $\gamma_n$ on ${F}^{-1}_{U,\xi}$, and let
$h^{\uparrow}_{\xi,n}=\mu+\sigma\frac{\gamma_{\xi,n}^\uparrow-\hat{a}_{\xi,n}}{\hat{b}_{\xi,n}}$, where $\hat{a}_{\xi,n}=\mathbb E(\gamma_{\xi,n}^\uparrow(V))$ and $\hat{b}_{\xi,n}=\sqrt{Var(\gamma_{\xi,n}^\uparrow(V))}=\sqrt{\int_0^1\left(\gamma_{\xi,n}^\uparrow(u)-\hat{a}_{\xi,n}\right)^2\d u}$ with $V\sim U(0,1)$. Let $\|\cdot\|_2$ denote the $L_2$ norm on $\mathcal F^{-1}_{U,\xi}$.
\begin{proposition}\label{Prop:approx} If $\gamma^{\uparrow}_{\xi}$ and $\gamma^{\uparrow}_{\xi,n}$ are not constants,  we have
$$\| \gamma^\uparrow_{\xi,n}  - \gamma^\uparrow_{\xi} \|_2\leq \|\gamma_n-\gamma\|_2,~~\|h^\uparrow_{\xi,n}  - h^\uparrow_{\xi} \|_2\leq \left(2+\frac{2\|\gamma\|_2+\|\gamma_n\|_2}{\hat{b}_{\xi}}\right)\frac{\sigma}{\hat{b}_\xi}\|\gamma_n-\gamma\|_2,$$
and
$$
|\rho_g(h^{\uparrow}_\xi)-\rho_g(h^{\uparrow}_{\xi,n})|\leq \left(2+\frac{2\|\gamma\|_2+\|\gamma_n\|_2}{\hat{b}_{\xi}}\right)\frac{\sigma \|\gamma\|_2}{\hat{b}_\xi}\|\gamma_n-\gamma\|_2.$$
\end{proposition}

The inequalities in Proposition \ref{Prop:approx} show that the errors can be controlled by the distance between the step function $\gamma_n$ and original function $\gamma$, which  guarantees the efficiency of our approximation method using step functions to approximate the worst-case quantiles and the worst-case distortion risk metrics.


Next, we see some concrete examples to find the projection of $\gamma$.
\begin{example}\label{Example2}
\begin{enumerate}[(i)]
\item For $\GD$, $\gamma(u)=1-2u,~u\in (0,1)$. Hence, $\gamma^\uparrow_\xi(u)=1-2u,~u\in (0,1)$.
\item For $\MMD$, $g(t)=t\wedge(1-t),~t\in [0,1]$. Hence, $\gamma(u)=-\id_{(0,1/2)}+\id_{(1/2,1)}$. If $\xi=1/2$, then 
$\gamma^\uparrow_\xi(u)=3(u-1/2),~u\in (0,1)$.

\end{enumerate}
\end{example}

Next, we consider the uncertainty set $\mathcal F_{U,\xi}(\mu, \sigma, \epsilon)$ with $F\in\mathcal M^2$, $\mu\in\R$, $\sigma>0$ and $\epsilon>0$. For $\lambda>0$, let $k_\lambda(u)=\gamma(u)+\lambda F^{-1}(u)$, and $k_{\lambda,\xi}^\uparrow$ be the $L_2$-projection of $k_\lambda$ on $\mathcal F_{U,\xi}^{-1}$ for $\xi\in [0,1]$. Moreover, for $\lambda>0$, let \begin{equation}\label{Eq:h}h_{\lambda,\xi}=\mu+\frac{k_{\lambda,\xi}^\uparrow-a_{\lambda,\xi}}{b_{\lambda,\xi}}\sigma,\end{equation}
where  $a_{\lambda,\xi}=E(k_{\lambda,\xi}^\uparrow(V))=g(1)+\lambda\mu_F$ and $b_{\lambda,\xi}=\sqrt{Var(k_{\lambda,\xi}^\uparrow(V))}$ with $V \sim U(0,1)$. 
Moreover, let $c_1=Corr(F^{-1}(V), \gamma_{\xi}^\uparrow(V))$. 

Next, we discuss the range of $\epsilon$ for the problem \eqref{eq:problem sup_11} that is meaningful to study.
Let $F_\xi^{-1,\uparrow}$ be the $L_2$-projection of $F^{-1}$ on $\mathcal F_{U,\xi}^{-1}$ and $\hat{c}_0=Corr(F^{-1}(V),F_\xi^{-1,\uparrow}(V))$ if $F_\xi^{-1,\uparrow}$ is not a constant.  Clearly, $E(F_\xi^{-1,\uparrow}(V))=\mu_F$. 
\begin{lemma}\label{empty} Suppose $F_\xi^{-1,\uparrow}$ is not a constant. If $\epsilon<(\mu_F-\mu)^2+(\sigma_F-\sigma)^2+2\sigma_F\sigma(1-\hat{c}_0)$, then
$\mathcal F_{U,\xi}^{-1}(\mu,\sigma, \epsilon)=\varnothing$; If $\epsilon=(\mu_F-\mu)^2+(\sigma_F-\sigma)^2+2\sigma_F\sigma(1-\hat{c}_0)$, then  $\mathcal F_{U,\xi}^{-1}(\mu,\sigma)=\left\{\frac{F_\xi^{-1,\uparrow}-\mu_{F}}{\sigma_F^\uparrow}\sigma+\mu\right\}$, where $\sigma_F^\uparrow=\sqrt{Var(F_\xi^{-1,\uparrow}(V))}$.
\end{lemma}
 Due to Lemma \ref{empty}, in what follows, we only focus on the case $\epsilon>(\mu_F-\mu)^2+(\sigma_F-\sigma)^2+2\sigma_F\sigma(1-\hat{c}_0)$, where $\mathcal F_{U,\xi}$ contains infinitely many distributions.

The following lemma shows the continuity of $Corr(F^{-1}(V), k_{\lambda,\xi}^\uparrow(V))$ with respect to $\lambda$, which is crucial to our main result in this subsection.
\begin{lemma}\label{le:continuity} Suppose $F^{-1}\in \mathcal M^2$, $F_\xi^{-1,\uparrow}$ is not a constant,  and $k_{\lambda,\xi}^\uparrow$ are not constants for all $\lambda>0$. Then we have  $Corr(F^{-1}(V), k_{\lambda,\xi}^\uparrow(V))$ is continuous with respect to $\lambda$ on $[0,\infty)$, and
$$\lim_{\lambda\to\infty} Corr(F^{-1}(V), k_{\lambda,\xi}^\uparrow(V))=Corr(F^{-1}(V),F_\xi^{-1,\uparrow}(V)).$$
\end{lemma}

Note that in Lemma \ref{le:continuity}, the value of $\hat{c}_0$ depends on whether $F^{-1}\in \mathcal F^{-1}_{U,\xi}$. If $F^{-1}\in \mathcal F^{-1}_{U,\xi}$, then $\hat{c}_0=1$; If $F^{-1}\notin \mathcal F^{-1}_{U,\xi}$, one can easily check that
$\hat{c}_0<1.$ 
\begin{theorem}\label{Th:unimodal}
Assume $F^{-1}\in \mathcal M^2$, $F_\xi^{-1,\uparrow}$ is not a constant, and  $k_{\lambda,\xi}^\uparrow$ are not constants for all $\lambda\geq 0$.
\begin{enumerate}
\item[(i)] If $(\mu_F-\mu)^2+(\sigma_F-\sigma)^2+2\sigma_F\sigma(1-\hat{c}_0)<\epsilon<(\mu_F-\mu)^2+(\sigma_F-\sigma)^2+2\sigma_F\sigma(1-c_1)$, then
 $h_{\lambda_\epsilon,\xi}(u)$ given by \eqref{Eq:h} is the unique worst-case quantile
	to  problem \eqref{eq:problem sup_11} and $\sup_{G \in \mathcal F_{U,\xi}(\mu, \sigma,\epsilon)} \rho_g(G)=\mu g(1)+\frac{\sigma}{b_{\lambda_\epsilon,\xi}}\left(\int_0^1 k_{\lambda_\epsilon,\xi}^\uparrow(u)\gamma(u)\d u-g(1)(g(1)+\lambda_\epsilon\mu_F)\right)$, where $\lambda_\epsilon>0$ is the solution of $d_{W}(h_{\lambda,\xi}, F^{-1})=\sqrt{\epsilon}$. 
\item[(ii)]
If $\epsilon\geq (\mu_F-\mu)^2+(\sigma_F-\sigma)^2+2\sigma_F\sigma(1-c_1)$, then the unique worst-case quantile to problem \eqref{eq:problem sup_11} is $h^{\uparrow}_\xi(u)= \mu+ \sigma \left(\frac{\gamma^\uparrow_\xi-\hat{a}_\xi}{\hat{b}_\xi} \right)$.
\end{enumerate}	 
\end{theorem}

For the case with absolutely continuous $g$,  Theorem \ref{Th:unimodal} improves the results of Theorem \ref{Th:main} by imposing unimodality with a fixed inflection point on the underlying distributions.  This implies that the worst-case value is reduced by adding this additional information in the uncertainty sets. The computation of the worst-case values is not straightforward due to the complexity of the projection and no explicit  expression for $\lambda_\epsilon$. The numerical method  heavily relies on Propositions \ref{stepfunction} and \ref{Prop:approx}.


\subsection{Unknown inflection points}\label{Sec:unknown}
Different from $\mathcal{F}^{-1}_{U,\xi}$, the set $\mathcal{F}^{-1}_{U,[\xi_1,\xi_2]}$ with $0\leq \xi_1<\xi_2\leq 1$ is not convex. Hence, Theorem 2.1 of \cite{B65} cannot guarantee the existence and uniqueness of the projection. To compute the worst-case distortion risk metrics, we can apply Proposition \ref{Prop:2} and Theorem \ref{Th:unimodal} and the following relations:
\begin{align}\sup_{G \in \mathcal{F}_{U,[\xi_1,\xi_2]}(\mu, \sigma)} \rho_g(G)&=\sup_{\xi\in [\xi_1,\xi_2]}\sup_{G \in \mathcal{F}_{U,\xi}(\mu, \sigma)} \rho_g(G),\label{eq:generalformula}\\
\sup_{G \in \mathcal{F}_{U,[\xi_1,\xi_2]}(\mu, \sigma,\epsilon)} \rho_g(G)&=\sup_{\xi\in [\xi_1,\xi_2]}\sup_{G \in \mathcal{F}_{U,\xi}(\mu, \sigma,\epsilon)} \rho_g(G).\label{eq:generalformula1}
\end{align}
The above expressions, though complex, are amenable to numerical evaluation. 
Luckily, the problem \eqref{eq:problem sup_2} can be solved using the same method as the case with given inflection point although $\mathcal{F}^{-1}_{U,[\xi_1,\xi_2]}$ is not convex.  The existence of the $L_2$-projection of a function over $(0,1)$ on the set $\mathcal{F}^{-1}_{U,[\xi_1,\xi_2]}$ can be shown as below.
\begin{lemma}\label{lem:unimodal} For any $\gamma$ satisfying $\int_0^1 | \gamma (u)|^2 \mathrm{d}u  < + \infty$, there exists $\gamma_{\xi_1,\xi_2}^\uparrow\in \mathcal{F}^{-1}_{U,[\xi_1,\xi_2]}$ such that $\gamma_{\xi_1,\xi_2}^\uparrow\in \arg\min_{h \in \mathcal F^{-1}_{U,[\xi_1,\xi_2]}}\|\gamma-h\|_2$. 
\end{lemma}

Note that in Lemma \ref{lem:unimodal}, the uniqueness of the projection is not stated because the uniqueness may not hold due to the lack of convexity for the set $\mathcal{F}^{-1}_{U,[\xi_1,\xi_2]}$.

Next, we study bounds for distortion risk metrics in case that the inflection point is not known. By Lemma \ref{lem:unimodal},  the projection of $\gamma$ onto ${F}^{-1}_{U,[\xi_1,\xi_2]}$ is well-defined. Denote by $\gamma_{\xi_1,\xi_2}^\uparrow$ as one of the $L_2$-projection of $\gamma$ on  $\mathcal{F}^{-1}_{U,[\xi_1,\xi_2]}$, and let $a_{\xi_1,\xi_2}=E(\gamma_{\xi_1,\xi_2}^\uparrow(V))$ and $b_{\xi_1,\xi_2}=\sqrt{Var(\gamma_{\xi_1,\xi_2}^\uparrow(V))}$ with $V \sim U(0,1)$.

\begin{proposition}[Bounds for distortion risk measures for unimodal distribution functions]\label{Prop:3}  Suppose $\gamma_{\xi_1,\xi_2}^\uparrow$ is not a constant.  Then  $h_{\xi_1,\xi_2}^{\uparrow}(u):= \mu+ \sigma \left(\frac{\gamma_{\xi_1,\xi_2}^\uparrow-a_{\xi_1,\xi_2}}{b_{\xi_1,\xi_2}} \right)$ is a 
	worst-case quantile to the problem \eqref{eq:problem sup_2}.
\end{proposition}
Note that the worst-case distribution of Problem \eqref{eq:problem sup_2}  given in Proposition \ref{Prop:3} may not be unique due to the non-convexity of $\mathcal{F}^{-1}_{U,[\xi_1,\xi_2]}$. This fact can also been seen from \eqref{eq:generalformula}, where there might be two different $\xi\in [\xi_1,\xi_2]$ such that the largest value is obtained at the quantile functions with two different inflection points respectively.
Moreover, the conclusion in Proposition \ref{Prop:3} covers some result in \cite{LSWY18} and \cite{BKV2020}, where a special case of $\gamma$, $ \gamma(u)=\frac{1}{\beta-\alpha}\id_{(\alpha,\beta)}(u)$ with $0<\alpha<\beta\leq 1$, was considered. 

Due to the non-convexity of $\mathcal{F}^{-1}_{U,[\xi_1,\xi_2]}$, it is difficult to show the similar result as in Lemma \ref{le:continuity} for the case with Wasserstein constraint and unknown inflection point.  Hence, the problem \eqref{eq:problem sup_22} cannot be solved using the same method as that of Theorem \ref{Th:unimodal}. However, we can apply Theorem \ref{Th:unimodal} and \eqref{eq:generalformula1} to compute \eqref{eq:problem sup_22}.

\section{Robust portfolio optimization}\label{Sec:portfoliooptimisation}

In this section, we specialize some of our results to robust portfolio optimization under an uncertainty set defined by mean–variance constraints combined with a probability constraint based on the Wasserstein metric for the random return vector.

\subsection{Mean-variance and Wasserstein distance constraints}

As in this paper positive outcomes represent losses, we let $X_i$, $i=1,\dots,n$, denote the negative returns from investing one dollar in each of $n$ different assets in the market, and let   $(w_1,\dots,w_n)\in \mathcal A \subseteq \Delta_n$ represent the portfolio allocation, where $\Delta_n=\{(w_1,\dots,w_n): w_i\geq 0,~\sum_{i=1}^n w_i=1\}$. The negative return of the portfolio is therefore given by $\sum_{i=1}^n w_iX_i$. The Wasserstein metric between two $n$-dimensional distributions $F$ and $G$ is defined by  
$$d_{W}^{(n)}(F,G)=\inf_{\mathbf Z_1\sim F,\mathbf Z_2\sim G}\left(\mathbb E(\|\mathbf Z_1-\mathbf Z_2\|_2^2)\right)^{1/2};$$
see, for example, \cite{BCZ22}. 

Assume that only partial information about the vector $\mathbf X= (X_1, \cdots, X_n)$ is available, namely the means $\mu_i$, the positive definite covariance matrix $\Sigma_0$, and the Wasserstein distance between $F_{\mathbf X}$ and a reference distribution $F_{\mathbf X_0}$, satisfying $d_{W}^{(n)}(F_{\mathbf X},F_{\mathbf X_0})\leq\sqrt{\epsilon}$ for some $\epsilon>0$. 

To enhance practical relevance, we assume that the underlying distribution and the reference distribution share the same mean vector and covariance matrix; that is, $\mathbb E(\mathbf X_{0})=\boldsymbol \mu=(\mu_1,\dots,\mu_n)$ and $Cov(\mathbf X_0)=\Sigma_0$. 

The uncertainty set for the negative return of the portfolio $\sum_{i=1}^n w_iX_i$ can then be defined as
$$\M_{\mathbf w,\epsilon}=\left\{F_{\sum_{i=1}^n w_iX_i}:  E(\mathbf X)=\boldsymbol\mu, Cov(\mathbf X)=\Sigma_0\right\}\cap \left\{F_{\sum_{i=1}^n w_iX_i}: d_{W}^{(n)}(F_{\mathbf X},F_{\mathbf X_0})\leq \sqrt{\epsilon}\right\}.$$

The robust portfolio optimization problem is to solve
\begin{align}\label{PS1}
\argmin_{\mathbf w\in \mathcal A}\sup_{G\in\M_{\mathbf w,\epsilon}} \rho_g(G).
\end{align}

An example of $\mathcal A$ is $\mathcal A=\{\mathbf w\in \Delta_n: -\mathbf w^\top \boldsymbol{\mu}\geq a \}$ for some $a>0$, which requires the portfolio return to be greater than $a$.

The uncertainty set $\M_{\mathbf w,\epsilon}$ is novel and differs from those considered in the existing literature, such as \cite{BCZ22}, \cite{BPV24}, \cite{PWW20}, and \cite{MWW24}. Specifically, in $\M_{\mathbf w,\epsilon}$, the Wasserstein probability constraint is imposed on the multivariate distribution of the negative return vector $\mathbf X$, whereas in \cite{BPV24} it is imposed on the univariate distribution of the portfolio return $\sum_{i=1}^n w_iX_i$.

We can recast $\M_{\mathbf w,\epsilon}$ as an uncertainty set involving only the univariate distribution $F_{\sum_{i=1}^n w_iX_i}$, which plays an important role in applying our results derived in Section \ref{Sec:general}.
\begin{proposition}\label{prop:uncertainty}
We have
$\M_{\mathbf w,\epsilon}=\M_{\epsilon\|\mathbf w\|_2^2}(\mathbf w^\top\boldsymbol\mu,\sqrt{\mathbf w^\top\Sigma_0\mathbf w})$ with $F=F_{\mathbf w^\top \mathbf X_0 }$.
\end{proposition}


Proposition \ref{prop:uncertainty} implies that problem \eqref{PS1} is equivalent to 
$$\argmin_{\mathbf w\in \mathcal A}\sup_{G\in\M_{\epsilon\|\mathbf w\|_2^2}(\mathbf w^\top\boldsymbol\mu,\sqrt{\mathbf w^\top\Sigma_0\mathbf w})} \rho_g(G),$$
which allows us to apply the results in Theorems \ref{Th:main} and \ref{Th:main1}. We therefore obtain the following result. For $\lambda\geq 0$ and $\boldsymbol w\in \mathcal A$, let $V_{\mathbf w, \lambda}:=Var\left((g_\lambda^*)'(V)\right)$, where $g_\lambda(t)=g(t)+\lambda \int_{1-t}^1 F_{\mathbf w^\top \mathbf X_0}^{-1}(s)\d s,~t\in (0,1)$.
\begin{proposition}\label{Prop:PS} Problem \eqref{PS1} can be solved as follows.
    \begin{enumerate}[(i)] 
    \item If $g$ is concave and $(g^*)'$ is not a constant, then the solution to problem \eqref{PS1} is given by
    $$\argmin_{\mathbf w\in \mathcal A}\left\{\mathbf w^\top\boldsymbol \mu g(1)+\frac{\sqrt{\mathbf w^\top\Sigma_0\mathbf w}}{b_{\mathbf w}}\left(V_g+\lambda_{\mathbf w}C_{g,F_{\mathbf w^\top \mathbf X_0}}\right)\right\},$$
    where $V_{g}=Var(g'(1-V))$, $C_{g,F_{\mathbf w^\top \mathbf X_0}}=Cov(F_{\mathbf w^\top \mathbf X_0}^{-1}(V),g'(1-V))$, and 
$$b_{\mathbf w}=\sqrt{V_g+2\lambda_{\mathbf w}C_{g,F_{\mathbf w^\top X_0}}+\lambda_{\mathbf w}^2\mathbf w^\top\Sigma_0\mathbf w}, ~~\lambda_{\mathbf w}=\frac{-C_{g,F_{\mathbf w^\top \mathbf X_0}}+\sqrt{\frac{\left(C_{g,F_{\mathbf w^\top \mathbf X_0}}^2-V_{g}\mathbf w^\top\Sigma_0\mathbf w\right) A_{\mathbf w}^2}{A_{\mathbf w}^2-\left(\mathbf w^\top\Sigma_0\mathbf w\right)^2}}}{\mathbf w^\top\Sigma_0\mathbf w}$$
with $A_{\mathbf w}=\left(\mathbf w^\top\Sigma_0\mathbf w-\frac{\epsilon\|\mathbf w\|_2^2}{2}\right)\vee\left(\sqrt{\mathbf w^\top\Sigma_0\mathbf w/V_g}C_{g,F_{\mathbf w^\top \mathbf X_0}}\right)$.
    \item For $\rho=\IQD_{\alpha}^+$ or $\IQD_{\alpha}^-$ with $\alpha\in (0,1/2)$, the optimal robust portfolio position is 
   $$\argmin_{\mathbf w\in \mathcal A}\left\{\frac{\sqrt{\mathbf w^\top\Sigma_0\mathbf w}}{\sqrt{V_{\mathbf w,\lambda_{\mathbf w}}}}\left(\frac{1+\lambda_{\mathbf w}\int_{1-\alpha}^{1-t_{\alpha,\lambda_{\mathbf w}}} F_{\mathbf w^\top \mathbf X_0}^{-1}(s)\d s}{\alpha-t_{\alpha,\lambda_{\mathbf w}}}-\frac{\lambda_{\mathbf w}\int_{1-\hat{t}_{\alpha, \lambda_{\mathbf w}}}^{\alpha} F_{\mathbf w^\top \mathbf X_0}^{-1}(s)\d s-1}{\hat{t}_{\alpha, \lambda_{\mathbf w}}-1+\alpha}\right)\right\} ,$$
where $\lambda_{\mathbf w}$ is the solution to $d_{W}(F_{\mathbf w^\top \mathbf X_0},H_\lambda)=\sqrt{\epsilon}\|{\mathbf w}\|_2$ if $d_{W}(F_{\mathbf w^\top \mathbf X_0},H_0)>\sqrt{\epsilon}\|{\mathbf w}\|_2$; otherwise, $\lambda_{\mathbf w}=0$.
\item For $\rho=\VaR_\alpha$ or $\VaR_\alpha^+$ with $\alpha\in (0,1)$, the optimal robust portfolio position is given by 
$$\argmin_{\mathbf w\in \mathcal A}\left\{\mathbf w^\top\boldsymbol \mu+\frac{\sqrt{\mathbf w^\top\Sigma_0\mathbf w}}{\sqrt{V_{\mathbf w,\lambda_{\mathbf w}}}}\left(\frac{1+\lambda_{\mathbf w}\int_{\alpha}^{1-t_{1-\alpha,\lambda_{\mathbf w}}} F_{\mathbf w^\top \mathbf X_0}^{-1}(s)\d s}{1-\alpha-t_{1-\alpha,\lambda_{\mathbf w}}}-1-\lambda_{\mathbf w}\mathbf w^\top\boldsymbol \mu\right)\right\} ,$$
where  $\lambda_{\mathbf w}$ is the solution to $d_{W}(F_{\mathbf w^\top \mathbf X_0},H_\lambda)=\sqrt{\epsilon}\|{\mathbf w}\|_2$ if $d_{W}(F_{\mathbf w^\top \mathbf X_0},H_0)>\sqrt{\epsilon}\|{\mathbf w}\|_2$; otherwise, $\lambda_{\mathbf w}=0$.
\item For $\rho=\GlueVaR_{\beta,\alpha}^{h_1,h_2}$ with $0<\alpha< \beta<1$ and $0<h_1<h_2<1$ satisfying $\frac{h_1}{1-\beta}\geq \frac{h_2 - h_1}{\beta - \alpha}$, the solution to problem \eqref{PS1} is given by
\begin{align*}&\argmin_{\mathbf w\in \mathcal A}\left\{\mathbf w^\top\boldsymbol \mu-\frac{\sqrt{\mathbf w^\top\Sigma_0\mathbf w}}{\sqrt{V_{\mathbf w,\lambda_{\mathbf w}}}}\left(1+\lambda_{\mathbf w}\mathbf w^\top\boldsymbol \mu-(1-h_2)c_{\alpha,\beta,\lambda_{\mathbf w}}^{h_1,h_2}\right)\right.\\
&+\frac{\sqrt{\mathbf w^\top\Sigma_0\mathbf w}}{\sqrt{V_{\mathbf w,\lambda_{\mathbf w}}}}\frac{h_1}{1-\beta}\left(c_{\alpha,\beta,\lambda_{\mathbf w}}^{h_1,h_2}(1-u_{\alpha,\beta,\lambda_{\mathbf w}}^{h_1,h_2}-\beta)_++\frac{h_1((1-\beta)\wedge u_{\alpha,\beta,\lambda_{\mathbf w}}^{h_1,h_2})}{1-\beta}\right.\\
&\quad\left.+\lambda_{\mathbf w} \int_{\beta\vee (1-u_{\alpha,\beta,\lambda_{\mathbf w}}^{h_1,h_2})}^1 F_{\mathbf w^\top \mathbf X_0 } ^{-1}(s)\d s\right)\\
&+\frac{\sqrt{\mathbf w^\top\Sigma_0\mathbf w}}{\sqrt{V_{\mathbf w,\lambda_{\mathbf w}}}}\frac{h_2-h_1}{\beta-\alpha}\left(c_{\alpha,\beta,\lambda_{\mathbf w}}^{h_1,h_2}(\beta\wedge(1-u_{\alpha,\beta,\lambda_{\mathbf w}}^{h_1,h_2})-\alpha)+\frac{(h_2-h_1)(\beta-1+ u_{\alpha,\beta,\lambda_{\mathbf w}}^{h_1,h_2})_+}{\beta-\alpha}\right.\\
&~~~~~~~~~~~~~~~~~~~~~~\left.\left.+\lambda_{\mathbf w} \int_{\beta\wedge (1-u_{\alpha,\beta,\lambda_{\mathbf w}}^{h_1,h_2})}^\beta F_{\mathbf w^\top \mathbf X_0}^{-1}(s)\d s\right)\right\},
\end{align*}
where $c_{\alpha,\beta,\lambda}^{h_1,h_2}$ is given in (iii) of Corollary \ref{Cor:2}, and $\lambda_{\mathbf w}$ is the solution to $d_{W}(F_{\mathbf w^\top \mathbf X_0},H_\lambda)=\sqrt{\epsilon}\|{\mathbf w}\|_2$ if $d_{W}(F_{\mathbf w^\top \mathbf X_0},H_0)>\sqrt{\epsilon}\|{\mathbf w}\|_2$; otherwise, $\lambda_{\mathbf w}=0$.
\end{enumerate} 
\end{proposition}
\subsection{Elliptical reference distributions}
In Proposition \ref{Prop:PS}, the optimal portfolio position $\mathbf w$ depends on the reference distribution $F_{\mathbf X_0}$. To simplify the analysis and enhance practical relevance, we assume throughout this subsection that the reference distribution is elliptical, i.e., $F_{\mathbf X_0}\sim E_n(\boldsymbol \mu, \Sigma_0,\psi)$, where $\boldsymbol \mu$ is the mean, $\Sigma_0$ is the covariance matrix and $\psi$ is the characteristic generator. Note that elliptical distributions include the family of multivariate normal distributions, multivariate $t$-distributions and multivariate normal variance mixture distributions as special cases. It follows that $F_{\mathbf w^\top \mathbf X_0}\sim E_1(\mathbf w^\top\boldsymbol\mu,\mathbf w^\top\Sigma_0\mathbf w,\psi)$. Let $F_0=F_{\frac{\mathbf w^\top \mathbf X_0-\mathbf w^\top \boldsymbol\mu}{\sqrt{\mathbf w^\top\Sigma_0\mathbf w}}}$ and then we have $F_0\sim E_1(0,1,\psi)$. Hence,
$C_{g,F_{\mathbf w^\top \mathbf X_0}}=\sqrt{\mathbf w^\top\Sigma_0\mathbf w}C_0$ with $C_0=Cov(F_0^{-1}(V),g'(1-V))$.
For this special $F_{\mathbf X_0}$, we can simplify Proposition \ref{Prop:PS} as follows.

\begin{align}\label{PS2}&\argmin_{\mathbf w\in \mathcal A}\sup_{G\in\M_{\epsilon\|\mathbf w\|_2^2}(\mathbf w^\top\boldsymbol\mu,\sqrt{\mathbf w^\top\Sigma_0\mathbf w})} \rho_g(G)\nonumber\\
&=\argmin_{\mathbf w\in \mathcal A}\sup_{G\in\M_{\epsilon_{\mathbf w}}(0,1)}\left\{ \mathbf w^\top\boldsymbol \mu g(1)+\sqrt{\mathbf w^\top\Sigma_0\mathbf w}\rho_g(G)\right\},
\end{align}
where $\epsilon_{\mathbf w}=\frac{\epsilon\|\mathbf w\|_2^2}{\mathbf w^\top\Sigma_0\mathbf w}$, and the reference distribution in $\M_{\epsilon_{\mathbf w}}(0,1)$ is $F_0\sim E_1(0,1,\psi)$.


\begin{corollary}\label{cor:3} Suppose $F_{\mathbf X_0}\sim E_n(\boldsymbol \mu, \Sigma_0,\psi)$. If $g$ is concave and $(g^*)'$ is not a constant, then the solution to problem \eqref{PS2} is given by
    \begin{equation}\label{PS:Numerical}\argmin_{\mathbf w\in \mathcal A}\left\{\mathbf w^\top\boldsymbol \mu g(1)+\frac{\sqrt{\mathbf w^\top\Sigma_0\mathbf w}}{\sqrt{V_g+2C_0B_{\mathbf w}+B_{\mathbf w}^2}}\left(V_g+C_0B_{\mathbf w}\right)\right\},
 \end{equation}
where $$B_{\mathbf w}=-C_0+\sqrt{\frac{ (C_0^2-V_{g})A_{\mathbf w}^2}{A_{\mathbf w}^2-\left(\mathbf w^\top\Sigma_0\mathbf w\right)^2}}$$
with $A_{\mathbf w}=\left(\mathbf w^\top\Sigma_0\mathbf w-\frac{\epsilon\|\mathbf w\|_2^2}{2}\right)\vee\left(C_0\mathbf w^\top\Sigma_0\mathbf w/\sqrt{V_g}\right)$.
\end{corollary}

As shown in Corollary \ref{cor:3}, the expected negative returns disappear from the objective function of \eqref{PS:Numerical} if $g(1)=0$. However, they can still influence investors' decisions through $\mathcal A$ when $\mathcal A$ is defined in terms of $\boldsymbol\mu$.

For portfolio optimization using $\IQD$ and $\VaR$, we introduce the following notation.
For $\alpha\in (0,1)$, $\lambda\geq 0$ and $\mathbf w\in \mathcal A$, let
\begin{equation*}
t_{\alpha, \mathbf w,\lambda}=\inf\left\{t\in [0,\alpha): \frac{1/\sqrt{\mathbf w^\top\Sigma_0\mathbf w}+\lambda\int_{1-\alpha}^{1-t} F_0^{-1}(s)\d s}{\alpha-t}\geq\lambda F_0^{-1}(1-t)\right\},
\end{equation*}
and \begin{align*}
 \hat{t}_{\alpha, \mathbf w,\lambda}=\sup\left\{t\in (1-\alpha,1]: \frac{\lambda\int_{1-t}^{\alpha} F_0^{-1}(s)\d s-1/\sqrt{\mathbf w^\top\Sigma_0\mathbf w}}{t-\alpha+1}\leq\lambda F_0^{-1}(1-t)\right\}.
 \end{align*}
\begin{corollary}\label{cor:4} Suppose $F_{\mathbf X_0}\sim E_n(\boldsymbol \mu, \Sigma_0,\psi)$.
If $\rho=\IQD_{\alpha}^+$ or $\IQD_{\alpha}^-$ with $\alpha\in (0,1/2)$, then the solution to problem \eqref{PS2} is given by
   \begin{align*}&\argmin_{\mathbf w\in \mathcal A}\left\{\sqrt{\frac{\mathbf w^\top\Sigma_0\mathbf w}{V_{\mathbf w,\lambda_{\mathbf w}}}}\left(\frac{1+\lambda_{\mathbf w}\sqrt{\mathbf w^\top\Sigma_0\mathbf w}\int_{1-\alpha}^{1-t_{\alpha,\mathbf w,\lambda_{\mathbf w}}} F_{0}^{-1}(s)\d s}{\alpha-t_{\alpha,\mathbf w,\lambda_{\mathbf w}}}\right.\right.\\
   &\left.\left.~~~~~~~~~~~~~~~~~~~~~~~~~-\frac{\lambda_{\mathbf w}\sqrt{\mathbf w^\top\Sigma_0\mathbf w}\int_{1-\hat{t}_{\alpha,\mathbf w, \lambda_{\mathbf w}}}^{\alpha} F_{0}^{-1}(s)\d s-1}{\hat{t}_{\alpha,\mathbf w, \lambda_{\mathbf w}}-1+\alpha}\right)\right\},
   \end{align*}
where 
 $$\lambda_{\mathbf w}=\eta_{\mathbf w}\id_{\left\{2\mathbf w^\top\Sigma_0\mathbf w\left(1-\frac{\int_{1-\alpha}^1 F_0^{-1}(t)\d t-\int_0^\alpha F_0^{-1}(t)\d t}{\sqrt{2\alpha}}\right)>\epsilon\|{\mathbf w}\|_2^2\right\}}$$ with $\eta_{\mathbf w}\in (0,\infty)$ being the solution of
\begin{align*}\left(1-\frac{\epsilon\|\mathbf w\|_2^2}{2\mathbf w^\top\Sigma_0\mathbf w}\right)\sqrt{V_{\mathbf w, \lambda}}&=\frac{1+\lambda\sqrt{\mathbf w^\top\Sigma_0\mathbf w}\int_{1-\alpha}^{1-t_{\alpha,\mathbf w,\lambda}} F_0^{-1}(s)\d s }{\alpha-t_{\alpha,\mathbf w,\lambda}}\int_{1-\alpha}^{1-t_{\alpha,\mathbf w,\lambda}}F_0^{-1}(t)\d t\\
&+\frac{\lambda\sqrt{\mathbf w^\top\Sigma_0\mathbf w}\int_{1-\hat{t}_{\alpha,\mathbf w,\lambda}}^{\alpha}F_0^{-1}(s)\d s-1}{\hat{t}_{\alpha,\mathbf w,\lambda}-1+\alpha}\int_{1-\hat{t}_{\alpha,\mathbf w,\lambda}}^{\alpha}F_0^{-1}(t)\d t\\
 &+\lambda \sqrt{\mathbf w^\top\Sigma_0\mathbf w}\int_{(0,1-\hat{t}_{\alpha,\mathbf w,\lambda})\cup(\alpha,1-\alpha)\cup (1-t_{\alpha,\mathbf w,\lambda},1)}(F_0^{-1}(t))^2\d t.
\end{align*}
\end{corollary}

 \begin{corollary}\label{cor:5} Suppose $F_{\mathbf X_0}\sim E_n(\boldsymbol \mu, \Sigma_0,\psi)$.
If $\rho=\VaR_\alpha$ or $\VaR_\alpha^+$ with $\alpha\in (0,1)$, then the solution to problem \eqref{PS2} is given by
   \begin{align*}&\argmin_{\mathbf w\in \mathcal A}\left\{\mathbf w^\top\boldsymbol \mu+\sqrt{\frac{\mathbf w^\top\Sigma_0\mathbf w}{V_{\mathbf w,\lambda_{\mathbf w}}}}\left(\frac{1+\lambda_{\mathbf w}\sqrt{\mathbf w^\top\Sigma_0\mathbf w}\int_{\alpha}^{1-t_{1-\alpha,\mathbf w,\lambda_{\mathbf w}}} F_{0}^{-1}(s)\d s}{1-\alpha-t_{1-\alpha,\mathbf w,\lambda_{\mathbf w}}}-1\right)\right\},
   \end{align*}
where 
 $$\lambda_{\mathbf w}=\eta_{\mathbf w}\id_{\left\{2\mathbf w^\top\Sigma_0\mathbf w\left(1-\frac{\int_{\alpha}^1 F_0^{-1}(t)\d t}{\sqrt{\alpha(1-\alpha)}}\right)>\epsilon\|{\mathbf w}\|_2^2\right\}}$$ with $\eta_{\mathbf w}\in (0,\infty)$ being the solution of
\begin{align*}\left(1-\frac{\epsilon\|\mathbf w\|_2^2}{2\mathbf w^\top\Sigma_0\mathbf w}\right)\sqrt{V_{\mathbf w, \lambda}}&=\frac{1+\lambda\sqrt{\mathbf w^\top\Sigma_0\mathbf w}\int_{\alpha}^{1-t_{1-\alpha,\mathbf w,\lambda}} F_0^{-1}(s)\d s }{1-\alpha-t_{1-\alpha,\mathbf w,\lambda}}\int_{\alpha}^{1-t_{1-\alpha,\mathbf w,\lambda}}F_0^{-1}(t)\d t\\
 &+\lambda \sqrt{\mathbf w^\top\Sigma_0\mathbf w}\int_{(0,\alpha)\cup (1-t_{1-\alpha,\mathbf w,\lambda},1)}(F_0^{-1}(t))^2\d t.
\end{align*}
\end{corollary}

For portfolio optimization using $\GlueVaR$, we introduce the following notation.
For $0<\alpha<\beta<1$, $0<h_1<h_2<1$, $\lambda\geq 0$ and $\mathbf w\in \mathcal A$, let 
 \begin{align*}
 u_{\alpha,\beta,\mathbf w,\lambda}^{h_1,h_2}&=\inf\left\{t\in [0,1-\alpha): \frac{1-g_{\alpha,\beta}^{h_1,h_2}(t)+\lambda\sqrt{\mathbf w^\top\Sigma_0\mathbf w}\int^{1-t}_{\alpha} F_0^{-1}(s)\d s}{1-\alpha-t}\right.\nonumber\\
 &~~~~~~~~~~~~~\left.\geq \frac{h_1}{1-\beta}\id_{(0,1-\beta)}(t)+\frac{h_2 - h_1}{\beta - \alpha}\id_{[1-\beta,1-\alpha)}(t)+\lambda \sqrt{\mathbf w^\top\Sigma_0\mathbf w}F_0^{-1}(1-t)\right\}
 \end{align*}
 and $$d_{\mathbf w,\lambda}=1-h_2+\frac{h_1}{1-\beta}(1-u_{\alpha,\beta,\mathbf w, \lambda}^{h_1,h_2}-\beta)_+
+\frac{h_2-h_1}{\beta-\alpha}(\beta\wedge(1-u_{\alpha,\beta,\mathbf w, \lambda}^{h_1,h_2})-\alpha).$$

 \begin{corollary}\label{cor:6} Suppose $F_{\mathbf X_0}\sim E_n(\boldsymbol \mu, \Sigma_0,\psi)$.
     If $\rho=\GlueVaR_{\beta,\alpha}^{h_1,h_2}$ with $0<\alpha< \beta<1$ and $0<h_1<h_2<1$ satisfying $\frac{h_1}{1-\beta}\geq \frac{h_2 - h_1}{\beta - \alpha}$, then the solution to problem \eqref{PS2} is given by
\begin{align*}&\argmin_{\mathbf w\in \mathcal A}\left\{\mathbf w^\top\boldsymbol \mu-\frac{\sqrt{\mathbf w^\top\Sigma_0\mathbf w}}{\sqrt{V_{\mathbf w,\lambda_{\mathbf w}}}}\left(1-\frac{h_1^2((1-\beta)\wedge u_{\alpha,\beta,\mathbf w, \lambda_{\mathbf w}}^{h_1,h_2})}{(1-\beta)^2}-\frac{(h_2-h_1)^2(\beta-1+ u_{\alpha,\beta,\mathbf w, \lambda_{\mathbf w}}^{h_1,h_2})_+}{(\beta-\alpha)^2}\right)\right.\\
&+\frac{\sqrt{\mathbf w^\top\Sigma_0\mathbf w}}{\sqrt{V_{\mathbf w,\lambda_{\mathbf w}}}}\frac{1-g_{\alpha,\beta}^{h_1,h_2}(u_{\alpha,\beta,\mathbf w,\lambda}^{h_1,h_2})}{1-\alpha-u_{\alpha,\beta,\mathbf w,\lambda}^{h_1,h_2}}d_{\mathbf w,\lambda_{\mathbf w}}+\frac{\lambda_{\mathbf w}\mathbf w^\top\Sigma_0\mathbf w}{\sqrt{V_{\mathbf w,\lambda_{\mathbf w}}}}\frac{\int^{1-u_{\alpha,\beta,\mathbf w,\lambda_{\mathbf w}}^{h_1,h_2}}_{\alpha} F_0^{-1}(s)\d s}{1-\alpha-u_{\alpha,\beta,\mathbf w,\lambda_{\mathbf w}}^{h_1,h_2}}d_{\mathbf w,\lambda_{\mathbf w}}\\
&\left.+\frac{\lambda_{\mathbf w}\mathbf w^\top\Sigma_0\mathbf w}{\sqrt{V_{\mathbf w,\lambda_{\mathbf w}}}}\left(\frac{h_1}{1-\beta}\int_{\beta\vee (1-u_{\alpha,\beta,\mathbf w, \lambda_{\mathbf w}}^{h_1,h_2})}^1 F_0 ^{-1}(s)\d s+\frac{h_2-h_1}{\beta-\alpha}\int_{\beta\wedge (1-u_{\alpha,\beta,\mathbf w, \lambda_{\mathbf w}}^{h_1,h_2})}^\beta F_0^{-1}(s)\d s\right)\right\},
\end{align*}
where
 $$\lambda_{\mathbf w}=\eta_{\mathbf w}\id_{\left\{2\mathbf w^\top\Sigma_0\mathbf w\left(1-\frac{\left(\frac{1}{1-\alpha}\bigvee\frac{h_1}{1-\beta}\right)\int_{\beta}^1 F_0^{-1}(t)\d t+\left(\frac{1}{1-\alpha}\bigwedge\frac{1-h_1}{\beta-\alpha}\right)\int_{\alpha}^{\beta} F_0^{-1}(t)\d t}{\sqrt{\frac{1-\beta}{(1-\alpha)^2}\bigvee \frac{h_1^2}{1-\beta}+\frac{\beta-\alpha}{(1-\alpha)^2}\bigwedge\frac{(1-h_1)^2}{\beta-\alpha}-1}}\right)>\epsilon\|{\mathbf w}\|_2^2\right\}}$$ with $\eta_{\mathbf w}\in (0,\infty)$ being the solution of
\begin{align*}&\left(1-\frac{\epsilon\|\mathbf w\|_2^2}{2\mathbf w^\top\Sigma_0\mathbf w}\right)\sqrt{V_{\mathbf w, \lambda}}\\&=\frac{1-g_{\alpha,\beta}^{h_1,h_2}(u_{\alpha,\beta,\mathbf w,\lambda}^{h_1,h_2})}{1-\alpha-u_{\alpha,\beta,\mathbf w,\lambda}^{h_1,h_2}}\int_{\alpha}^{1-u_{\alpha,\beta,\mathbf w,\lambda}^{h_1,h_2}}F_0^{-1}(t)\d t+\frac{h_1}{1-\beta}\int_{\beta\vee (1-u_{\alpha,\beta,\mathbf w,\lambda}^{h_1,h_2})}^1F_0^{-1}(t)\d t\\
&\quad+\frac{h_2-h_1}{\beta-\alpha}\int_{\beta\wedge (1-u_{\alpha,\beta,\mathbf w,\lambda}^{h_1,h_2})}^{\beta}F_0^{-1}(t)\d t
 +\lambda \sqrt{\mathbf w^\top\Sigma_0\mathbf w}\int_{(0,\alpha)\cup (1-u_{\alpha,\beta,\mathbf w,\lambda}^{h_1,h_2},1)}(F_0^{-1}(t))^2\d t\\
 &\quad +\frac{\lambda\sqrt{\mathbf w^\top\Sigma_0\mathbf w}}{1-\alpha-u_{\alpha,\beta,\mathbf w,\lambda}^{h_1,h_2}}\left(\int^{1-u_{\alpha,\beta,\mathbf w,\lambda}^{h_1,h_2}}_{\alpha} F_0^{-1}(s)\d s\right)^2.
\end{align*}
 \end{corollary}

The above corollaries show that the robust optimisation problem \eqref{PS2} is reduced to a standard optimisation problem which however is generally nonconvex implying that it is difficult to obtain a solution in practice, in particular when the dimensionality is high. To deal with this practical issue, we propose the following procedure: 

Let $\mathcal A=\{\mathbf w\in \Delta_n: -\mathbf w^\top \boldsymbol{\mu}\geq a \}$ for some $a\in [-\max_{i\in [n]}\mu_i,-\min_{i\in [n]}\mu_i]$.  Note that the feasible range for  $x=\mathbf w^\top\boldsymbol \mu$ is $ x\in [\min_{i\in [n]}\mu_i, -a]$.
Then the algorithm of numerical study is displayed as below.
\begin{itemize}
    \item Step 1: For any chosen $x\in [\min_{i\in [n]}\mu_i,-a]$, 
    solve 
\begin{eqnarray*}
y_{\min }(x):=\min_{\mathbf w^\top\boldsymbol \mu=x, \mathbf w\in \A}\mathbf{w}^{\top }\Sigma _{0}\mathbf{w}, \quad
y_{\max }(x):=\max_{\mathbf w^\top\boldsymbol \mu=x, \mathbf w\in \A}\mathbf{w}^{\top }\Sigma _{0}\mathbf{w}.
\end{eqnarray*}
    \item Step 2: For any point $(x,y)$, $y\in [y_{\min }(x), y_{\max }(x)]$, 
    solve  $$z(x,y,\epsilon )=\min_{\{\mathbf w^\top\boldsymbol \mu=x, \mathbf w^\top\Sigma_0\mathbf w=y, \mathbf w\in \mathcal A\}}\frac{\epsilon\|\mathbf w\|_2^2}{y}.$$
    
    \item Step 3: 
    Combining the preceding steps yields the following
two-dimensional search region: 

$$\widetilde{\mathcal{S}}=\left\{ (x,y):x\in [\min_{i\in [n]}\mu_i,-a],\
y\in \lbrack y_{\min }(x),y_{\max }(x)]\right\} , 
$$
and \eqref{PS2} is reduced to 
\begin{equation}\label{step3}
\min_{(x,y)\in \widetilde{\mathcal{S}}}H(x,y,z(x,y,\epsilon)),
\end{equation}
where $$H(x,y,z):=xg(1)+\sqrt{y}\sup_{G\in\M_{z}(0,1)}\rho_g(G).$$
\item Step 4: Choose a grid over the feasible interval for $x$. For each grid point $x$, obtain $y_{\min}(x)$ and $y_{\max}(x)$ as in Step 1, and then choose a grid on $[y_{\min}(x), y_{\max}(x)]$. For each pair $(x,y)$ on this nested grid, compute $H(x,y,z(x,y,\epsilon))$ and search for a minimizer, denoted by $(x^\ast,y^\ast)$. The approximate solution $\mathbf w^\ast$ to \eqref{step3} is then obtained from Step 2 with $x=x^\ast$ and $y=y^\ast$.


\end{itemize}

\begin{remark}
The minimizer $(x^\ast,y^\ast)$ theoretically 
exists for our concerned distortion risk metrics. This is due to the fact that $H(x,y,z)$ is a continuous function if $\rho_{\hat{g}}(H_\lambda)$ is continuous in $\lambda$ over $(0,\infty)$, which is shown in  the  proof of Theorem \ref{Th:main1} in the E-companion. This condition is satisfied by $\VaR$, $\IQD$, $\GlueVaR$ and $\rho_g$ with concave $g$. 
\end{remark}

\begin{remark}
The minimization problem in Step 1 is a convex quadratic program whereas the corresponding maximization problem is nonconvex. However, the maximum is attained at an extreme point of the feasible polytope (Corollary 32.3.2 Rockafellar 1997). Step 2 is a quadratically constrained quadratic program (QCQP) which due to the variance equality constraint is however nonconvex. 
If all entries of $$\mathbf{w}^{\ast }_L=(I+\Sigma _{0})^{-1}A^{\top }[A(I+\eta ^{\ast }\Sigma
_{0})A^{\top }]^{-1}b$$ where $I$ is an identity matrix, $A=\binom{\mathbf{1}^{\top }}{\mathbf{\mu }^{\top }}$, $b=\binom{1}{x}$, and $\eta ^{\ast }$ is a scale parameter determined via $(\mathbf{w}^{\ast }_L)^{\top } \Sigma _{0}\mathbf{w}^{\ast }_L=y$, are non-negative, then $\mathbf{w}^{\ast }_L$ is the explicit solution to \eqref{PS2}. When $\mathbf{w}^{\ast }_L$ contains negative components, then there is no explicit solution. 
In this case we suggest to adopt an ADMM-type algorithm\footnote{For details on the ADMM method, we refer to } for finding a  global solution. 
\end{remark}

\subsection{Numerical examples}
Our main results reduce robust portfolio optimization under an ambiguity set characterized by mean, variance and a Wasserstein ball to minimizing a deterministic objective.
In this section, we present numerical results to illustrate the impact of model uncertainty on portfolio optimisation for different risk metrics: $\GD$, $\MMD$, $\IQD$, $\VaR$, $\ES$ and $\GlueVaR$. Specifically, we solve the optimization problem in Corollary \ref{cor:3} for $\GD$, $\MMD$ and $\ES$,
Corollary \ref{cor:4} for $\IQD$, Corollary \ref{cor:5} for $\VaR$ and Corollary \ref{cor:6} for $\GlueVaR$.

\begin{figure}[htbp]
    \centering
    \includegraphics[width=1\linewidth]{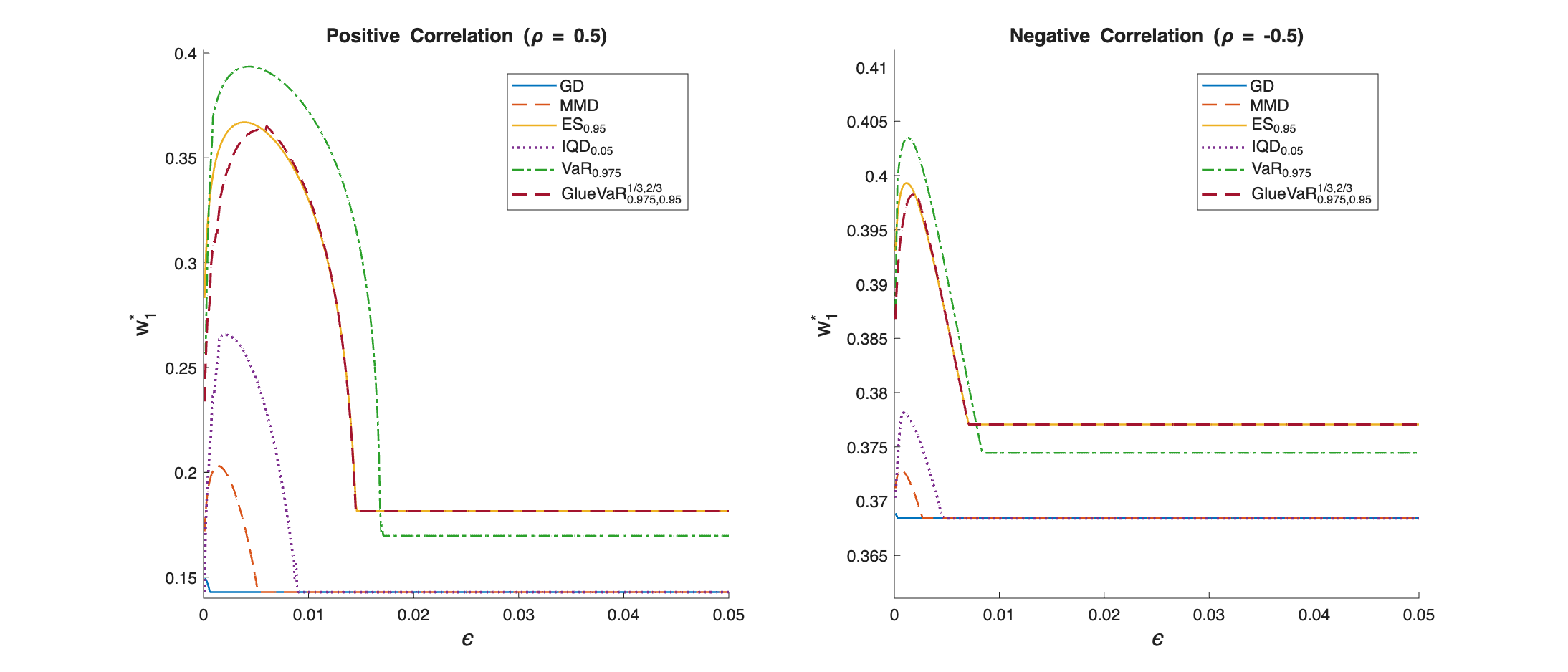}
    \caption{Optimal weight $w_1^*$ under different Wasserstein radius $\epsilon$ with positive and negative correlations under normal distribution.}
    \label{fig:epsilon_analysis_v42}
\end{figure}

We assume that $\mathcal A=\Delta_n$ and that the reference distribution $\mathbf{X}_0\sim N(\boldsymbol \mu, \Sigma_0)$ represents the negative returns from investing one dollar in each of $n$ different assets in the market. To simplify the analysis, we consider the case $n=2$. For the reference mean vector, we set $\boldsymbol \mu = (-0.06, -0.03)^\top$. 
We consider both positively and negatively correlated covariance matrices $\Sigma_0$, as follows:
     $$\Sigma_0^{(1)} =\begin{bmatrix}
        0.15^2, & 0.5\times 0.15\times 0.1 \\
        0.5\times 0.15\times 0.1 &  0.1^2
        \end{bmatrix},\quad \Sigma_0^{(2)} =\begin{bmatrix}
      0.15^2, & -0.5\times 0.15\times 0.1 \\
        -0.5\times 0.15\times 0.1 &  0.1^2
        \end{bmatrix}.$$
The uncertainty is controlled by the Wasserstein radius $\epsilon$. 
The impact of model uncertainty on portfolio optimization under these distortion risk metrics can be seen clearly in Figure \ref{fig:epsilon_analysis_v42}.  In fact, as $\epsilon$ changes, $w_1$ follows a similar pattern across all six distortion risk metrics. For small $\epsilon$ (low uncertainty), more weight is allocated to the first asset; for intermediate $\epsilon$ (moderate uncertainty), the weight allocated to the first asset begins to decline; and for large $\epsilon$, it becomes constant (corresponding to the case in which the Wasserstein constraint becomes inactive). The distortion risk metrics can be naturally divided into two groups in Figure \ref{fig:epsilon_analysis_v42}: the optimal weights of the first asset for $\GD$, $\MMD$, and $\IQD$ (variability measures) are always smaller than those for $\VaR$, $\ES$ and $\GlueVaR$ (tail-risk measures), and the first group is less sensitive than the second group in portfolio optimization. In particular, the optimal weights for $\GD$ remains nearly constant, showing that model uncertainty has only a small impact on portfolio selection under this criterion. These qualitative findings are driven by the fact that $g(1)=0$ for $\GD$, $\MMD$ and $\IQD$, implying that $\boldsymbol{\mu}$ does not contribute to the portfolio optimization; consequently, all these criteria behave similarly to variance.  In contrast, for the tail-risk measures $\VaR$, $\ES$ and $\GlueVaR$, substantially more weight is allocated to the most risky asset, and these criteria exhibit more sensitivity to model uncertainty. Indeed, note that $g(1)=1$ for $\VaR$, $\ES$ and $\GlueVaR$, so that  $\boldsymbol{\mu}$ contributes to portfolio optimization under these distortion risk metrics. Moreover, these distortion risk metrics focus on the tail risk beyond a threshold, which is qualitatively different from $\GD$, $\MMD$ and $\IQD$.

 
\begin{figure}[htbp]
    \centering
    \includegraphics[width=1\linewidth]{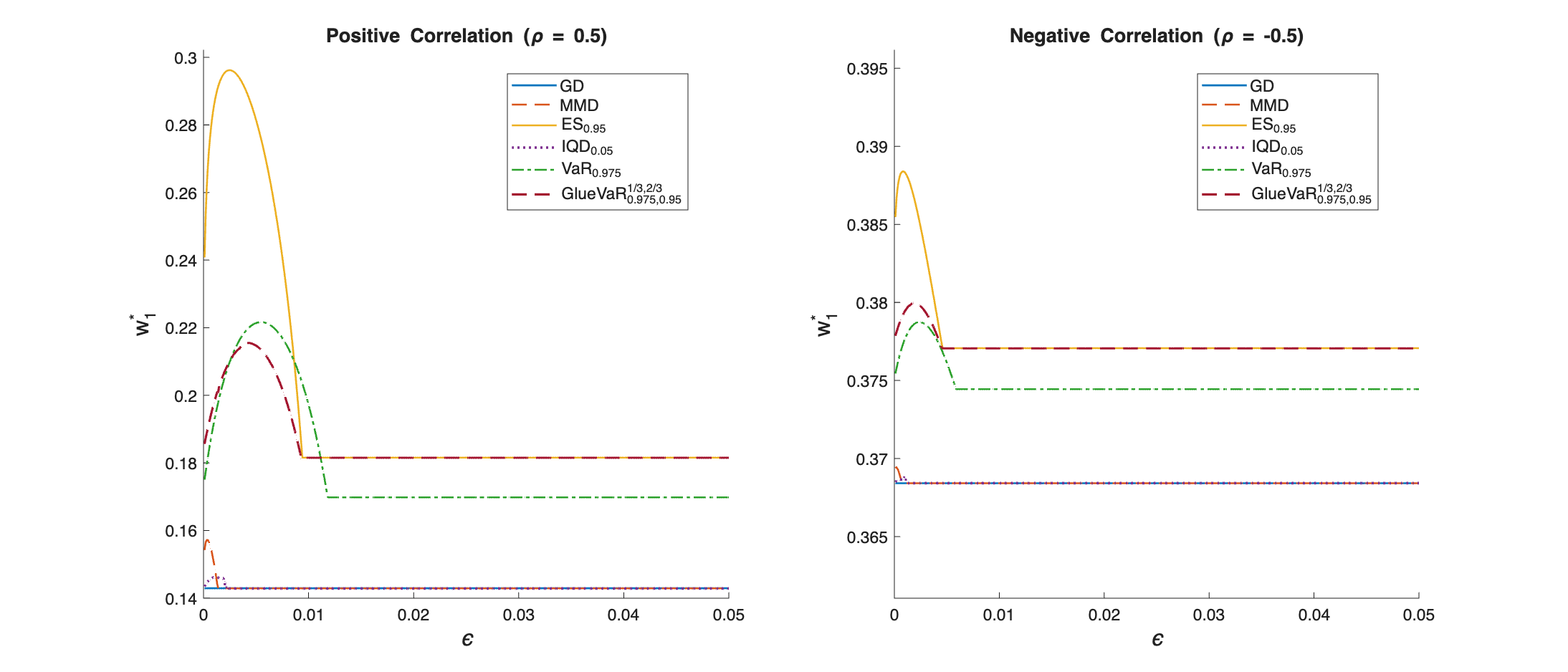}
    \caption{Optimal weight $w_1^*$ under different Wasserstein radius $\epsilon$ with positive and negative correlations under Student's t distribution.}
    \label{fig:epsilon_analysis_studentt}
\end{figure}

Next, we suppose the reference distribution follows Student's t-distribution, i.e., $\mathbf{X}_0\sim \mathbf{t}_p(\boldsymbol\mu,\Sigma_0,\nu)$  with $\nu=5$ and $p=2$. We set $\boldsymbol \mu = (-0.06, -0.03)^\top$ and $\Sigma_0=\Sigma_0^{(1)}$ or $\Sigma_0^{(2)}$.
The uncertainty is controlled by the Wasserstein radius $\epsilon$. The impact of model uncertainty on portfolio optimization can be seen in Figure~\ref{fig:epsilon_analysis_studentt}

\section{Conclusion}

In this paper, we apply  a variant of the concave envelope approach to derive the sharp upper and lower bounds for distortion risk metrics over uncertainty sets characterized by fixed mean, variance and bounded Wasserstein distance from a reference distribution. We emphasize that the proof is highly non-trivial and requires several new arguments.  Our findings provide a unified framework that recovers many existing bounds in the literature and yields new sharp bounds for some commonly studied distortion risk metrics, including $\GD, \MMD, \IQD$ and $\GlueVaR$. 

Using the projection method, we further establish the sharp bounds for distortion risk metrics with absolutely continuous distortion functions under uncertainty sets incorporating unimodality constraints, substantially extends the existing results. From computational perspective, we  develop an algorithm to approximate the sharp bounds and the corresponding worst-case quantiles. 

Finally, we investigate the applications to the robust portfolio optimization for various  distortion risk metrics, illustrating the practical relevance and effectiveness of the proposed framework. The framework may also be applied to robust assessment of risk under model uncertainty, which we leave for future work.  


\appendix

\label{Appendix:proofs}

\section{Proof of Section \ref{Sec:general}}\label{Sec3:proof}
In this section, we present detailed proofs of all results stated in  Section \ref{Sec:general}. 

{\bf Proof of Lemma \ref{lem:1}}.   By definition,  we have
\begin{align*}Corr(F^{-1}(V), (g_{\lambda}^*)'(1-V))&=\frac{\mathbb E(F^{-1}(V)(g_{\lambda}^*)'(1-V))-\mathbb E(F^{-1}(V))\mathbb E((g_{\lambda}^*)'(1-V))}{\sigma_{F}\sqrt{Var((g_{\lambda}^*)'(V))}}\\
&=\frac{\mathbb E(F^{-1}(V)(g_{\lambda}^*)'(1-V))-\mu_F(g(1)+\lambda\mu_F)}{\sigma_{F}\sqrt{\mathbb E((g_{\lambda}^*)'(V))^2-(g(1)+\lambda\mu_F)^2}}
\end{align*}
We fix $\lambda_0\in [0,\infty)$ and show that $(g_{\lambda}^*)'(t)$ is continuous with respect to $\lambda$ at $\lambda_0$ if  $(g_{\lambda_0}^*)'(t)$ is continuous at $t$.
 For $\lambda_1,\lambda_2\in [0,\infty)$, $|g_{\lambda_2}(t)-g_{\lambda_1}(t)|\leq C|\lambda_2-\lambda_1|$ with $C=\int_{0}^{1} |F^{-1}(s)|\d s$. By definition, $|g_{\lambda_2}^*(t)-g_{\lambda_1}^*(t)|\leq  C|\lambda_2-\lambda_1|$. This implies $\sup_{t\in (0,1)}|g_{\lambda_1}^*(t)-g_{\lambda_2}^*(t)|\leq C|\lambda_2-\lambda_1|\to 0$ as $|\lambda_2-\lambda_1|\to 0$. Next, we suppose by contradiction that there exists some continuous point of $(g_{\lambda_0}^*)'$, $t\in (0,1)$, such that $(g_{\lambda}^*)'(t)\to (g_{\lambda_0}^*)'(t)$ as $\lambda\to \lambda_0$ does not hold. Without loss of generality, we suppose there exist $\lambda_n\to\lambda_0$ as $n\to\infty$ such that $\lim_{n\to\infty}(g_{\lambda_n}^*)'(t)=c>(g_{\lambda_0}^*)'(t)$. Let us denote $d=c-(g_{\lambda_0}^*)'(t)>0$. By the continuity of $(g_{\lambda_0}^*)'$ at $t$, there exists $\epsilon>0$ such that $(g_{\lambda_0}^*)'(s)\leq (g_{\lambda_0}^*)'(t)+d/3$ for $s\in (t-\epsilon,t]$. Moreover, there exists $n_0$ such that $(g_{\lambda_n}^*)'(t)>c-d/3$ for all $n\geq n_0$. Note that $(g_{\lambda_n}^*)'$ is decreasing. Hence, $(g_{\lambda_n}^*)'(s)\geq c-d/3$ for $s\in(t-\epsilon,t]$ and $n\geq n_0$. Consequently, we have $\int_{t-\epsilon}^{t} (g_{\lambda_n}^*)'(s)\d s\geq \int_{t-\epsilon}^{t} (g_{\lambda_0}^*)'(s)\d s+(d\epsilon)/3$,
  which can be rewritten as $g_{\lambda_n}^*(t)-g_{\lambda_n}^*(t-\epsilon)\geq g_{\lambda_0}^*(t)-g_{\lambda_0}^*(t-\epsilon)+(d\epsilon)/3$ for all $n\geq n_0$. This contradicts the fact that $\sup_{t\in (0,1)}|g_{\lambda_n}^*(t)-g_{\lambda_0}^*(t)|\leq C|\lambda_n-\lambda_0|\to 0$ as $n\to\infty$. Hence, $(g_{\lambda}^*)'(t)$ is continuous with respect to $\lambda$ at $\lambda_0$ if  $(g_{\lambda_0}^*)'(t)$ is continuous at $t$. Note that $(g_{\lambda_0}^*)'(t)$ is continuous over $(0,1)$ except countable points. Hence, we have $(g_{\lambda}^*)'(t)\to (g_{\lambda_0}^*)'(t)$ a.e. as $\lambda\to\lambda_0$.

 Define $X_\lambda=(g_{\lambda}^*)'(1-V)$ and $Y_{\lambda}=(g^*)'(1-V)+\lambda F^{-1}(V)$. Direct calculation shows $\ES_{\alpha}(X_\lambda)=\frac{g_{\lambda}^*(1-\alpha)}{1-\alpha}$ and $\ES_{\alpha}(Y_\lambda)=\frac{g^*(1-\alpha)+\lambda\int_{\alpha}^{1}F^{-1}(t)\d t}{1-\alpha}$. Using the fact $g_{\lambda}^*(t)\leq g^*(t)+\lambda\int_{1-t}^{1}F^{-1}(s)\d s$, we have $\ES_{\alpha}(X_\lambda)\leq \ES_{\alpha}(Y_\lambda)$ for all $\alpha\in (0,1)$. Moreover, we have $\mathbb E(X_\lambda)=g(1)+\lambda\mu_F=\mathbb E(Y_{\lambda})$. Hence, in light of Corollary 2.61 and  Theorem 2.57 of \cite{FS16}, we have $X_\lambda\leq_{\mathrm{cx}} Y_{\lambda}$, which means $X_\lambda\leq Y_\lambda$ in convex order. This implies $\mathbb E(X_{\lambda}^2)\leq \mathbb E(Y_{\lambda}^2)$, i.e., $\int_{0}^{1}((g_{\lambda}^*)'(t))^2\d t\leq \int_{0}^{1}((g^*)'(t)+\lambda F^{-1}(1-t))^2\d t\leq 2\int_{0}^{1}((g^*)'(t))^2\d t +2\lambda^2\int_{0}^{1}(F^{-1}(t))^2\d t<\infty$.

 For any $\epsilon>0$, let $f_{\lambda}(t)=g_{\lambda}^*(t)\id_{\{0\leq t\leq \epsilon\}}+\left(g_{\lambda}^*(\epsilon)+\frac{g_{\lambda}^*(1)-g_{\lambda}^*(\epsilon)}{1-\epsilon}(t-\epsilon)\right)\id_{\{\epsilon<t\leq 1\}}$ and $k_{\lambda}(t)=r_{\lambda}(t)\id_{\{0\leq t\leq \epsilon\}}+\left(r_{\lambda}(\epsilon)+\frac{r_{\lambda}(1)-r_{\lambda}(\epsilon)}{1-\epsilon}(t-\epsilon)\right)\id_{\{\epsilon<t\leq 1\}}$, where $r_{\lambda}(t)=g^*(t)+\lambda \int_{1-t}^{1} F^{-1}(s)\d s$. Note that both $f_{\lambda}$ and $k_{\lambda}$ are continuous concave functions on $[0,1]$ and $f_{\lambda}\leq k_{\lambda}$. Moreover, $f_{\lambda}(0)=k_{\lambda}(0)=0$ and $f_{\lambda}(1)=k_{\lambda}(1)$. Similarly as the above argument, one can easily check that  $f_\lambda'(1-V)\leq_{\mathrm{cx}} k_\lambda'(1-V)$. Hence, we have
 $\int_{0}^{1}(f_{\lambda}'(t))^2\d t\leq \int_{0}^{1}(k_{\lambda}'(t))^2\d t$. It can be rewritten as \begin{align*}\int_{0}^{\epsilon}((g_{\lambda}^*)'(t))^2\d t&\leq \int_{0}^{\epsilon}((g^*)'(t)+\lambda F^{-1}(1-t))^2\d t+\frac{(r_{\lambda}(1)-r_{\lambda}(\epsilon))^2}{1-\epsilon}-\frac{(g_{\lambda}^*(1)-g_{\lambda}^*(\epsilon))^2}{1-\epsilon}\\
 &\leq  2\int_{0}^{\epsilon}((g^*)'(t))^2\d t+2\lambda^2 \int_{0}^{\epsilon}(F^{-1}(1-t))^2\d t\\
 &~+\frac{|r_{\lambda}(1)-r_{\lambda}(\epsilon)+g_{\lambda}^*(1)-g_{\lambda}^*(\epsilon)\|r_{\lambda}(\epsilon)-g_{\lambda}^*(\epsilon)|}{1-\epsilon}.
 \end{align*}
 Note that $g_{\lambda}^*(1)=r_{\lambda}(1)=g_{\lambda}(1)$ and $g_{\lambda}(\epsilon)\leq g_{\lambda}^*(\epsilon)\leq r_{\lambda}(\epsilon)$. It follows that \begin{align*}\frac{|r_{\lambda}(1)-r_{\lambda}(\epsilon)+g_{\lambda}^*(1)-g_{\lambda}^*(\epsilon)\|r_{\lambda}(\epsilon)-g_{\lambda}^*(\epsilon)|}{1-\epsilon}\leq
 \frac{2(|g_\lambda(1)|+|r_{\lambda}(\epsilon)|+|g_{\lambda}(\epsilon)|)|g^*(\epsilon)-g(\epsilon)|}{1-\epsilon}.
 \end{align*}
 We fix $\lambda_0>0$. Using the above arguments and the fact that $g(0)=g(0+)$,  for any $\eta>0$, there exists $\epsilon_0>0$ such that if $\epsilon\in (0,\epsilon_0)$, then
 \begin{align*}\sup_{0\leq \lambda\leq \lambda_0}\int_{0}^{\epsilon}((g_{\lambda}^*)'(t))^2\d t\leq 2\int_{0}^{\epsilon}((g^*)'(t))^2\d t+2(\lambda_0+1)^2 \int_{0}^{\epsilon}(F^{-1}(1-t))^2\d t+M|g^*(\epsilon)-g(\epsilon)|<\eta.
 \end{align*}
 Using the similar argument as above, we can show that for any $\eta>0$, there exists $\epsilon_1>0$ such that if $\epsilon\in (0,\epsilon_1)$, then
 $\sup_{0\leq \lambda\leq \lambda_0}\int_{1-\epsilon}^{1}((g_{\lambda}^*)'(t))^2\d t<\eta.
 $
 Note also that $(g_{\lambda}^*)'$ is monotone over $(0,1)$.
 Hence, $\{((g_{\lambda}^*)'(t))^2, 0\leq \lambda\leq \lambda_0\}$ is uniformly integrable for any $\lambda_0>0$.
 Using H\"{o}lder's inequality and the above conclusions, we have \begin{align*}
 &|\mathbb E(F^{-1}(V)(g_{\lambda}^*)'(1-V))-\mathbb E(F^{-1}(V)(g_{\lambda_0}^*)'(1-V))|\\
 &=\left|\int_{0}^{1}F^{-1}(t)((g_{\lambda}^*)'(1-t)-(g_{\lambda_0}^*)'(1-t))\d t\right|\\
 &\leq \left(\int_{0}^{1}(F^{-1}(t))^2\d t\right)^{1/2}\left(\int_{0}^{1}((g_{\lambda}^*)'(t)-(g_{\lambda_0}^*)'(t))^2\d t\right)^{1/2}\to 0
 \end{align*}
 as $\lambda\to\lambda_0$, and
 \begin{align*}
 |\mathbb E((g_{\lambda}^*)'(V))^2-\mathbb E((g_{\lambda_0}^*)'(V))^2|&=\left|\int_{0}^{1}((g_{\lambda}^*)'(t))^2-((g_{\lambda_0}^*)'(t))^2\d t\right|\\
 &\leq \left(\int_{0}^{1}((g_{\lambda}^*)'(t)+(g_{\lambda_0}^*)'(t))^2\d t\right)^{1/2}\left(\int_{0}^{1}((g_{\lambda}^*)'(t)-(g_{\lambda_0}^*)'(t))^2\d t\right)^{1/2}\\
 &\to 0~\text{as}~\lambda\to\lambda_0.
 \end{align*}
 Hence, $Corr(F^{-1}(V), (g_{\lambda}^*)'(1-V))$ is continuous for $\lambda\in [0,\infty)$.

 Finally, we show that  $\lim_{\lambda\to\infty} Corr(F^{-1}(V), (g_{\lambda}^*)'(1-V))=1.$  Let $l_{\lambda}=\frac{g_\lambda}{\lambda}$ for $\lambda>0$. Then by definition, we have $l_{\lambda}^*=\frac{g_{\lambda}^*}{\lambda}$. Direct computation gives \begin{align*}Corr(F^{-1}(V), (g_{\lambda}^*)'(1-V))&=Corr(F^{-1}(V), (l_{\lambda}^*)'(1-V))\\
 &=\frac{\mathbb E(F^{-1}(V)(l_{\lambda}^*)'(1-V))-\mu_F(g(1)/\lambda+\mu_F)}{\sigma_{F}\sqrt{\mathbb E((l_{\lambda}^*)'(U))^2-(g(1)/\lambda+\mu_F)^2}}.
 \end{align*}
  We denote $\int_{1-t}^{1} F^{-1}(s)\d s$ by $l_{\infty}(t)$. Then $\sup_{t\in [0,1]}|l_{\lambda}(t)-l_{\infty}(t)|\leq \sup_{t\in [0,1]}|g(t)|/\lambda$, which implies
  $\sup_{t\in [0,1]}|l_{\lambda}^*(t)-l_{\infty}^*(t)|\leq \sup_{t\in [0,1]}|g(t)|/\lambda$. Using the similar argument as $(g_{\lambda}^*)'$, we have $(l_{\lambda}^*)'(t)\to (l_{\infty}^*)'(t)$ a.e. on $(0,1)$ as $\lambda\to\infty$ and $\{(l_{\lambda}^*)'(t),~1\leq \lambda<\infty\}$ is uniformly integrable. Therefore, we have $\lim_{\lambda\to\infty}\mathbb E(F^{-1}(V)(l_{\lambda}^*)'(1-V))=\mathbb E(F^{-1}(V)(l_{\infty}^*)'(1-V))=\mathbb E((F^{-1}(V))^2)$ and $\lim_{\lambda\to\infty}\mathbb E((l_{\lambda}^*)'(V))^2=\mathbb E((l_{\infty}^*)'(V))^2=\mathbb E((F^{-1}(V))^2)$. Hence, $\lim_{\lambda\to \infty}Corr(F^{-1}(V), (g_{\lambda}^*)'(1-V))=1$. \qed

Combing Theorem 5 and Remark 2 of \cite{PWW20}, we immediately arrive at the following result for $\epsilon=\infty$, which will play an important role to prove Theorem \ref{Th:main}.
\begin{lemma}\label{lem:2} For $g\in\mathcal H$, we have
$$\sup_{G\in \mathcal M_{\infty}(\mu, \sigma)} \rho_g(G)=\rho_{\hat{g}}(H_0),$$
where the supremum is uniquely attained at $H_0$ if $g=\hat{g}$.
\end{lemma}

{\bf Proof of Theorem \ref{Th:main}}.  First, note that for any $G\in \mathcal M_\epsilon(\mu, \sigma)$, we have $(\mu_F-\mu)^2+(\sigma_F-\sigma)^2\leq d_W^2(F,G)\leq \epsilon$. Note that $d_W^2(F,G)=(\mu_F-\mu)^2+(\sigma_F-\sigma)^2$ implies $G^{-1}(t)=\mu+\sigma\frac{F^{-1}(t)-\mu_F}{\sigma_F}$, whose distribution is denoted by $H_{\infty}$.  For the case $(\mu_F-\mu)^2+(\sigma_F-\sigma)^2<d_W^2(F,G)\leq \epsilon$, by Lemma \ref{lem:1}, there exists  $\lambda\geq 0$ such that
$d_W(F,H_{\lambda})=d_W(F,G)$. This implies $\int_{0}^{1} F^{-1}(t) h_{\lambda}(t)\d t=\int_{0}^{1} F^{-1}(t) G^{-1}(t)\d t$, which is equivalent to \begin{align}\label{eq:1}\rho_{g_{\lambda}-g}(H_\lambda)=\rho_{g_\lambda-g}(G).
\end{align} In light of Lemma \ref{lem:2}, we have $\sup_{G\in \mathcal M_\infty(\mu, \sigma)} \rho_{g_\lambda}(G)=\rho_{g_\lambda}(H_\lambda)$ and $H_{\lambda}$ is the unique maximizer.  Hence, for $G\in \mathcal M_{\infty} (\mu, \sigma)$, if $G\neq H_\lambda$, then $\rho_{g_\lambda}(G)<\rho_{g_\lambda}(H_\lambda)$, which can be rewritten as $\rho_{g}(G)+\rho_{g_\lambda-g}(G)<\rho_{g}(H_\lambda)+ \rho_{g_\lambda-g}(H_\lambda)$. It follows from \eqref{eq:1} that $\rho_{g}(G)<\rho_{g}(H_\lambda)$ if $d_W(F,H_{\lambda})=d_W(F,G)$ and $G\neq H_\lambda$. This means that the optimal solution has the form of $H_\lambda$ for $\lambda\in (0,\infty]$.

For $d_{W}(F, H_{\lambda_1})<d_{W}(F, H_{\lambda_2})$, we have $\rho_{g_{\lambda_2}-g}(H_{\lambda_2})<\rho_{g_{\lambda_2}-g}(H_{\lambda_1})$. Moreover, by Lemma \ref{lem:2}, we have $\sup_{G\in \mathcal M_\infty(\mu, \sigma)} \rho_{g_{\lambda_2}}(G)=\rho_{g_{\lambda_2}}(H_{\lambda_2})$. This implies  $\rho_{g}(H_{\lambda_1})+\rho_{g_{\lambda_2}-g}(H_{\lambda_1})=\rho_{g_{\lambda_2}}(H_{\lambda_1})\leq \rho_{g_{\lambda_2}}(H_{\lambda_2})=\rho_{g}(H_{\lambda_2})+\rho_{g_{\lambda_2}-g}(H_{\lambda_2})$. Hence, $\rho_{g}(H_{\lambda_1})<\rho_{g}(H_{\lambda_2})$.  Consequently,  $H_{\lambda_\epsilon}$ is the unique maximizer with $\lambda_\epsilon$ satisfying $d_{W}(F, H_{\lambda})=\sqrt{\epsilon}$.  We proved the statement of (i).

 We next consider scenario (ii).  If $\epsilon\geq (\mu_F-\mu)^2+(\sigma_F-\sigma)^2+2\sigma_F\sigma(1-c_0)$,  then by Lemma \ref{lem:2}, we have $\sup_{G\in \mathcal M_{\epsilon}(\mu, \sigma)} \rho_g(G)\leq \sup_{G\in \mathcal M_{\infty}(\mu, \sigma)} \rho_g(G)=\rho_g(H_0).$ As $H_0\in \mathcal M_{\epsilon}(\mu, \sigma)$, we have $\rho_g(H_0)\leq\sup_{G\in \mathcal M_{\epsilon}(\mu, \sigma)}\rho_g(G)$. Hence, $\sup_{G\in \mathcal M_{\epsilon}(\mu, \sigma)}\rho_g(G)=\rho_g(H_0)$ and $H_0$ is the unique maximizer. \qed


{\bf Proof of Corollary \ref{Cor:1}}. (i) Note that if $g$ is concave, then $(g_\lambda^*)'(1-t)=g'(1-t)+\lambda F^{-1}(t)$ for $t\in (0,1)$. Applying Theorem \ref{Th:main}, we obtain the desired result.

(ii) Note that for $\VaR_{\alpha}^+$, we have $g(t)=\id_{[1-\alpha,1]}(t)$ and $g_\lambda(t)=\id_{[1-\alpha,1]}(t)+\lambda \int_{1-t}^{1} F^{-1}(s)\d s$. 
Using $t_{1-\alpha,\lambda}$, we have
\begin{align*}g_\lambda^*(t)&=\lambda \int_{1-t}^{1} F^{-1}(s)\d s\id_{[0,t_{1-\alpha,\lambda})}(t) +\left(\frac{g_\lambda(1-\alpha)-g_{\lambda}(t_{1-\alpha,\lambda})}{1-\alpha-t_{1-\alpha,\lambda}}(t-t_{1-\alpha,\lambda})+g_{\lambda}(t_{1-\alpha,\lambda})\right)\id_{[t_{1-\alpha,\lambda},1-\alpha]}(t)\\
&+\left(1+\lambda \int_{1-t}^{1} F^{-1}(s)\d s\right)\id_{(1-\alpha,1]}(t),
\end{align*} 
Direct computation shows  $$(g_\lambda^*)'(1-t)=\lambda F^{-1}(t)\id_{(0,\alpha] \cup (1-t_{1-\alpha,\lambda},1)}(t)+\frac{1+\lambda\int_{\alpha}^{1-t_{1-\alpha,\lambda}}F^{-1}(s)\d s}{1-\alpha-t_{1-\alpha,\lambda}}\id_{(\alpha,1-t_{1-\alpha,\lambda}]},~t\in (0,1),$$
which implies 
 $$\VaR_\alpha^+(H_\lambda)=\mu+\sigma\frac{\frac{1+\lambda\int_{\alpha}^{1-t_{1-\alpha,\lambda}}F^{-1}(s)\d s}{1-\alpha-t_{1-\alpha,\lambda}}-a_{\lambda}}{b_{\lambda}}.$$

 (iii) For $\IQD_{\alpha}^+$, we have  $g(t)=\id_{[\alpha, 1-\alpha]}(t)$ and
 $g_\lambda(t)=\id_{[\alpha, 1-\alpha]}(t)+\lambda \int_{1-t}^{1} F^{-1}(s)\d s$.
 It follows from the definition that
 \begin{align*}g_\lambda^*(t)&=\left(\frac{g_{\lambda}(\alpha)-g_{\lambda}(t_{\alpha,\lambda})}{\alpha-t_{\alpha,\lambda}}(t-t_{\alpha,\lambda})+g_\lambda(t_{\alpha,\lambda})\right)\id_{[t_{\alpha,\lambda},\alpha]}(t)+g_{\lambda}(t)\id_{[0,t_{\alpha,\lambda})\cup (\alpha, 1-\alpha]\cup (\hat{t}_{\alpha,\lambda},1]}(t)\\
 &+\left(\frac{g_{\lambda}(\hat{t}_{\alpha,\lambda})-g_{\lambda}(1-\alpha)}{\hat{t}_{\alpha,\lambda}-1+\alpha}(t-\hat{t}_{\alpha,\lambda})+g_\lambda(\hat{t}_{\alpha,\lambda})\right)\id_{(1-\alpha,\hat{t}_{\alpha,\lambda}]}(t).
 \end{align*}
 Direct computation gives 
 \begin{align*}(g_\lambda^*)'(1-t)&=\frac{g_{\lambda}(\alpha)-g_{\lambda}(t_{\alpha,\lambda})}{\alpha-t_{\alpha,\lambda}}\id_{(1-\alpha, 1-t_{\alpha,\lambda})}(t)+\frac{g_{\lambda}(\hat{t}_{\alpha,\lambda})-g_{\lambda}(1-\alpha)}{\hat{t}_{\alpha,\lambda}-1+\alpha}\id_{(1-\hat{t}_{\alpha,\lambda},\alpha)}(t)\\
 &+\lambda F^{-1}(t)\id_{(0,1-\hat{t}_{\alpha,\lambda})\cup(\alpha,1-\alpha)\cup (1-t_{\alpha,\lambda},1)}.
 \end{align*}
 Applying Theorem \ref{Th:main}, we have $$\IQD^+(H_\lambda)=\frac{\frac{g_{\lambda}(\alpha)-g_{\lambda}(t_{\alpha,\lambda})}{\alpha-t_{\alpha,\lambda}}-\frac{g_{\lambda}(\hat{t}_{\alpha,\lambda})-g_{\lambda}(1-\alpha)}{\hat{t}_{\alpha,\lambda}-1+\alpha}}{b_\lambda}\sigma.$$
(iv)  For $g(t)=\frac{t}{1-\alpha_1}\wedge 1-\id_{(1-\alpha_2,1]}(t)$, we have $g_\lambda(t)=\frac{t}{1-\alpha_1}\wedge 1+\lambda \int_{1-t}^{1} F^{-1}(s)\d s-\id_{(1-\alpha_2,1]}(t)$. By the definition, we have
\begin{align*}g_\lambda^*(t)&=\left(g_\lambda(1-\alpha_2)+
\frac{g_\lambda(u_{\alpha_1,\alpha_2,\lambda})-g_\lambda(1-\alpha_2)}{u_{\alpha_1,\alpha_2,\lambda}-1+\alpha_2}(t-1+\alpha_2)\right)\id_{[1-\alpha_2,u_{\alpha_1,\alpha_2,\lambda}]}(t)\\
&~g_\lambda(t)\id_{(0,1-\alpha_2)\cup (u_{\alpha_1,\alpha_2,\lambda},1)}(t).
\end{align*}
Direct computation shows 
\begin{align*}(g_\lambda^*)'(1-t)&=\left(\frac{1}{1-\alpha_1}\id_{(\alpha_1,1)}(t)+\lambda F^{-1}(t)\right)\id_{(0,1-u_{\alpha_1,\alpha_2,\lambda})\cup(\alpha_2,1)}(t)\\
&~+c_{\alpha_1,\alpha_2,\lambda}\id_{(1-u_{\alpha_1,\alpha_2,\lambda},\alpha_2)}(t),~t\in (0,1).
 \end{align*}
 Applying Theorem \ref{Th:main}, we obtain the desired result.
\qed

{\bf Proof of Theorem \ref{Th:main1}}.  In light of the conclusion in Theorem \ref{Th:main}, it suffices to show that $\sup_{G\in \mathcal M_{\epsilon}(\mu, \sigma)} \rho_g(G)=\sup_{G\in \mathcal M_{\epsilon}(\mu, \sigma)} \rho_{\hat{g}}(G)$.   As $g\leq \hat{g}$, we have $\sup_{G\in \mathcal M_{\epsilon}(\mu, \sigma)} \rho_g(G)\leq \sup_{G\in \mathcal M_{\epsilon}(\mu, \sigma)} \rho_{\hat{g}}(G)$. We next show the inverse inequality. 

 First, suppose $g(t)\neq \hat{g}(t)$ at finite number of points denoted by $\{t_1,\dots, t_m\}$ with $t_1<t_2<\dots<t_m$. Note that we have either  $\hat{g}(t_i)=\lim_{t\downarrow t_i}g(t)$ or  $\hat{g}(t_i)=\lim_{t\uparrow t_i}g(t)$. Let $\mathcal D_1=\{i: \hat{g}(t_i)=\lim_{t\downarrow t_i}g(t)\}$ and  $\mathcal D_2=\{i: \hat{g}(t_i)=\lim_{t\uparrow t_i}g(t)\}\setminus \mathcal D_1$.  For $G\in \mathcal M_{\epsilon}(\mu, \sigma)$ and $n\geq 1$, let
\begin{align*}G_n^{-1}(t)&=G^{-1}(t)\id_{\left((0,1)\setminus\cup_{i\in \mathcal D_1} (1-t_i-1/n^2, 1-t_i+1/n)\right)\setminus\cup_{i\in \mathcal D_2} (1-t_i-1/n, 1-t_i+1/n^2)}(t)\\
&+\sum_{i\in\mathcal D_1}\frac{n^2}{n+1}\int_{1-t_i-1/n^2}^{1-t_i+1/n}G^{-1}(s)\d s\id_{(1-t_i-1/n^2, 1-t_i+1/n)}(t)\\
&+\sum_{i\in\mathcal D_2}\frac{n^2}{n+1}\int_{1-t_i-1/n}^{1-t_i+1/n^2}G^{-1}(s)\d s\id_{(1-t_i-1/n, 1-t_i+1/n^2)}(t),~t\in (0,1).
\end{align*}
Note that if $n>\max(1/t_1, 1/(1-t_m), \max_{i=1}^{m-1} \frac{2}{t_{i+1}-t_{i}})$, then $(1-t_i-1/n^2, 1-t_i+1/n),~i\in \mathcal D_1$ and $(1-t_i-1/n, 1-t_i+1/n^2),~i\in \mathcal D_2$ are disjoint subintervals of $(0,1)$.
We denote the standard deviation of $G_n$ by $\sigma_n$ and let 
\begin{align}\label{eq:G}\widehat{G}_n^{-1}(t)=\mu+\frac{G_n^{-1}(t)-\mu}{\sigma_n}\sigma.
\end{align}
Note that $\lim_{n\to\infty}d_{W}(\widehat{G}_n,F)\leq\lim_{n\to\infty}(d_{W}(\widehat{G}_n,G_n)+d_{W}(G_n,F))\leq\sqrt{\epsilon}$. Hence, for any $\eta>0$, there exists $n_0>0$ such that $d_{W}(\widehat{G}_n,F)\leq \sqrt{\epsilon+\eta}$ for all $n\geq n_0$. This implies $\widehat{G}_{n}\in \mathcal M_{\epsilon+\eta}(\mu, \sigma)$ for all $n\geq n_0$.  Moreover, we have \begin{align*}\rho_{\hat{g}}(G)&=\rho_{\hat{g}}(G_n)+\sum_{i\in\mathcal D_1}\left(\int_{G^{-1}(1-t_i-1/n^2)}^{G^{-1}(1-t_i+1/n)} \hat{g}(1-G(x))\d x-\int_{G^{-1}(1-t_i-1/n^2)}^{G^{-1}(1-t_i+1/n)}\hat{g}(1-G_n(x))\d x\right)\\
&+\sum_{i\in\mathcal D_2}\left(\int_{G^{-1}(1-t_i-1/n)}^{G^{-1}(1-t_i+1/n^2)} \hat{g}(1-G(x))\d x-\int_{G^{-1}(1-t_i-1/n)}^{G^{-1}(1-t_i+1/n^2)}\hat{g}(1-G_n(x))\d x\right).
\end{align*}
Direct computation gives 
\begin{align*}\lim_{n\to\infty}\int_{G^{-1}(1-t_i-1/n^2)}^{G^{-1}(1-t_i+1/n)} \hat{g}(1-G(x))\d x&=\lim_{n\to\infty}\int_{G^{-1}(1-t_i-1/n)}^{G^{-1}(1-t_i+1/n^2)} \hat{g}(1-G(x))\d x\\
&=\hat{g}(t_i)(G_+^{-1}(1-t_i)-G^{-1}(1-t_i)).
\end{align*}
Moreover, it follows that  for $i\in\mathcal D_1$, $c_{i,n}:=\frac{n^2}{n+1}\int_{1-t_i-1/n^2}^{1-t_i+1/n}G^{-1}(s)\d s\to G_+^{-1}(1-t_i)$ as $n\to\infty$ and for $i\in\mathcal D_2$, $c_{i,n}:=\frac{n^2}{n+1}\int_{1-t_i-1/n}^{1-t_i+1/n^2}G^{-1}(s)\d s\to G^{-1}(1-t_i)$ as $n\to\infty$.
This implies that if $G_+^{-1}(1-t_i)>G^{-1}(1-t_i)$, then for  $i\in\mathcal D_1$,
\begin{align*}&\int_{G^{-1}(1-t_i-1/n^2)}^{G^{-1}(1-t_i+1/n)}\hat{g}(1-G_n(x))\d x\\
&=\hat{g}(t_i+1/n^2)(c_{i,n}-G^{-1}(1-t_i-1/n^2))+\hat{g}(t_i-1/n)(G^{-1}(1-t_i+1/n)-c_{i,n})\\
&~\to \hat{g}(t_i)(G_+^{-1}(1-t_i)-G^{-1}(1-t_i)),~n\to\infty,
\end{align*}
and for $i\in\mathcal D_2$,
\begin{align*}&\int_{G^{-1}(1-t_i-1/n)}^{G^{-1}(1-t_i+1/n^2)}\hat{g}(1-G_n(x))\d x\\
&=\hat{g}(t_i+1/n)(c_{i,n}-G^{-1}(1-t_i-1/n))+\hat{g}(t_i-1/n^2)(G^{-1}(1-t_i+1/n^2)-c_{i,n})\\
&~\to \hat{g}(t_i)(G_+^{-1}(1-t_i)-G^{-1}(1-t_i)),~n\to\infty.
\end{align*}
Note that the above conclusion also holds if $G_+^{-1}(1-t_i)=G^{-1}(1-t_i)$.
 Consequently, we have $\rho_{\hat{g}}(G)=\lim_{n\to\infty}\rho_{\hat{g}}(G_n)$. By the definition of $\widehat{G}_n$ and the properties of $\rho_g$, we have $\rho_g(\widehat{G}_n)=\rho_{\hat{g}}(\widehat{G}_n)$ and $\rho_{\hat{g}}(G_n)=\frac{\rho_{\hat{g}}(\widehat{G}_n)-\mu g(1)}{\sigma}\sigma_n+\mu g(1)$. Hence, we have \begin{align}\label{eq:Glimit}\rho_{\hat{g}}(G)=\lim_{n\to\infty}\rho_{\hat{g}}(G_n)
=\lim_{n\to\infty}\frac{\rho_g(\widehat{G}_n)-\mu g(1)}{\sigma}\sigma_n+\mu g(1)=\lim_{n\to\infty}\rho_g(\widehat{G}_n)\leq \sup_{G\in \mathcal M_{\epsilon+\eta}(\mu, \sigma)} \rho_g(G).
\end{align}
Therefore, we conclude that $\sup_{G\in \mathcal M_{\epsilon}(\mu, \sigma)} \rho_{\hat{g}}(G)\leq \sup_{G\in \mathcal M_{\epsilon+\eta}(\mu, \sigma)} \rho_g(G)$ for any $\eta>0$.

 Next, we consider the case $g\neq \hat{g}$ on infinite number of points denoted by $\{t_i,~i\geq 1\}$. Let $g_m(t)=\hat{g}(t)\id_{\{t_1,\dots, t_m\}}(t)+g(t)\id_{[0,1]\setminus\{t_1,\dots, t_m\}}(t)$. Note that $g\neq g_m$ on finite number of points $\{t_1,\dots, t_m\}$ and $g_m$ is either left- or right-continuous on those points.  For any  $G\in  \mathcal M_{\epsilon}(\mu, \sigma)$, let $\widehat{G}^{(m)}_{n}$ be as in \eqref{eq:G}. Applying the argument in \eqref{eq:Glimit}, we have for any $\eta>0$, $$\rho_{g_m}(G)=\lim_{n\to\infty}\rho_{g}(\widehat{G}^{(m)}_{n})\leq \sup_{G\in \mathcal M_{\epsilon+\eta}(\mu, \sigma)} \rho_g(G).$$ Note that $g_m(t)\uparrow \hat{g}(t)$ as $m\to\infty$ for all $t\in [0,1]$. If $\rho_g(G)>-\infty$, using the monotone convergence theorem, we have $\lim_{m\to\infty}\rho_{g_m}(G)=\rho_{\hat{g}}(G)$. Consequently, we have $\rho_{\hat{g}}(G)\leq \sup_{G\in \mathcal M_{\epsilon+\eta}(\mu, \sigma)} \rho_g(G)$. If $\rho_{\hat{g}}(G)=-\infty$, the previous conclusion holds obviously. Next, we focus on the case $\rho_{\hat{g}}(G)>-\infty$ and $\rho_{g}(G)=-\infty$. Let $G_n(x)=G(x)\id_{\{x>-n\}}$ for $n\geq 1$.
  We denote the mean and standard deviation of $G_n$ by $\mu_n$ and $\sigma_n$, respectively. Let $\overline{G}_n^{-1}(t)=\mu+\frac{G_n^{-1}(t)-\mu_n}{\sigma_n}\sigma$. For any $\eta>0$, there exists $n_1\geq 1$ such that $\overline{G}_n\in \mathcal M_{\epsilon+\eta/2}(\mu, \sigma)$ holds for all $n\geq n_1$. Using the above argument, we have $\rho_{g_m}(\overline{G}_n)\leq \sup_{G\in \mathcal M_{\epsilon+\eta}(\mu, \sigma)} \rho_g(G).$ Using the monotone convergence theorem, we have $\rho_{\hat{g}}(\overline{G}_n)=\lim_{m\to\infty}\rho_{g_m}(\overline{G}_n)\leq \sup_{G\in \mathcal M_{\epsilon+\eta}(\mu, \sigma)} \rho_g(G).$
  Letting $n\to\infty$, it follows that $\rho_{\hat{g}}(G)\leq \sup_{G\in \mathcal M_{\epsilon+\eta}(\mu, \sigma)} \rho_g(G).$
  
  By the arbitrary of $G$, we have
  $\sup_{G\in \mathcal M_{\epsilon}(\mu, \sigma)} \rho_{\hat{g}}(G)\leq \sup_{G\in \mathcal M_{\epsilon+\eta}(\mu, \sigma)} \rho_g(G)$ for any $\eta>0$. Therefore, we can conclude that
 $$\sup_{G\in \mathcal M_{\epsilon}(\mu, \sigma)} \rho_g(G)\leq \sup_{G\in \mathcal M_{\epsilon}(\mu, \sigma)} \rho_{\hat{g}}(G)\leq \lim_{\eta\downarrow 0}\sup_{G\in \mathcal M_{\epsilon+\eta}(\mu, \sigma)} \rho_g(G).$$
Let $l(\epsilon):=\sup_{G\in \mathcal M_{\epsilon}(\mu, \sigma)} \rho_{\hat{g}}(G)$. Then we have $l$ is increasing and \begin{align*}\lim_{\eta\downarrow 0}l(\epsilon-\eta)\leq \sup_{G\in \mathcal M_{\epsilon}(\mu, \sigma)} \rho_{g}(G)\leq l(\epsilon). \end{align*}

By  Lemma \ref{lem:1}, $d_W^2(F, H_\lambda)$ is continuous with respect to $\lambda$ over $(0,\infty)$ and direct computation shows that $d_W^2(F, H_\lambda)\geq (\mu_F-\mu)^2+(\sigma_F-\sigma)^2$ for $\lambda>0$, and $\lim_{\lambda\to \infty }d_W^2(F, H_\lambda)=(\mu_F-\mu)^2+(\sigma_F-\sigma)^2$. Let $\lambda_0=\sup\{\lambda>0: d_W^2(F, H_\lambda)\geq (\mu_F-\mu)^2+(\sigma_F-\sigma)^2+2\sigma_F\sigma(1-c_0)\}$. Then we have $$\{d_W^2(F, H_\lambda): \lambda>\lambda_0\}=[\mu_F-\mu)^2+(\sigma_F-\sigma)^2, (\mu_F-\mu)^2+(\sigma_F-\sigma)^2+2\sigma_F\sigma(1-c_0))$$ or $$((\mu_F-\mu)^2+(\sigma_F-\sigma)^2, (\mu_F-\mu)^2+(\sigma_F-\sigma)^2+2\sigma_F\sigma(1-c_0)).$$
If  $l$  is not continuous at $\epsilon_0$ for $(\mu_F-\mu)^2+(\sigma_F-\sigma)^2<\epsilon_0<(\mu_F-\mu)^2+(\sigma_F-\sigma)^2+2\sigma_F\sigma(1-c_0)$, then the monotonicity of $l$ implies  $\{l(\epsilon):(\mu_F-\mu)^2+(\sigma_F-\sigma)^2<\epsilon<(\mu_F-\mu)^2+(\sigma_F-\sigma)^2+2\sigma_F\sigma(1-c_0)\}$ is not an interval, which implies  $\rho_{\hat{g}}(H_\lambda)$ is not continuous over $(\lambda_0,\infty)$. Hence,
 the continuity of $\rho_{\hat{g}}(H_\lambda)$ with respect to $\lambda$ over $(0,\infty)$ implies the continuity of $l$ for $(\mu_F-\mu)^2+(\sigma_F-\sigma)^2<\epsilon<(\mu_F-\mu)^2+(\sigma_F-\sigma)^2+2\sigma_F\sigma(1-c_0)$. Hence, we have $\sup_{G\in \mathcal M_{\epsilon}(\mu, \sigma)} \rho_g(G)=\sup_{G\in \mathcal M_{\epsilon}(\mu, \sigma)} \rho_{\hat{g}}(G)$. Applying Theorem \ref{Th:main}, we obtain the conclusion of (i). 

For $\epsilon>(\mu_F-\mu)^2+(\sigma_F-\sigma)^2+2\sigma_F\sigma(1-c_0)$, we have $l(\epsilon)$ is a constant. Hence, $\sup_{G\in \mathcal M_{\epsilon}(\mu, \sigma)} \rho_g(G)=\sup_{G\in \mathcal M_{\epsilon}(\mu, \sigma)} \rho_{\hat{g}}(G)$ holds. For $\epsilon=(\mu_F-\mu)^2+(\sigma_F-\sigma)^2+2\sigma_F\sigma(1-c_0)$, the conclusion follows from Theorem 5 and Remark 2 of \cite{PWW20}.  Applying Theorem \ref{Th:main}, we obtain the results in (ii).
We complete the proof. \qed

 
{\bf Proof of Corollary \ref{Cor:2}}. (i) By Corollary \ref{Cor:1}, we have $$\VaR_\alpha^+(H_\lambda)=\mu+\sigma\frac{\frac{1+\lambda\int_{\alpha}^{1-t_{1-\alpha,\lambda}}F^{-1}(s)\d s}{1-\alpha-t_{1-\alpha,\lambda}}-a_{\lambda}}{b_{\lambda}}.$$
 Clearly, $a_\lambda$ is continuous for $\lambda\in (0,\infty)$. Note that $b_\lambda=\sqrt{\int_0^1((g_\lambda^*)'(t))^2\d t-(g(1)+\lambda\mu_F)^2}$. Hence, the continuity of $b_{\lambda}$ is implied
 by the uniform integrability of $\{((g_{\lambda}^*)'(t))^2, 0\leq \lambda\leq \lambda_0\}$ for any $\lambda_0>0$ and the fact that $(g_{\lambda}^*)'(t)\to (g_{\lambda_0}^*)'(t)$ a.e. as $\lambda\to\lambda_0$, showed in  the proof of Lemma \ref{lem:1}.  Using the expression of $(g_{\lambda}^*)'(t)$ given in (ii) of Corollary \ref{Cor:1}, the fact $(g_{\lambda}^*)'(t)\to (g_{\lambda_0}^*)'(t)$ a.e. as $\lambda\to\lambda_0$ also implies  $$\frac{1+\lambda\int_{\alpha}^{1-t_{1-\alpha,\lambda}}F^{-1}(s)\d s}{1-\alpha-t_{1-\alpha,\lambda}}\to \frac{1+\lambda\int_{\alpha}^{1-t_{1-\alpha,\lambda_0}}F^{-1}(s)\d s}{1-\alpha-t_{1-\alpha,\lambda_0}},~\text{as}~\lambda\to\lambda_0.$$ Hence, $\VaR_\alpha^+(H_\lambda)$ is continuous for $\lambda\in (0,\infty)$.  In light of Theorem \ref{Th:main1}, we obtain the desired result. 

 (ii) By Corollary \ref{Cor:1}, we have $\IQD^+(H_\lambda)=\frac{\frac{g_{\lambda}(\alpha)-g_{\lambda}(t_{\alpha,\lambda})}{\alpha-t_{\alpha,\lambda}}-\frac{g_{\lambda}(\hat{t}_{\alpha,\lambda})-g_{\lambda}(1-\alpha)}{\hat{t}_{\alpha,\lambda}-1+\alpha}}{b_\lambda}\sigma$. The continuity of  $b_{\lambda}$ is discussed in (i). Using the expression of $(g_{\lambda}^*)'(t)$ given in (iii) of Corollary \ref{Cor:1}, the continuity of  $\frac{g_{\lambda}(\alpha)-g_{\lambda}(t_{\alpha,\lambda})}{\alpha-t_{\alpha,\lambda}}$ and $\frac{g_{\lambda}(\hat{t}_{\alpha,\lambda})-g_{\lambda}(1-\alpha)}{\hat{t}_{\alpha,\lambda}-1+\alpha}$ are implied by the fact that $(g_{\lambda}^*)'(t)\to (g_{\lambda_0}^*)'(t)$ a.e. as $\lambda\to\lambda_0$ for any $\lambda_0>0$, showed in  the proof of Lemma \ref{lem:1}. Hence, $\IQD^+(H_\lambda)$ is continuous for $\lambda$ over $(0,\infty)$. Applying Theorem \ref{Th:main1}, we obtain the desired conclusion.

(iii) For $g=\hat{g}^{h_1,h_2}_{\beta,\alpha}$, we have  
$g_\lambda(t)=g^{h_1,h_2}_{\beta,\alpha}(t)\wedge h_2+\lambda \int_{1-t}^{1} F^{-1}(s)\d s+(1-h_2)\id_{[1-\alpha,1]}(t)$. The condition $\frac{h_1}{1-\beta}\geq \frac{h_2 - h_1}{\beta - \alpha}$ guarantees that $g^{h_1,h_2}_{\beta,\alpha}(t)\wedge h_2$ is concave over $(0,1)$. Hence, by the definition, we have
\begin{align*}g_\lambda^*(t)&=\left(g_\lambda(u_{\alpha,\beta,\lambda}^{h_1,h_2})+
\frac{g_\lambda(1-\alpha)-g_\lambda(u_{\alpha,\beta,\lambda}^{h_1,h_2})}{1-\alpha-u_{\alpha,\beta,\lambda}^{h_1,h_2}}(t-u_{\alpha,\beta,\lambda}^{h_1,h_2})\right)\id_{[u_{\alpha,\beta,\lambda}^{h_1,h_2}, 1-\alpha]}(t)\\
&~+g_\lambda(t)\id_{(0,u_{\alpha,\beta,\lambda}^{h_1,h_2})\cup (1-\alpha,1)}(t),
\end{align*}
which implies
\begin{align*}(g_\lambda^*)'(1-t)&=c_{\alpha,\beta,\lambda}^{h_1,h_2}\id_{(\alpha,1-u_{\alpha,\beta,\lambda}^{h_1,h_2})}(t)+\frac{h_1}{1-\beta}\id_{(\beta\vee (1-u_{\alpha,\beta,\lambda}^{h_1,h_2}),1)}(t)+\frac{h_2-h_1}{\beta-\alpha}\id_{(\beta\wedge (1-u_{\alpha,\beta,\lambda}^{h_1,h_2}),\beta)}(t)\\
 &+\lambda F^{-1}(t)\id_{(0,\alpha)\cup (1-u_{\alpha,\beta,\lambda}^{h_1,h_2},1)}(t),~t\in (0,1).
 \end{align*}
 Applying Theorem \ref{Th:main}, we can obtain the expression of $\sup_{G\in \mathcal M_{\epsilon}(\mu, \sigma)} \rho_{\hat{g}_{\beta,\alpha}^{h_1,h_2}}(G)$. Note that $\rho_{\hat{g}_{\beta,\alpha}^{h_1,h_2}}=w_1\ES_{\alpha}+w_2\ES_{\beta}+w_3\VaR_\alpha^+$ with some $w_1,w_2,w_3\geq 0$ satisfying $w_1+w_2+w_3=1$. Hence, \begin{align*}\sup_{G\in \mathcal M_{\epsilon}(\mu, \sigma)} \rho_{\hat{g}_{\beta,\alpha}^{h_1,h_2}}(G)&=\frac{\sigma}{b_\lambda}\left(\frac{w_1}{1-\alpha}\int_\alpha^1(g_\lambda^*)'(1-t)\d t+\frac{w_2}{1-\beta}\int_\beta^1(g_\lambda^*)'(1-t)\d t +w_3c_{\alpha,\beta,\lambda}^{h_1,h_2}\right)\\
 &+\mu-\frac{\sigma(1+\lambda\mu_F)}{b_{\lambda}}.
 \end{align*}
 The continuity of $b_{\lambda}$, $c_{\alpha,\beta,\lambda}^{h_1,h_2}$,  $\int_\alpha^1(g_\lambda^*)'(1-t)\d t$ and $\int_\beta^1(g_\lambda^*)'(1-t)\d t$ are implied
 by the uniform integrability of $\{((g_{\lambda}^*)'(t))^2, 0\leq \lambda\leq \lambda_0\}$ for any $\lambda_0>0$ and the fact that $(g_{\lambda}^*)'(t)\to (g_{\lambda_0}^*)'(t)$ a.e. as $\lambda\to\lambda_0$, showed in  the proof of Lemma \ref{lem:1}. 
  It follows from Theorem \ref{Th:main1} that $\sup_{G\in \mathcal M_{\epsilon}(\mu, \sigma)} \GlueVaR_{\beta,\alpha}^{h_1,h_2}(G)=\sup_{G\in \mathcal M_{\epsilon}(\mu, \sigma)} \rho_{\hat{g}_{\beta,\alpha}^{h_1,h_2}}(G)$. This completes the proof.
\qed

{\bf Proof of Proposition \ref{Prop:1}}. 
    If $g$ is concave distortion function, then $$g_{\lambda}^*(t)=g_{\lambda}(t)=g(t)+\lambda\int_{1-t}^{1} F^{-1}(s)ds,$$
    which implies $$(g_{\lambda}^*)'(t)= g'(t)+\lambda F^{-1}(1-t),~
    h_\lambda(t)
    =\mu+\sigma\frac{g'(1-t)+\lambda F^{-1}(t)-a_\lambda}{b_\lambda},$$
where $a_{\lambda}=g(1)+\lambda\mu_F$ and $b_{\lambda}=\sqrt{\int_{0}^{1}((g_{\lambda}^*)'(t))^2\d t-(g(1)+\lambda\mu_F)^2}$. By definition, $d_W(F,H_\lambda)=\sqrt{\epsilon}$ can be rewritten as  $$\epsilon = \mu^2_F+\sigma^2_F+\mu^2+\sigma^2-2 Cov(F^{-1}(V),h_\lambda(V))-2\mu\mu_F,$$
which implies $$Cov(F^{-1}(V),h_\lambda(V))=\frac{\mu_F^2+\sigma_F^2+\mu^2+\sigma^2-2\mu\mu_F-\epsilon}{2}=C_{\epsilon,F}\geq 0.$$  Using the expression of $h_\lambda$, we have \begin{align*}
    Cov(F^{-1}(V),h_\lambda(V)) =\frac{\sigma}{b_\lambda}Cov(F^{-1}(V), g'(1-V))+\lambda F^{-1}(V))
 =\frac{\sigma}{b_\lambda}(C_{g,F}+\lambda\sigma^2_F).
\end{align*}
Hence, we have $b_\lambda=\frac{\sigma (C_{g,F}+\lambda\sigma^2_F)}{C_{\epsilon, F}}$.
  Moreover, by definition, $b_\lambda=\sqrt{V_{g} +2\lambda C_{g,F}+\lambda^2\sigma_F^2}$. Hence, we have $
    C_{\epsilon,F}^2(V_{g} +2\lambda C_{g,F}+\lambda^2\sigma_F^2) =(\sigma C_{g,F}+\lambda \sigma \sigma_F^2)^2$, which can be simplified as
    $\lambda^2\sigma_F^2++2\lambda C_{g,F} +\frac{V_{g} C_{\epsilon,F}^2-\sigma^2 C_{g,F}^2}{C_{\epsilon,F}^2-\sigma^2\sigma_F^2} =0$.
Solving the quadratic equation, we have $$\lambda_\epsilon=\frac{-C_{g,F}+\sqrt{C_{g,F}^2-\sigma_F^2\frac{V_{g} C_{\epsilon,F}^2-\sigma^2 C_{g,F}^2}{C_{\epsilon,F}^2-\sigma^2\sigma_F^2}}}{\sigma_F^2}.$$
\qed
\section{Proofs of results in Section \ref{Sec:unimodel}}
In this section, we offer the proofs for all results in Section \ref{Sec:unimodel}. Let $\langle\cdot,\cdot\rangle$ denote the inner product of two functions in $\mathcal{F}_{U,\xi}^{-1}$.

{\bf Proof of Proposition \ref{Prop:2}}. By definition, $\gamma^\uparrow_\xi(u)\in\mathcal{F}_{U,\xi}^{-1}$. Since $\gamma^\uparrow_\xi$ is not constant, this implies  that $h^{\uparrow}_\xi(u)
\in \mathcal{F}_{U,\xi}^{-1}(\mu, \sigma)$. Consider any $h(u) \in \mathcal{F}_{U,\xi}^{-1}(\mu, \sigma)$, then $k(u) = \hat{b}_{\xi}\left(h(u)-\mu \right)/ \sigma + \hat{a}_{\xi} \in \mathcal{F}_{U,\xi}^{-1}$. Moreover, it holds that $\|\gamma^\uparrow_\xi\|_2 = \| k\|_2$. Thus, the stated assertion follows from the following equivalent inequalities:
\begin{align*}
	\| \gamma - \gamma^\uparrow_\xi\|_2  \leq \| \gamma - k\|_2
	\quad &\Leftrightarrow  \quad
	\langle \gamma , ~ \gamma^\uparrow_\xi \rangle \geq \langle \gamma , ~k \rangle
	\\
	&\Leftrightarrow  \quad \langle \gamma, ~h^\uparrow_\xi \rangle \geq  \langle \gamma, ~h \rangle\
	\\
	&\Leftrightarrow  \quad \rho_g(h^\uparrow_\xi)  \geq  \rho_g(h).
\end{align*}
Note that unless $\gamma_{\xi}^\uparrow = k$ the inequalities are strict, which implies that the solution is unique.  Using the above conclusion, we have
\begin{align*}\sup_{G^{-1} \in \mathcal{F}^{-1}_{U,\xi}(\mu, \sigma)} \rho_g(G^{-1})&=\left\langle\gamma, \mu+ \sigma \left(\frac{\gamma^\uparrow_\xi-\hat{a}_\xi}{\hat{b}_\xi} \right)\right\rangle\\
&=\mu g(1)+\frac{\sigma}{\hat{b}_\xi}\langle \gamma,\gamma^\uparrow_\xi-\hat{a}_\xi\rangle.
\end{align*}
By Corollary 2.3 of \cite{B65}, we have $\langle \gamma-\gamma^\uparrow_\xi,\gamma^\uparrow_\xi\rangle=0$ and $\langle \gamma-\gamma^\uparrow_\xi, c\rangle=0$ for any $c\in \R$. Therefore, we have
$$\langle \gamma,\gamma^\uparrow_\xi-\hat{a}_\xi\rangle=\langle \gamma-\gamma^\uparrow_\xi+\gamma^\uparrow_\xi,\gamma^\uparrow_\xi-\hat{a}_\xi\rangle=\langle\gamma^\uparrow_\xi,\gamma^\uparrow_\xi-\hat{a}_\xi\rangle=\langle\gamma^\uparrow_\xi-\hat{a}_\xi,\gamma^\uparrow_\xi-\hat{a}_\xi\rangle=\hat{b}_{\xi}^2.$$
Hence, $\sup_{G^{-1} \in \mathcal{F}^{-1}_{U,\xi}(\mu, \sigma)} \rho_g(G^{-1})\mu g(1)+\sigma \hat{b}_\xi.$
This completes the proof.
\qed

{\bf Proof of Proposition \ref{stepfunction}}.  Let $\gamma(u)=\sum_{i=1}^n y_i\id_{(x_{i-1},x_i)}(u)$ with $0=x_0<x_1<\dots<x_n=1$ and $y_i\in\R$. Without loss of generality, suppose $\xi\in (x_{i_0-1},x_{i_0})$. Then we choose one interval from $\{(x_{i-1},x_{i}), i\neq i_0\}\cup \{(x_{i_0-1},\xi), (\xi,x_{i_0})\}$ and denote it by $(a,b)$. Moreover, the value of $\gamma$ over this interval is denoted by $y$. Suppose $b\leq \xi$. Then $\gamma^\uparrow_\xi$ is concave over $(a,b)$. 

If $\gamma^\uparrow_\xi(a)\geq y$, 
define $\gamma_1(u)=\frac{\gamma^\uparrow_\xi(b)-\gamma^\uparrow_\xi(a)}{b-a}(u-a)+\gamma^\uparrow_\xi(a)$ for $u\in (a,b)$ and otherwise $\gamma_1(u)=\gamma^\uparrow_\xi(u)$. Clearly, $\gamma_1\in {F}^{-1}_{U,\xi}$ and  $\|\gamma_1-\gamma\|_2\leq \|\gamma^\uparrow_\xi-\gamma\|_2$. The uniqueness of $\gamma^\uparrow_\xi$ implies $\gamma^\uparrow_\xi=\gamma_1$. Hence, $\gamma^\uparrow_\xi$ is  linear over $(a,b)$.

If $\gamma^\uparrow_\xi(a)<y<\gamma^\uparrow_\xi(b)$, then there exists $a<c<b$ such that $\gamma^\uparrow_\xi(c)=0$. Define $$\gamma_2(u)=\left(\frac{\gamma^\uparrow_\xi(b)-\gamma^\uparrow_\xi(c)}{b-c}(u-b)+\gamma^\uparrow_\xi(b)\right)\bigwedge\left( (\gamma^\uparrow_\xi)_+'(a) (u-a)+\gamma^\uparrow_\xi(a)\right)$$ for $u\in (a,b)$ and otherwise $\gamma_2(u)=\gamma^\uparrow_\xi(u)$, where $(\gamma^\uparrow_\xi)_+'(a)$ is the right derivative of $\gamma^\uparrow_\xi$ at $u=a$.  Clearly, $\gamma_2\in {F}^{-1}_{U,\xi}$ and  $\|\gamma_2-\gamma\|_2\leq \|\gamma^\uparrow_\xi-\gamma\|_2$. Using the uniqueness of $\gamma^\uparrow_\xi$, we have $\gamma^\uparrow_\xi=\gamma_2$. Hence, $\gamma^\uparrow_\xi$ is piecewise linear over $(a,b)$ with two pieces.

If $\gamma^\uparrow_\xi(b)\leq y$, then define $$\gamma_3(u)=\left((\gamma^\uparrow_\xi)_+'(a) (u-a)+\gamma^\uparrow_\xi(a)\right)\bigwedge\left( (\gamma^\uparrow_\xi)_-'(b) (u-b)+\gamma^\uparrow_\xi(b)\right)$$ for $u\in (a,b)$ and otherwise $\gamma_3(u)=\gamma^\uparrow_\xi(u)$, where $(\gamma^\uparrow_\xi)_-'(b)$ is the left derivative of $\gamma^\uparrow_\xi$ at $u=b$.  Clearly, $\gamma_3\in {F}^{-1}_{U,\xi}$ and  $\|\gamma_3-\gamma\|_2\leq \|\gamma^\uparrow_\xi-\gamma\|_2$. Using the same argument as above, we conclude that $\gamma^\uparrow_\xi$ is piecewise linear over $(a, b)$ with two pieces.

Using the similar arguments, we can show the conclusion also holds for $a\geq \xi$, i.e., the case that $\gamma^\uparrow_\xi$ is convex over $(a,b)$.

The above proof shows that if $\xi=x_{i_0}$ for some $i_0=0,1,\cdots,n$, then $\gamma^\uparrow_\xi$ satisfies \eqref{stepprojection}. Moreover, it follows from Corollary 2.3 of \cite{B65} that $\langle \gamma-\gamma^\uparrow_\xi, 1\rangle=0$, which implies $\int_0^1\gamma^\uparrow_\xi(t)\d t=\int_0^1\gamma(t)\d t=g(1)$. Hence, we have \eqref{step0}.  Direct computation shows that
\begin{align*}
     &\int_0^1(\gamma(u)-\gamma^\uparrow_\xi(u))^2\d u\\
     &=\sum_{i=1}^n\left(\frac{(e_i^+)^3-(e_i^-)^3}{3c_i^-}+\frac{(e_{i}^++c_i^+(x_i-a_i))^3-(e_i^+)^3}{3c_i^+}\right)\\
     &=\sum_{i=1}^n\left(\frac{(a_i-x_{i-1})((e_i^+)^2+e_i^+e_i^-+(e_i^-)^2)}{3}+\frac{(x_i-a_i)((e_{i}^++c_i^+(x_i-a_i))^2+(e_{i}^++c_i^+(x_i-a_i))e_i^++(e_i^+)^2)}{3}\right),
 \end{align*}
 where $e_i^-=g(1)+\sum_{j=1}^n \frac{c_j^-(a_j-x_{j-1})^2+c_j^+(x_j-a_j)^2}{2}-\sum_{j=i}^{n} \left(c_j^-(a_j-x_{j-1})+c_j^+(x_j-a_j)\right)-y_i$ and $e_i^+=e_i^-+c_i^-(a_i-x_{i-1}),~i=1,\dots,n$. Hence, the optimal parameters are the minimizer of the above quantity over $\mathcal D_n$. This completes the proof.
\qed

{\bf Proof of Proposition \ref{Prop:approx}}. Since the projection operator is distance reducing with respect to the $L^2$-norm (Theorem 2.3 of \cite{B65}), it follows that  
$$
 \| \gamma^\uparrow_{\xi,n}  - \gamma^\uparrow_{\xi} \|_2\leq \|\gamma_n-\gamma\|_2.
$$
Moreover, note that $\mathcal F^{-1}_{U,\xi}$ is a closed convex cone and
$$\hat{b}_\xi^2-\hat{b}_{\xi,n}^2=\langle\gamma_\xi^\uparrow, \gamma_\xi^\uparrow\rangle-\langle\gamma_{\xi,n}^\uparrow, \gamma_{\xi,n}^\uparrow\rangle-(\hat{a}_{\xi}^2-\hat{a}_{\xi,n}^2).$$
By Corollary 2.3 of \cite{B65}, we have
$$\left|\langle\gamma_\xi^\uparrow, \gamma_\xi^\uparrow\rangle-\langle\gamma_{\xi,n}^\uparrow, \gamma_{\xi,n}^\uparrow\rangle\right|=|\langle\gamma, \gamma_\xi^\uparrow\rangle-\langle\gamma, \gamma_{\xi,n}^\uparrow\rangle|=|\langle \gamma, \gamma_\xi^\uparrow-\gamma_{\xi,n}^\uparrow\rangle|\leq \|\gamma\|_2\|\gamma_n-\gamma\|_2.$$
Moreover, direct computation shows
$$|\hat{a}_{\xi}^2-\hat{a}_{\xi,n}^2|\leq (\|\gamma\|_2+ \|\gamma_n\|_2)\|\gamma_n-\gamma\|_2.$$
Hence, we have $$|\hat{b}_\xi-\hat{b}_{\xi,n}|=\frac{\left|\hat{b}_\xi^2-\hat{b}_{\xi,n}^2\right|}{\hat{b}_\xi+\hat{b}_{\xi,n}}\leq \frac{(2\|\gamma\|_2+\|\gamma_n\|_2)\|\gamma_n-\gamma\|_2}{\hat{b}_{\xi}}.$$
Hence, we have
\begin{align*}
\|h^\uparrow_{\xi,n}  - h^\uparrow_{\xi} \|_2&=\sigma\left\|\frac{\gamma_{\xi,n}^\uparrow-\hat{a}_{\xi,n}}{\hat{b}_{\xi,n}}-\frac{\gamma_{\xi}^\uparrow-\hat{a}_\xi}{\hat{b}_{\xi}}\right\|_2\leq \sigma\frac{\| \gamma^\uparrow_{\xi,n}  - \gamma^\uparrow_{\xi} \|_2+|\hat{a}_{\xi,n}-\hat{a}_{\xi}|}{\hat{b}_\xi}+\sigma\frac{\hat{b}_{\xi,n}}{\hat{b}_{\xi,n}\hat{b}_{\xi}}|\hat{b}_{\xi,n}-\hat{b}_{\xi}|\\
&\leq \left(2+\frac{2\|\gamma\|_2+\|\gamma_n\|_2}{\hat{b}_{\xi}}\right)\frac{\sigma}{\hat{b}_\xi}\|\gamma_n-\gamma\|_2,
\end{align*}
and 
\begin{align*}
    |\rho_g(h^{\uparrow}_\xi)-\rho_g(h^{\uparrow}_{\xi,n})|\leq \|\gamma\|_2\|h^{\uparrow}_\xi-h^{\uparrow}_{\xi,n}\|_2 \leq \left(2+\frac{2\|\gamma\|_2+\|\gamma_n\|_2}{\hat{b}_{\xi}}\right)\frac{\sigma \|\gamma\|_2}{\hat{b}_\xi}\|\gamma_n-\gamma\|_2.
\end{align*}
This completes the proof. \qed

{\bf Proof of Example \ref{Example2} }
 Note that (i) is trivial. Next, we consider (ii).  If $\xi=1/2$, then $\gamma^\uparrow_\xi$ has the form of 
 $\left(c_1(u-1/2)+a-1\right)\id_{(0,1/2]}(u)+\left(c_2(u-1/2)+a-1\right)\id_{(1/2,1)}(u)$ with $a\in [0,2]$ and $c_1\geq 2a$ and $c_2\geq 2(2-a)$. Direct computation shows that 
 \begin{align*}\int_0^{1}(\gamma^\uparrow_\xi(u)-\gamma(u))^2\d u&=c_1^2\left(\int_0^{a/c_1}u^2\d u+\int_0^{1/2-a/c_1}u^2\d u\right)+c_2^2\left(\int_0^{(2-a)/c_2}u^2\d u+\int_0^{1/2-(2-a)/c_2}u^2\d u\right)\\
 &=\frac{c_1^2}{3}\left(\left(\frac{a}{c_1}\right)^3+\left(\frac{1}{2}-\frac{a}{c_1}\right)^3\right)+\frac{c_2^2}{3}\left(\left(\frac{2-a}{c_2}\right)^3+\left(\frac{1}{2}-\frac{2-a}{c_2}\right)^3\right):=G(c_1,c_2,a)
 \end{align*}
 with the convention $\frac{0}{0}=0$.

Let $f(c):=c^2\left(\left(\frac{a}{c}\right)^3+\left(b-\frac{a}{c}\right)^3\right)$, with $a>0, b>0$ and $c\geq \frac{a}{b}$. Note that $\frac{f(c)}{a^3}=\frac{1+\left(\frac{b}{a}c-1\right)^3}{c}$. Then the solution of 
$$\frac{\partial f(c)/a^3}{\partial c}=\frac{3\left(\frac{b}{a}c-1\right)^2\frac{b}{a}c-1-\left(\frac{b}{a}c-1\right)^3}{c^2}=\frac{\left(\frac{b}{a}c\right)^2(\frac{2b}{a}c-3)}{c^2}=0$$
 is $c=\frac{3}{2}\frac{a}{b}$, which is the unique minimizer of $f(c)$ over $c\geq \frac{a}{b}$.
 
 If $a$ is fixed, then the minimum of $G(c_1,c_2,a)$ is attained uniquely at $c_1=3a$ and $c_2=3(2-a)$. Then we have $G(3a,3(2-a),a)=\frac{1}{8}(a^2+(2-a)^2)$. The minimum is attained at $a=1$. Hence, we have $\gamma^\uparrow_\xi(u)=3(u-1/2),~u\in (0,1)$.
\qed

{\bf Proof of Lemma \ref{empty}}. Direct computation gives for $G^{-1}\in \mathcal F_{U,\xi}^{-1}(\mu,\sigma)$, 
\begin{align*}d_W^2(F^{-1},G^{-1})=(\mu-\mu_F)^2+(\sigma-\sigma_F)^2+2\sigma\sigma_F+2\mu\mu_F-2 \langle F^{-1},G^{-1}\rangle.
\end{align*}
For $G^{-1}\in\mathcal F_{U,\xi}^{-1}(\mu_F,\sigma_F^{\uparrow})$, by Theorem 2.2 of \cite{B65}, we have $$\langle F^{-1}-F_\xi^{-1,\uparrow}, F_\xi^{-1,\uparrow}-G^{-1}\rangle\geq 0,$$
which implies $$\langle F^{-1}, G^{-1}\rangle\leq \langle F^{-1}, F_\xi^{-1,\uparrow}\rangle+\langle F_\xi^{-1,\uparrow}, G^{-1}\rangle-\langle F_\xi^{-1,\uparrow}, F_\xi^{-1,\uparrow}\rangle.$$
It follows from Cauchy–Schwarz inequality that $\langle F_\xi^{-1,\uparrow}, G^{-1}\rangle\leq \langle F_\xi^{-1,\uparrow}, F_\xi^{-1,\uparrow}\rangle$. Hence, for $G^{-1}\in\mathcal F_{U,\xi}^{-1}(\mu_F,\sigma_F^{\uparrow})$, we have $\langle F^{-1}, G^{-1}\rangle\leq \langle F^{-1}, F_\xi^{-1,\uparrow}\rangle$, which implies for $G^{-1}\in \mathcal F_{U,\xi}^{-1}(\mu,\sigma)$,
$ \langle F^{-1},G^{-1}\rangle\leq  \langle F^{-1}, \frac{F_\xi^{-1,\uparrow}-\mu_{F}}{\sigma_F^\uparrow}\sigma+\mu\rangle$. Consequently, for $G^{-1}\in \mathcal F_{U,\xi}^{-1}(\mu,\sigma)$,
\begin{align} d_W^2(F^{-1},G^{-1})&\geq (\mu-\mu_F)^2+(\sigma-\sigma_F)^2+2\sigma\sigma_F+2\mu\mu_F-2 \langle F^{-1}, \frac{F_\xi^{-1,\uparrow}-\mu_{F}}{\sigma_F^\uparrow}\sigma+\mu\rangle\nonumber\\
&= (\mu-\mu_F)^2+(\sigma-\sigma_F)^2+2\sigma\sigma_F(1-Corr(F^{-1}(V),F_\xi^{-1,\uparrow}(V)).\label{boundss}
\end{align}
Let $\hat{c}_0=Corr(F^{-1}(V),F_\xi^{-1,\uparrow}(V))$.
We notice that, due to \eqref{boundss}, if $\epsilon<(\mu_F-\mu)^2+(\sigma_F-\sigma)^2+2\sigma_F\sigma(1-\hat{c}_0)$, then
$\mathcal F_{U,\xi}^{-1}(\mu,\sigma, \epsilon)=\varnothing$. If $\epsilon=(\mu_F-\mu)^2+(\sigma_F-\sigma)^2+2\sigma_F\sigma(1-\hat{c}_0)$, then  $\mathcal F_{U,\xi}^{-1}(\mu,\sigma)=\left\{\frac{F_\xi^{-1,\uparrow}-\mu_{F}}{\sigma_F^\uparrow}\sigma+\mu\right\}$, which is a singleton.  \qed

{\bf Proof of Lemma \ref{le:continuity}}. In light of Theorem 2.3 of \cite{B65}, we have $\|k_{\lambda_2,\xi}^\uparrow-k_{\lambda_1,\xi}^\uparrow\|_2^2\leq |\lambda_2-\lambda_1|^2(\mu_F^2+\sigma_F^2)$. This implies the continuity of $Corr(F^{-1}(V), k_{\lambda,\xi}^\uparrow(V))$ with respect to $\lambda$ over $[0,\infty)$. Note that $$\lim_{\lambda\to\infty}Corr(F^{-1}(V), k_{\lambda,\xi}^\uparrow(V))=\lim_{\lambda\downarrow 0}Corr(F^{-1}(V), \hat{k}_{\lambda,\xi}^\uparrow(V)),$$
where $\hat{k}_{\lambda,\xi}=\lambda\gamma+F^{-1}$ and $\hat{k}_{\lambda,\xi}^\uparrow$ is the $L_2$-projection of $\hat{k}_{\lambda,\xi}$ on $\mathcal F_{U,\xi}^{-1}$. Note that $\hat{k}_{0,\xi}^\uparrow=F_\xi^{-1,\uparrow}$. Applying Theorem 2.3 of \cite{B65} again, we have $\|\hat{k}_{\lambda,\xi}^\uparrow-F_\xi^{-1,\uparrow}\|_2^2\leq \lambda^2\|\gamma\|_2^2$. It follows  that 
$$\lim_{\lambda\downarrow 0}Corr(F^{-1}(V), \hat{k}_{\lambda,\xi}^\uparrow(V))=Corr(F^{-1}(V), F_\xi^{-1,\uparrow}(V)).$$
We complete the proof. \qed

{\bf Proof of Theorem \ref{Th:unimodal}}.  (i). In light of Lemma \ref{le:continuity}, for any $G^{-1}\in {F}^{-1}_{U,\xi}(\mu, \sigma, \epsilon)$, there exists $h_{\lambda,\xi}^{\uparrow}$ with $\lambda>0$ such that 
$d_W(h_{\lambda,\xi}^{\uparrow},F^{-1})=d_W(G^{-1},F^{-1})$. This implies $ \langle F^{-1}, h_{\lambda,\xi}^{\uparrow}\rangle=\langle F^{-1}, G^{-1}\rangle$.  Applying Proposition \ref{Prop:2}, we have $\sup_{G^{-1} \in {F}^{-1}_{U,\xi}(\mu, \sigma)} \langle\gamma+\lambda F^{-1}, G^{-1}\rangle=\langle\gamma+\lambda F^{-1}, h_{\lambda,\xi}^{\uparrow}\rangle$ and $h_{\lambda,\xi}^{\uparrow}$ is the unique maximizer. Hence, we have $\langle \gamma+\lambda F^{-1},h_{\lambda,\xi}^{\uparrow}\rangle>\langle \gamma+\lambda F^{-1}, G^{-1}\rangle$ if $G^{-1}\neq h_{\lambda,\xi}^{\uparrow}$, which implies $\rho_g(h_{\lambda,\xi}^{\uparrow})>\rho_g(G^{-1})$ if $G^{-1}\neq h_{\lambda,\xi}^{\uparrow}$ and $d_W(h_{\lambda,\xi}^{\uparrow},F^{-1})=d_W(G^{-1},F^{-1})$.

For $d_W(h_{\lambda_1,\xi}^{\uparrow},F^{-1})<d_W(h_{\lambda_2,\xi}^{\uparrow},F^{-1})$, we have $ \langle F^{-1}, h_{\lambda_2,\xi}^{\uparrow}\rangle<\langle F^{-1}, h_{\lambda_1,\xi}^{\uparrow}\rangle$. In light of Proposition \ref{Prop:2}, it follows that $\sup_{G^{-1} \in {F}^{-1}_{U,\xi}(\mu, \sigma)} \langle\gamma+\lambda_2 F^{-1}, G^{-1}\rangle=\langle\gamma+\lambda_2 F^{-1}, h_{\lambda_2,\xi}^{\uparrow}\rangle$, which implies $\langle\gamma+\lambda_2 F^{-1}, h_{\lambda_2,\xi}^{\uparrow}\rangle\geq \langle\gamma+\lambda_2 F^{-1}, h_{\lambda_1,\xi}^{\uparrow}\rangle$. Hence, we have $\rho_g(h_{\lambda_2,\xi}^{\uparrow})>\rho_g(h_{\lambda_1,\xi}^{\uparrow})$. Consequently, the conclusion in (i) holds.

(ii) Note that $\gamma_\xi^\uparrow$ is not a constant. Then we have $h^{\uparrow}_\xi\in {F}^{-1}_{U,\xi}(\mu, \sigma, \epsilon)$.  Applying Proposition \ref{Prop:2}, we obtain the first conclusion in (ii).  \qed

{\bf Proof of Lemma \ref{lem:unimodal}}.  Clearly, there exists a sequence of $h_n\in \mathcal{F}^{-1}_{U, [\xi_1,\xi_2]},~n\geq 1$ such that $\lim_{n\to\infty}\|\gamma-h_n\|_2=\inf_{h \in \mathcal{F}^{-1}_{U,[\xi_1,\xi_2]}}\|\gamma-h\|_2$. For any $u\in (0,1)$, we have $\{h_n(u), n\geq 1\}$ is a bounded sequence.  Hence, there is a subsequence $\{h_{n_m}(u), m\geq 1\}$ such that $\lim_{m\to\infty}h_{n_m}(u)$ exists. This implies that we could find a subsequence also denoted by $\{h_{n_m}(u), m\geq 1\}$ such that $\lim_{m\to\infty}h_{n_m}(u)$ exists for any $u\in (0,1)\cap \mathbb Q$, where $\mathbb Q$ is the set of all rational numbers. Define $h^*(u)=\lim_{m\to\infty}h_{n_m}(u)$ for all $u\in (0,1)\cap \mathbb Q$. Note that $h^*(u)$ is increasing on $(0,1)\cap \mathbb Q$. Define $\gamma_{\xi_1,\xi_2}^\uparrow(u)=\inf_{u'\in (u,1)\cap \mathbb Q}h^*(u')$ for $u\in (0,1)$. Then $\gamma_{\xi_1,\xi_2}^\uparrow$ is increasing and right-continuous on $(0,1)$.

Next, we show that $\gamma_{\xi_1,\xi_2}^\uparrow\in \mathcal{F}^{-1}_{U,[\xi_1,\xi_2]}$. For $u\in (0,1)$, if $\gamma_{\xi_1,\xi_2}^\uparrow$ is continuous at $u$, then by definition, we have  $\gamma_{\xi_1,\xi_2}^\uparrow(u)=\inf_{u'\in (u,1)\cap \mathbb Q}h^*(u')=\sup_{u'\in (0,u)\cap \mathbb Q}h^*(u')$. Hence, there exist  sequences $u_l\uparrow u$ and $u_l'\downarrow u$ with $u_l, u_l'\in (0,1)\cap \mathbb Q$ such that  $\gamma_{\xi_1,\xi_2}^\uparrow(u)=\lim_{l\to\infty}h^*(u_l)=\lim_{l\to\infty}h^*(u_l')$. It follows that
\begin{align*}|h_{n_m}(u)-\gamma_{\xi_1,\xi_2}^\uparrow(u)|&\leq |h_{n_m}(u)-h_{n_m}(u_l)|+|h_{n_m}(u_l)-h^*(u_l)|+|h^*(u_l)-\gamma_{\xi_1,\xi_2}^\uparrow(u)|\\
&\leq  |h_{n_m}(u_l')-h_{n_m}(u_l)|+|h_{n_m}(u_l)-h^*(u_l)|+|h^*(u_l)-\gamma_{\xi_1,\xi_2}^\uparrow(u)|\\
&= |h^*(u_l')-h^*(u_l)|+|h^*(u_l)-\gamma_{\xi_1,\xi_2}^\uparrow(u)|~\text{as}~ m\to\infty,\\
&=0 ~\text{as}~ l\to\infty.
\end{align*}
Hence, $\lim_{m\to\infty}h_{n_m}(u)=\gamma_{\xi_1,\xi_2}^\uparrow(u)$ for all  $u\in C$, where $C$ is the collection of all continuous points of $\gamma_{\xi_1,\xi_2}^\uparrow$.
 The inflection point of $h_{n_m}$ is denoted by $\xi_{m}$. As $\{\xi_m,~m\geq 3\}$ is bounded, there exists a subsequence such that the limit exists. Without loss of generality, we suppose $\lim_{m\to\infty} \xi_m=\xi\in [\xi_1,\xi_2]$. Let us first focus on $(\xi,1)$. For $u_i\in (\xi,1),~i=1,2,3,4$ with  $u_1<u_2\leq u_3<u_4$, there exist $u_{i,l}\in (\xi,1)\cap C$ satisfying $u_{1,l}<u_{2,l}\leq u_{3,l}<u_{4,l}$ and $u_{i,l}\downarrow u_i$ as $l\to\infty$. Note that $h_{n_m}$ is convex over $(\xi_m,1)$ and $\xi_m<u_1$ for all $m\geq m_0$. Hence, we have $\frac{h_{n_m}(u_{2,l})-h_{n_m}(u_{1,l})}{u_{2,l}-u_{1,l}}\leq \frac{h_{n_m}(u_{4,l})-h_{n_m}(u_{3,l})}{u_{4,l}-u_{3,l}}$ for all $m\geq m_0$. Letting $m\to\infty$, it follows that $\frac{\gamma_{\xi_1,\xi_2}^\uparrow(u_{2,l})-\gamma_{\xi_1,\xi_2}^\uparrow(u_{1,l})}{u_{2,l}-u_{1,l}}\leq \frac{\gamma_{\xi_1,\xi_2}^\uparrow(u_{4,l})-\gamma_{\xi_1,\xi_2}^\uparrow(u_{3,l})}{u_{4,l}-u_{3,l}}$. The letting $l\to\infty$ and using the fact that $\gamma_{\xi_1,\xi_2}^\uparrow$ is right-continuous, we have $\frac{\gamma_{\xi_1,\xi_2}^\uparrow(u_2)-\gamma_{\xi_1,\xi_2}^\uparrow(u_1)}{u_2-u_1}\leq \frac{\gamma_{\xi_1,\xi_2}^\uparrow(u_4)-\gamma_{\xi_1,\xi_2}^\uparrow(u_3)}{u_4-u_3}$, which implies $\gamma_{\xi_1,\xi_2}^\uparrow$ is convex over $(\xi,1)$. Hence, $\gamma_{\xi_1,\xi_2}^\uparrow$ is also continuous over $(\xi,1)$. We can similarly show that $\gamma_{\xi_1,\xi_2}^\uparrow$ is  continuous and concave over $(0,\xi)$.

If $\xi=\xi_1=0$ or $\xi=\xi_2=1$, then clearly, $\gamma_{\xi_1,\xi_2}^\uparrow$ is continuous over $(0,1)$. Next, we consider the case $\xi\in (0,1)$. For any $u_1, u_2\in (0,1)$ satisfying  $\frac{\xi}{2}<u_1<\xi<u_2<\frac{\xi+1}{2}$, using the fact that $h_{n_m}\in \mathcal{F}^{-1}_{U,\xi}$, we have 
\begin{align*}
&h_{n_m}(u_2)-h_{n_m}(u_1)\\
&\leq \max\left\{\frac{4(h_{n_m}(\xi/2)-h_{n_m}(\xi/4))}{\xi}, \frac{4(h_{n_m}((3+\xi)/4)-h_{n_m}((1+\xi)/2))}{1-\xi}\right\}(u_2-u_1).
\end{align*}
For the above inequality, letting $m\to\infty$, we have
\begin{align*}&\gamma_{\xi_1,\xi_2}^\uparrow(u_2)-\gamma_{\xi_1,\xi_2}^\uparrow(u_1)\\
&\leq \max\left\{\frac{4(\gamma_{\xi_1,\xi_2}^\uparrow(\xi/2)-\gamma_{\xi_1,\xi_2}^\uparrow(\xi/4))}{\xi}, \frac{4(\gamma_{\xi_1,\xi_2}^\uparrow((3+\xi)/4)-\gamma_{\xi_1,\xi_2}^\uparrow((1+\xi)/2))}{1-\xi}\right\}(u_2-u_1).
\end{align*}
Letting $u_2\downarrow \xi$ and $u_1\uparrow \xi$, we have $\gamma_{\xi_1,\xi_2}^\uparrow(\xi+)-\gamma_{\xi_1,\xi_2}^\uparrow(\xi-)=0.$ Hence, $\gamma_{\xi_1,\xi_2}^\uparrow$ is continuous at $\xi$. Combing the above conclusions, we obtain the continuity of  $\gamma_{\xi_1,\xi_2}^\uparrow$ over $(0,1)$.  

Note that the continuity of $\gamma_{\xi_1,\xi_2}^\uparrow$ over $(0,1)$ implies $\lim_{m\to\infty}h_{n_{m}}(u)=\gamma_{\xi_1,\xi_2}^\uparrow(u)$ for all $u\in (0,1)$.
It follows from Fatou's lemma  that $$\|\gamma-\gamma_{\xi_1,\xi_2}^\uparrow\|_2\leq \liminf_{n\to\infty}\|\gamma-h_{n_m}\|_2=\inf_{h \in \mathcal {F}^{-1}_{U,[\xi_1,\xi_2]}}\|\gamma-h\|_2<\infty.$$ Hence, $\|\gamma_{\xi_1,\xi_2}^\uparrow\|_2\leq \|\gamma-\gamma_{\xi_1,\xi_2}^\uparrow\|_2+\|\gamma\|_2<\infty$. Consequently, we have $\gamma_{\xi_1,\xi_2}^\uparrow\in \mathcal{F}^{-1}_{U,[\xi_1,\xi_2]}$ and $\gamma_{\xi_1,\xi_2}^\uparrow\in \arg\min_{h \in \mathcal{F}^{-1}_{U,[\xi_1,\xi_2]}}\|\gamma-h\|_2$. \qed


{\bf Proof of Proposition \ref{Prop:3}}. 
The proof is exactly the same as that of Proposition \ref{Prop:1}. The details are omitted. \qed

\section{Proofs of results in Section \ref{Sec:portfoliooptimisation}}
{\bf Proof of Proposition \ref{prop:uncertainty}}. In light of \cite{P07}, we have $$\left\{F_{\sum_{i=1}^n w_iX_i}: \mathbb E(X_i)=\mu_i, Cov(\mathbf X)=\Sigma_0\right\}=\M_{\infty}\left(\mathbf w^\top\boldsymbol\mu,\sqrt{\mathbf w^\top\Sigma_0\mathbf w}\right).$$
Moreover, it follows from Theorem 5 of \cite{MWW24} that 
$$\left\{F_{\sum_{i=1}^n w_iX_i}: d_{W}^{(n)}(F_{\mathbf X},F_{\mathbf X_0})\leq \sqrt{\epsilon}\right\}=\{G: d_W(F_{\mathbf w^\top\mathbf X_0},G)\leq \sqrt{\epsilon \|\mathbf w\|_2^2}\}.$$
Combining the above results, we obtain 
$$\M_{\mathbf w,\epsilon}=\M_{\epsilon\|\mathbf w\|_2^2}\left(\mathbf w^\top\boldsymbol\mu,\sqrt{\mathbf w^\top\Sigma_0\mathbf w}\right).$$
\qed

{\bf Explicit expressions for $V_{\mathbf w,\lambda}$ in Corollaries \ref{cor:4}-\ref{cor:6}}.

Note that in Corollary \ref{cor:4}, $V_{\mathbf w,\lambda}$ can be expressed in a more explicit form for computation as follows:
\begin{align*}V_{\mathbf w,\lambda}&=\left(\frac{1+\lambda\sqrt{\mathbf w^\top\Sigma_0\mathbf w}\int_{1-\alpha}^{1-t_{\alpha,\mathbf w,\lambda}} F_0^{-1}(s)\d s }{\alpha-t_{\alpha,\mathbf w,\lambda}}\right)^2(\alpha-t_{\alpha,\mathbf w,\lambda})\\
&+\left(\frac{\lambda\sqrt{\mathbf w^\top\Sigma_0\mathbf w}\int_{1-\hat{t}_{\alpha,\mathbf w,\lambda}}^{\alpha}F_0^{-1}(s)\d s-1}{\hat{t}_{\alpha,\mathbf w,\lambda}-1+\alpha}\right)^2(\hat{t}_{\alpha,\mathbf w,\lambda}-1+\alpha)\\
 &+\lambda^2\mathbf w^\top\Sigma_0\mathbf w\int_{(0,1-\hat{t}_{\alpha,\mathbf w,\lambda})\cup(\alpha,1-\alpha)\cup (1-t_{\alpha,\mathbf w,\lambda},1)}(F_0^{-1}(t))^2\d t.
 \end{align*}
 In Corollary \ref{cor:5}, $V_{\mathbf w,\lambda}$ can be expressed as follows:
\begin{align*}V_{\mathbf w,\lambda}&=\frac{\left(1+\lambda\sqrt{\mathbf w^\top\Sigma_0\mathbf w}\int_{\alpha}^{1-t_{1-\alpha,\mathbf w,\lambda}} F_0^{-1}(s)\d s\right)^2 }{1-\alpha-t_{1-\alpha,\mathbf w,\lambda}}\\
&\quad+\lambda^2\mathbf w^\top\Sigma_0\mathbf w\int_{(0,\alpha)\cup (1-t_{1-\alpha,\mathbf w,\lambda},1)}(F_0^{-1}(t))^2\d t-1.
 \end{align*}
 In Corollary \ref{cor:6}, an explicit expression for $V_{\mathbf w,\lambda}$ for computation is given by 
 \begin{align*}
V_{\mathbf w,\lambda}&=\frac{1}{1-\alpha-u_{\alpha,\beta,\mathbf w,\lambda}^{h_1,h_2}}\left(1-g_{\alpha,\beta}^{h_1,h_2}(u_{\alpha,\beta,\mathbf w,\lambda}^{h_1,h_2})+\lambda\sqrt{\mathbf w^\top\Sigma_0\mathbf w}\int_{\alpha}^{1-u_{\alpha,\beta,\mathbf w,\lambda}^{h_1,h_2}}F_0^{-1}(t)\d t\right)^2\\
&\quad +\frac{h_1^2((1-\beta)\wedge u_{\alpha,\beta,\mathbf w,\lambda}^{h_1,h_2})}{(1-\beta)^2}+\frac{(h_2-h_1)^2(\beta-1+u_{\alpha,\beta,\mathbf w,\lambda}^{h_1,h_2})_+}{(\beta-\alpha)^2}\\
&\quad+\lambda^2 \mathbf w^\top\Sigma_0\mathbf w\int_{(0,\alpha)\cup (1-u_{\alpha,\beta,\mathbf w,\lambda}^{h_1,h_2},1)}(F_0^{-1}(t))^2\d t-1.
 \end{align*}

\begin{thebibliography}{}


\bibitem[\protect\citeauthoryear{Allais}{1953}]{A53} 
Allais, M. (1953): Le Comportement de l’Homme Rationnel Devant le Risque: Critique des Axiomes et Postulats de l’Ecole Américaine. \emph{Econometrica}, \textbf{21}(4), 503--546.

\bibitem[\protect\citeauthoryear{{Basel Committee on Banking
Supervision}}{{BCBS}}{2019}]{BASEL19}
{BCBS} (2019).
{\em  Minimum Capital Requirements for Market Risk.  February 2019.}
Basel Committee on Banking
Supervision. Basel: Bank for International Settlements. Document number d457.


\bibitem[\protect\citeauthoryear{Belles-Sampera et al.}{2013}]{BGS13}
Belles-Sampera, J., Guillén, M. and Santolino, M. (2013). Beyond value-at-risk: GlueVaR distortion risk measures. \emph{Risk Analysis}, \textbf{34} (1), 121--134.

\bibitem[\protect\citeauthoryear{Bellini et al.}{2022}]{BFWW22}
Bellini, F., Fadina, T., Wang, R. and Wei, Y. (2022). Parametric measures of variability induced by risk measures. \emph{Insurance: Mathematics and Economics}, \textbf{106}, 270--284.

\bibitem[\protect\citeauthoryear{Ben-Tal et al.}{Ben-Tal et al.}{2009}]{BT09}
Ben-Tal, A., El Ghaoui, L. and Nemirovski, A. (2009). \emph{Robust Optimization}. Princeton University Press, Princeton.



\bibitem[\protect\citeauthoryear{Bernard et al.}{2020}]{BKV2020} 
Bernard, C., Kazzi, R. and Vanduffel, S (2020). Range value-at-risk bounds for unimodal distributions under partial information. \emph{Insurance: Mathematics and Economics}, \textbf{94}, 9--24. 


\bibitem[\protect\citeauthoryear{Bernard}{Bernard et al.}{2024}]{BPV24}
Bernard, C,  Pesenti, S. M. and  Vanduffel, S. (2024). Robust Distortion Risk Measures. \emph{Mathematical Finance}, \textbf{34}(3), 774--818.





\bibitem[\protect\citeauthoryear{Blanchet et al.}{2022}]{BCZ22}
Blanchet, J., Chen, L. and Zhou, X. (2022). Distributionally robust mean-variance portfolio selection with
Wasserstein distances. \emph{Management Science}, \textbf{68}(9), 6382--6410.
\bibitem[\protect\citeauthoryear{Blanchet and Murthy}{Blanchet and Murthy}{2019}]{BM19}
Blanchet, J. and Murthy, K. (2019). Quantifying distributional model risk via optimal transport. \emph{Mathematics
of Operations Research}, \textbf{44}(2), 565--600.


\bibitem[\protect\citeauthoryear{Brennan and Solanki}{Brennan and Solanki}{1981}]{BS81}
Brennan, M. and Solanki, R. (1981). Optimal portfolio insurance. \emph{J. Financ. Quant. Anal.} \textbf{16}, 279–300.

\bibitem[\protect\citeauthoryear{Brunk}{Brunk}{1965}]{B65}
Brunk, H. (1965). Conditional expectation given a $\sigma$-lattice and applications. \emph{The
Annals of Mathematical Statistics}, \textbf{36}(5), 1339--1350.

\bibitem[\protect\citeauthoryear{Cai et al.}{Cai et al.}{2025}]{CLM23}
Cai, J., Li, J. Y. M. and Mao, T. (2025). Distributionally robust optimization under distorted expectations. \emph{Operations Research}, \textbf{73}(2), 969--985.




\bibitem[\protect\citeauthoryear{Chen et al.}{2011}]{CHZ11}
Chen, L., He, S. and Zhang, S. (2011). Tight bounds for some risk measures, with applications to robust portfolio selection. \emph{Operations Research}, \textbf{59} (4), 847-865.

\bibitem[\protect\citeauthoryear{Cont et al.}{2010}]{CONT10}
Cont, R., Deguest, R. and Scandolo, G. (2010). Robustness and sensitivity analysis of risk measurement procedures. \emph{Quantitative finance}, \textbf{10} (6), 593-606.

\bibitem[\protect\citeauthoryear{European Central Bank}{2017}]{ECB17}
European Central Bank. (2017). Guidance on model risk management. European Central Bank.



\bibitem[\protect\citeauthoryear{El Ghaoui et al.}{2003}]{E03}
El Ghaoui, L., Oks, M. and Oustry, F. (2003). Worst-case value-at-risk and robust portfolio optimization: A conic programming approach. \emph{Operations Research}, \textbf{51(4)}, 543-556.


\bibitem[\protect\citeauthoryear{Esfahani and Kuhn}{2018}]{EK18}
Esfahani, PM. and Kuhn, D. (2018). Data-driven distributionally robust optimization using the Wasserstein metric: Performance guarantees and
tractable reformulations. \emph{Math. Programming} \textbf{17} 1(1):115–166.


\bibitem[\protect\citeauthoryear{F\"ollmer and Schied}{F\"ollmer and Schied}{2016}]{FS16} 
F\"ollmer, H.~and Schied, A.~(2016). \emph{Stochastic Finance. An Introduction in Discrete Time}.  {Walter de Gruyter, Berlin}, Fourth Edition.

\bibitem[\protect\citeauthoryear{Gilboa and Schmeidler}{Gilboa and Schmeidler}{1987}]{GB1987}
Gilboa, I. and Schmeidler, D. (1989). Maxmin expected utility with non-unique prior. \emph{Journal of Mathematical Economics}, \textbf{18}(2), 141-153.


\bibitem[\protect\citeauthoryear{Goldstein}{Goldstein et al.}{2008}]
{GJS08} Goldstein, D., Johnson, E. and  Sharpe, W. (2008). Choosing outcomes versus choosing products: consumer-focused retirement investment advice. \emph{J. Consum. Res.} \textbf{35}, 440–456 (2008).

\bibitem[\protect\citeauthoryear{Hansen, L. P., and Sargent, T. J.}{Hansen and Sargent}{2001}]{HS01}
Hansen, L. P. and Sargent, T. J. (2001). Robust control and model uncertainty. \emph{American Economic Review}, \textbf{91} (2), 60-66.







\bibitem[\protect\citeauthoryear{Lauzier}{Lauzier et al.}{2025}]{LLW23}
Lauzier, J.G., Lin, L. and Wang, R. (2025).
Risk sharing, measuring variability, and distortion riskmetrics. \emph{Mathematical Finance}, \textbf{36}(2), 330--351.

\bibitem[\protect\citeauthoryear{Li}{Li}{2018}]{L18}
Li, Y. (2018). Closed-form solutions for worst-case law invariant risk measures with application to robust
portfolio optimization. \emph{Operations Research}, \textbf{66}(6), 1533--1541.


\bibitem[\protect\citeauthoryear{Li et al.}{Li et al.}{2018}]{LSWY18}
Li, L., Shao, H., Wang, R. and Yang, J. (2018). Worst-case Range Value-at-Risk with partial information.
\emph{SIAM Journal on Financial Mathematics}, \textbf{9}(1), 190--218.



\bibitem[\protect\citeauthoryear{Liu et al.}{Liu et al.}{2022}]{LMWW22}
Liu, F.,  Mao, T., Wang, R.  and Wei, L. (2022). Inf-convolution, optimal allocations, and model uncertainty for tail risk measures. \emph{Mathematics of Operations Research}, \textbf{47}(3), 2494--2519.

\bibitem[\protect\citeauthoryear{Mao et al.}{Mao et al.}{2025}]{MWW24}
Mao, T., Wang, R. and Wu, Q. (2025). Model Aggregation for Risk Evaluation and Robust Optimization. \emph{Management Science}, forthcoming.




\bibitem[\protect\citeauthoryear{Pesenti et al.}{2025}]{PWW20}
Pesenti, S. M., Wang, Q. and Wang, R. (2025). Optimizing distortion riskmetrics with distributional uncertainty.
\emph{Mathematical Programming}, \textbf{213}, 51--106.


\bibitem[\protect\citeauthoryear{Popescu}{2005}]{P05}
Popescu, I. (2005). A Semidefinite Programming Approach to Optimal-Moment Bounds for Convex Classes
of Distributions. \emph{Mathematics of Operations Research}, \textbf{30}(3), 632--657.

\bibitem[\protect\citeauthoryear{Popescu}{2007}]{P07}
Popescu, I. (2007). Robust mean-covariance solutions for stochastic optimization. \emph{Operations Research}, \textbf{55}(1), 98--112.

\bibitem[\protect\citeauthoryear{Quiggin}{1982}]{Q82}
Quiggin, J. (1982). A theory of anticipated utility. \emph{Journal of Economic Behavior \& Organization}, \textbf{3}(4), 323--343.

\bibitem[\protect\citeauthoryear{Roese and Olson}{1995}]{R095}
Roese, N. J. and Olson, J. M. (1995). Counterfactual thinking: A critical overview. \emph{Psychological Bulletin}, \textbf{118}(1), 1-19.


\bibitem[\protect\citeauthoryear{Scarf}{Scarf}{1958}]{S58}
Scarf, H. E. (1958). Studies in the mathematical theory of inventory and production. In: Arrow, K.J., Karlin,
S., Scarf, H.E. (eds.) A Min–Max Solution of an Inventory Problem, pp. 201–209. Stanford University
Press, Stanford.
\bibitem[\protect\citeauthoryear{Shalit and Yitzhaki}{1984}]{SY84}
Shalit, H. and Yitzhaki, S. (1984). Mean‐Gini, portfolio theory, and the pricing of risky assets. The Journal of Finance, \textbf{39}(5), 1449--1468.

\bibitem[\protect\citeauthoryear{Shao and Zhang}{2023}]{SZ23}
Shao, H. and Zhang, Z.G. (2023)
Distortion risk measure under parametric ambiguity. \emph{European Journal of Operations Research}, \textbf{331}, 1159--1172.

\bibitem[\protect\citeauthoryear{Shao and Zhang}{2024}]{SZ24}
Shao, H. and Zhang, Z.G. (2024)
Extreme-Case Distortion Risk Measures: A Unification and Generalization of Closed-Form Solutions. \emph{Mathematics of Operations Research}, \textbf{49}(4), 2341--2355.

\bibitem[\protect\citeauthoryear{Starmer}{Starmer}{2000}]{S00}
Starmer, C. (2000). Developments in non-expected utility theory: The hunt for a descriptive theory of choice under risk. \emph{Journal of Economic Literature}, \textbf{38}(2), 332-382. 

\bibitem[\protect\citeauthoryear{Tversky and Kahneman}{1973}]{KT73}
Tversky, A. and Kahneman, D. (1973). Availability: A heuristic for judging frequency and probability. \emph{Cognitive Psychology}, \textbf{5}(2), 207-232.

\bibitem[\protect\citeauthoryear{Tversky and Kahneman}{1992}]{TK92}
Tversky, A. and Kahneman, D. (1992). Advances in prospect theory: Cumulative representation of uncertainty. \emph{Journal of Risk and Uncertainty}, \textbf{5}, 297-323.


\bibitem[\protect\citeauthoryear{Wang et al.}{Wang et al.}{2020}]{WWW20a}
Wang, Q., Wang, R. and Wei, Y. (2020). Distortion riskmetrics on general spaces. \emph{ASTIN Bulletin: The Journal of the IAA}, \textbf{50}(3), 827-851.

\bibitem[\protect\citeauthoryear{Wang et al.}{Wang et al.}{2020}]{WWW20}
Wang, R., Wei, R. and  Willmot, G.E. (2020). Characterization, robustness, and aggregation of signed Choquet integrals. \emph{Mathematics of Operations Research}, \textbf{45}(3), 993-1015.


\bibitem[\protect\citeauthoryear{Yaari}{Yaari}{1987}]{Y87}
Yaari, M. E. (1987). The dual theory of choice under risk. \emph{Econometrica}, \textbf{55}(1), 95--115.



































































\bibitem[\protect\citeauthoryear{Zymler et al.}{Zymler et al.}{2013}]{zym13}
Zymler, S., Kuhn, D. and Rustem, B. (2013). Distributionally robust joint chance constraints with second-order moment information. \emph{Mathematical Programming}, \textbf{137}, 167-198.
 






\end{thebibliography}
\end{document}